\documentclass[a4paper,11pt]{article}
\usepackage{jheppub} % for details on the use of the package, please
\usepackage[T1]{fontenc} % if needed
\usepackage{graphicx}
\usepackage{tikz}
\usepackage{indentfirst}
\usepackage{slashed}
\usepackage{amsmath,amssymb,multirow,xspace,slashed,array,booktabs}
\usepackage{physics}
\usepackage{indentfirst}
\usepackage{url}
\usepackage{bm}
\usepackage{braket}
\usepackage[utf8]{inputenc}
\DeclareUnicodeCharacter{2212}{-}
\usepackage[colorlinks=true,urlcolor=blue,anchorcolor=blue,citecolor=blue,filecolor=blue,linkcolor=blue,menucolor=blue,pagecolor=blue]{hyperref}
\usepackage{natbib}
\setcitestyle{square, comma, numbers,sort&compress, super}
\usepackage{subfigure, placeins}
\usepackage{booktabs}
\usepackage{blindtext, rotating}
\usepackage{appendix}
\usepackage{afterpage}
\usepackage{enumitem}
\usepackage{marvosym}
\usepackage{bbold}
\usepackage{verbatim}
\usepackage{soul} %scratch out text
\usepackage[normalem]{ulem}
\usepackage{pifont}% http://ctan.org/pkg/pifont
\usepackage{cancel}
\newcommand{\KF}[1]{{\color{red}}}
\newcommand{\ynone}{y_{N_1}}
\newcommand{\yntwo}{y_{N_2}}
\newcommand{\ynth}{y_{N_3}}
\newcommand{\yeone}{y_{E_1}}

\newcommand{\yeth}{y_{E_3}}

\newcommand{\newc}{\newcommand}
\newc{\gsim}{\lower.7ex\hbox{$\;\stackrel{\textstyle>}{\sim}\;$}}
\newc{\lsim}{\lower.7ex\hbox{$\;\stackrel{\textstyle<}{\sim}\;$}}
\newc{\gev}{\,{\rm GeV}}
\newc{\mev}{\,{\rm MeV}}
\newc{\kev}{\,{\rm keV}}
\newc{\tev}{\,{\rm TeV}}
\newc{\MHT}{$H_T^{\text{miss}}$}
\newc{\MET}{$\slashed{E}_T$}
\newc{\MTT}{$M_{T2}$}

\def\ln{\mathop{\rm ln}}

\def\Tr{\mathop{\rm Tr}}
\def\Im{\mathop{\rm Im}}

\newc{\mz}{M_Z}
\newc{\mpl}{M_*}
\newc{\mw}{m_{\rm weak}}
\newc{\nr}[1]{N^c_R{}_{#1}}

\newcommand{\rmd}{\mathrm{d}}

\renewcommand{\dag}{\dagger}

\newcommand{\hvev}[1]{ \left\langle {#1} \right\rangle }

\title{Electroweak-like Baryogenesis with  New Chiral Matter}
\author[a]{Kohei Fujikura}
\author[b]{Keisuke Harigaya}
\author[c]{Yuichiro Nakai}
\author[d]{Isaac R. Wang}

\affiliation[a]{Department of Physics, Tokyo Institute of Technology,\\
2-12-1 Ookayama, Meguro-ku, Tokyo 152-8551, Japan}
\affiliation[b]{School of Natural Sciences, Institute for Advanced Study,\\ Princeton, New Jersey, 08540 USA}
\affiliation[c]{Tsung-Dao Lee Institute and School of Physics and Astronomy, \\Shanghai
Jiao Tong University, 800 Dongchuan Road, Shanghai, 200240 China}
\affiliation[d]{Department of Physics and Astronomy, Rutgers University,\\ Piscataway, New Jersey, 08854 USA}

\emailAdd{fuji@th.phys.titech.ac.jp}
\emailAdd{keisukeharigaya@ias.edu}
\emailAdd{ynakai0930@gmail.com}
\emailAdd{isaac.wang@rutgers.edu}
\abstract{
  We propose a framework where a phase transition associated with a gauge symmetry breaking that
  occurs (not far) above the electroweak scale sets a stage for baryogenesis
  similar to the electroweak baryogenesis in the Standard Model.
  A concrete realization utilizes the breaking of $SU(2)_R \times U(1)_X \rightarrow U(1)_Y$.
  New chiral fermions charged under the extended gauge symmetry have nonzero lepton numbers,
  which makes the $B-L$ symmetry anomalous.
  The new lepton sector contains a large flavor-dependent CP violation, similar to the
  Cabibbo-Kobayashi-Maskawa phase, without inducing sizable electric dipole moments of the Standard Model particles.
  A bubble wall dynamics associated with the first-order phase transition and $SU(2)_R$ sphaleron processes
  generate a lepton asymmetry, which is transferred into a baryon asymmetry via the ordinary electroweak sphaleron process.
  Unlike the Standard Model electroweak baryogenesis, the new phase transition can be of the strong first order
  and the new CP violation is not significantly suppressed by Yukawa couplings,
  so that the observed asymmetry can be produced. 
  The model can be probed by collider searches for new particles and the observation of gravitational waves.
  One of the new leptons becomes a dark matter candidate.
  The model can be also embedded into a left-right symmetric theory to solve the strong CP problem.
  }
\begin{document}
\maketitle
\flushbottom
\newpage
%%%%%%%%%%%%%%%%%%%%%%%%%%%%%%%%%%%%%%%%%%%%%%%%

%%%%%%%%%%%%%%%%%%%%%%%%%%%%%%%%%%%%%%%%%%%%%%%%
\section{Introduction}
%%%%%%%%%%%%%%%%%%%%%%%%%%%%%%%%%%%%%%%%%%%%%%%%

The question of why the Universe contains more matter than antimatter remains unsolved. 
The successful Big Bang Nucleosynthesis (BBN) and the observed spectrum of the cosmic microwave background require a baryon-to-entropy density ratio of order $10^{-10}$.
It is widely believed that the Universe experienced accelerated expansion called inflation that diluted a primordial baryon asymmetry if it existed; the observed baryon asymmetry must be created after the inflation and before the BBN.
Sakharov pointed out three conditions to be satisfied to create baryon asymmetry: baryon number violation, C and CP violation, and out-of-thermal equilibrium processes~\cite{Sakharov:1967dj}.

The Standard Model (SM) contains a baryon number violation process
from the quantum anomaly of the baryon number symmetry under the electroweak gauge interactions~\cite{tHooft:1976rip,tHooft:1976snw,Manton:1983nd,Klinkhamer:1984di,Kuzmin:1985mm}.
The SM quark sector also contains a CP-violating parameter,
the Cabibbo-Kobayashi-Maskawa (CKM) phase~\cite{Cabibbo:1963yz,Kobayashi:1973fv}.
Furthermore, the out-of-thermal equilibrium condition is realized if the electroweak phase transition is of the strong first order.
These facts naturally lead to the possibility of electroweak baryogenesis in the SM~\cite{Kuzmin:1985mm}.
Assuming a strong first order electroweak phase transition,
Farrar and Shaposhnikov estimated the baryon asymmetry produced
by a moving bubble wall on the thermal background, in which quarks and leptons obtain thermal masses
and behave as quasiparticles~\cite{Farrar:1993sp,Farrar:1993hn}.
QCD decoherence effects on the quasiparticles, which destroy quantum mechanical CP-violating processes, were included in~\cite{Gavela:1993ts,Huet:1994jb,Gavela:1994ds,Gavela:1994dt}.
It turned out that the resulting baryon asymmetry is suppressed by the tiny Yukawa couplings as well as the Glashow-Iliopoulos-Maiani (GIM) mechanism, and is much smaller than the observed value.
Moreover, the electroweak phase transition with the observed Higgs boson mass
is not of the strong first order as confirmed by lattice studies~\cite{Kajantie:1995kf,Kajantie:1996qd,Rummukainen:1998as}.
Therefore, the SM does not address the question of the baryon asymmetry in the Universe.

Physics beyond the SM may assist electroweak baryogenesis.
It may provide a new source of CP violation and new particles triggering a strong first-order electroweak phase transition.
(See e.g. refs.~\cite{Trodden:1998ym,Morrissey:2012db} for general reviews of electroweak baryogenesis.)
For example, the Minimal Supersymmetric Standard Model (MSSM) contains the superpartner of the top quark
that enables a first order phase transition, if it is sufficiently light,
and new CP-violating phases associated with supersymmetry breaking parameters~\cite{Carena:1996wj,Delepine:1996vn,Carena:1997ki,Quiros:1999tx,Cline:2000kb,Huber:2001xf,Carena:2002ss,Lee:2004we,Carena:2008rt,Carena:2008vj,Cirigliano:2009yd,Carena:2012np}.
In two-Higgs-doublet models, a new phase is induced by the Higgs potential,
and an additional Higgs field can make the phase transition of the strong first order~\cite{Turok:1990in,Turok:1990zg,Cohen:1991iu,Nelson:1991ab,Cohen:1992yh,Cline:1995dg,Cline:1996mga,Fromme:2006cm,Tranberg:2012jp,Andersen:2017ika}.
A great attraction/crisis of these electroweak baryogenesis scenarios in physics beyond the SM lies in their testability.
In fact, the Large Hadron Collider (LHC) has already excluded most of the parameter space in the MSSM that can realize electroweak baryogenesis~\cite{Curtin:2012aa}.
Furthermore, new sources of flavor-diagonal CP violation are severely constrained
by measurements of electric dipole moments (EDMs).
In particular, the ACME experiment~\cite{Andreev:2018ayy} has ruled out a wide range of new physics with CP violation
at energies even higher than what the LHC can test~\cite{Nakai:2016atk,Cesarotti:2018huy}.
This situation drives us to explore a possibility of baryogenesis
that utilizes a first order phase transition other than the electroweak one.\footnote{
Another approach is to make the electroweak phase transition occur at a higher temperature
\cite{Ishikawa:2014tfa,Baldes:2018nel,Glioti:2018roy}.}
Refs.~\cite{Shelton:2010ta,Carena:2019xrr,Hall:2019ank} discussed models of baryogenesis
through a phase transition in a dark sector that contains (usually light) new particles feebly interacting with the SM particles.
Searching for such dark sector particles belongs to {\it the high-intensity frontier}.

In this paper, we pursue a possibility that
a first-order phase transition associated with a new gauge symmetry breaking at an energy scale higher than the electroweak scale
sets a stage for baryogenesis and the sector responsible for baryogenesis couples to the SM with $O(1)$ couplings.
Ref.~\cite{Shu:2006mm} uses an $SU(2)\times SU(2)$ gauge symmetry to which $SU(2)_L$ is embedded and Ref.~\cite{Fornal:2017owa} considers an $SU(2)$ gauge symmetry under which leptons are charged. Ref.~\cite{Davoudiasl:2016yfa} also mentions a possibility to use an extra $SU(2)$ gauge symmetry for baryogenesis, although the amount of the produced baryon asymmetry is not estimated and it is not clear if their setup actually works.
If the electroweak scale is naturally obtained by some dynamics, it is plausible that the higher scale is determined by the same dynamics and not far above the electroweak scale. 
Some new particles are charged under the SM gauge symmetry and heavier than the SM particles; they may be within the LHC reach and at {\it the high-energy frontier}. 
We consider a minimal extension of the SM gauge symmetry with a new non-abelian gauge symmetry, $SU(3)_C\times SU(2)_L\times SU(2)_R\times U(1)_X$,
where the right-handed fermions in the SM (plus the right-handed neutrinos) transform as doublets under the $SU(2)_R$~\cite{Pati:1974yy,Mohapatra:1974gc,Senjanovic:1975rk}.
The spontaneous breaking of $SU(2)_R\times U(1)_X \rightarrow U(1)_Y$ is induced by a new Higgs field
whose potential is largely undetermined and
we may assume that the associated phase transition is of the strong first order. 
To generate the baryon asymmetry during the new phase transition while evading the wash-out by the electroweak sphaleron processes,
the $B-L$ symmetry must be (effectively) anomalous under the extended gauge interactions.
This is achieved by introducing a new chiral sector that has baryon or lepton numbers, is charged under the $SU(2)_R\times U(1)_X$, and has a structure different from the SM sector.
We introduce such a chiral lepton sector borrowing the setup in ref.~\cite{Kim:2017qxo},
discussed in the context of a dark sector,
which contains three generations of new chiral fermions.
We also provide a class of models where new particles are vector-like under the gauge symmetry but chiral under a discrete symmetry, so that the $B-L$ symmetry is effectively anomalous.
The new chiral sector accommodates a flavor-dependent CP-violating phase similar to the CKM phase.
This CP violation is not necessarily suppressed by Yukawa couplings unlike the SM quark sector.
We can apply the calculation technique of the SM electroweak baryogenesis to our model, appropriately incorporating damping effects on quasiparticles.
We find that the model can generate the observed baryon asymmetry in contrast to the SM.

In the setup we consider,
the left-right symmetry is explicitly broken since the new chiral sector is charged under $SU(2)_R$ but not under $SU(2)_L$, and the $SU(2)_R\times U(1)_X$ breaking Higgs has a quartic coupling much smaller than that of the SM Higgs. However, we show that the model can be embedded into a left-right symmetric theory with spontaneous breaking of the left-right symmetry. By combining the left-right symmetry with the space-time parity, the strong CP problem can be solved~\cite{Beg:1978mt,Mohapatra:1978fy}. This will be an advantage over other models based on dark gauge symmetries.

We may introduce a $Z_2$ symmetry to the new chiral sector, so that one of the new leptons is absolutely stable and a dark matter candidate. We briefly discuss how the correct dark matter abundance is obtained, leaving detailed discussions on dark matter phenomenology to a future work.

The rest of the paper is organized as follows.
Sec.~\ref{sec:framework} presents our model, where
the global $B-L$ symmetry is anomalous under the extended gauge symmetry
and the new chiral lepton sector provides a new source of CP violation.
The mechanism of baryogenesis is also described.
In sec.~\ref{baryogenesis}, after clarifying the sphaleron decoupling condition that must be fulfilled to generate a non-zero baryon asymmetry, we calculate the amount of the $B-L$ asymmetry
by considering the reflection problem of the new leptons by the bubble wall.
In sec.~\ref{pheno}, we discuss phenomenology of the model,
including collider searches, EDM measurements, and the production of gravitational waves (GWs).
We find that a GW signal
within the reach of future experiments may be produced for the model parameters that realize the observed baryon asymmetry.
Sec.~\ref{conclusions} is devoted to conclusions and discussions.

%%%%%%%%%%%%%%%%%%%%%%%%%%%%%%%%%%%%%%%%%%%%%%%%
\section{The model}
  \label{sec:framework}
%%%%%%%%%%%%%%%%%%%%%%%%%%%%%%%%%%%%%%%%%%%%%%%%

We here introduce our model and describe the baryogenesis mechanism.
It is also pointed out that there can be a dark matter candidate
and the model can be embedded into the left-right symmetric theory.
Finally, variants of the model are discussed.

%%%%%%%%%%%%%%%%%%%%%%%%%%%%%%%%%%%%%%%%%%%%%%%%
\subsection{The extended gauge symmetry}
  \label{extendedgauge}
%%%%%%%%%%%%%%%%%%%%%%%%%%%%%%%%%%%%%%%%%%%%%%%%

We consider a model where 
a first-order phase transition associated with a new gauge symmetry breaking at a higher scale sets a stage for baryogenesis.
The SM gauge symmetry can be embedded into $G_{\rm LR} \equiv SU(3)_C \times SU(2)_L \times SU(2)_R \times U(1)_X$
by adding three generations of right-handed neutrinos.
Under this extended gauge symmetry, the SM fermions (plus three right-handed neutrinos $\bar{\nu}$) transform as
\begin{equation}
\begin{split}
&q_i=\Biggl( \begin{matrix}
{u}_i \\ {d}_i 
\end{matrix}\Biggr) : ({\bf 3}, {\bf 2}, {\bf 1}, \frac{1}{6} )_{B=1/3} ,  \qquad \bar{q}_i = \Biggl( \begin{matrix}
\bar{u}_i \\ \bar{d}_i
\end{matrix}  \Biggr) : ({\bf \bar{3}}, {\bf 1}, {\bf 2}, -\frac{1}{6})_{B=-1/3} , \\
&\ell_i = \Biggl( \begin{matrix}
{\nu}_i \\ {e}_i
\end{matrix}  \Biggr): ({\bf 1}, {\bf 2}, {\bf 1}, -\frac{1}{2})_{L=1} ,  \qquad \bar{\ell}_i = \Biggl( \begin{matrix}
\bar{\nu}_i \\ \bar{e}_i
\end{matrix}  \Biggr) : ({\bf {1}}, {\bf 1}, {\bf 2}, \frac{1}{2})_{L=-1} ,
\end{split}
\end{equation}
where $i = 1,2,3$ is the generation index and $B$ and $L$ denote baryon and lepton numbers.
Throughout this paper we use the two-component left-handed Weyl spinor notation.
Note that the $B$ and $L$ symmetries are anomalous but the combination $B-L$ is anomaly-free.
We for now do not impose the left-right symmetry. In section~\ref{sec:LR}, we discuss the embedding of the model into a left-right symmetric theory and how the strong CP problem can be solved.

The spontaneous breaking of $G_{\rm LR} \rightarrow G_{\rm SM} \equiv SU(3)_C \times SU(2)_L \times U(1)_Y$
is driven by a new Higgs scalar field $H_R$, while the electroweak symmetry is broken by the SM Higgs field.
The representations of the Higgs fields under the extended gauge symmetry $G_{\rm LR}$ and the symmetry breaking chain are
\begin{align}
H_R({\bf 1}, {\bf 1}, {\bf 2}, \frac{1}{2}) & : G_{\rm LR} \rightarrow G_{\rm SM} \, , \\
H_L({\bf 1}, {\bf 2}, {\bf 1}, -\frac{1}{2})~{\rm or}~\Phi({\bf 1}, {\bf 2}, {\bf 2}, 0)& : G_{\rm SM}\rightarrow SU(3)_c\times U(1)_{\rm EM} \, .
\end{align}
The SM Higgs is $H_L$ or $\Phi$.
The $U(1)_Y$ gauge coupling $g^\prime$ is given by
\begin{align}
   \frac{1}{{g'}^2} = \frac{1}{g_X^2} + \frac{1}{g_R^2} \, , \label{eq:gauge couplings}
\end{align}
where
$g_X$ and $g_R$ are the $U(1)_X$ and $SU(2)_R$ gauge couplings respectively.

When the SM Higgs is $H_L$, we cannot write down the Yukawa couplings of the SM quarks and leptons at the renormalizable level.
The SM quark and lepton masses are obtained by the following dimension-5 operators,
\begin{equation}
\begin{split}
\mathcal{L}_{\rm Yukawa} =  &\,\, \frac{c_u^{ij}}{M_u}(H_L^\dag q_i) (H_R^\dag \bar{q}_j)
+ \frac{c_d^{ij}}{M_d}(H_L q_i) (H_R \bar{q}_j) +\frac{c_e^{ij}}{M_e}(H_L \ell_i) (H_R \bar{\ell}_j) \\[1ex]
&  + \frac{c_{\nu}^{ij}}{M_\nu}(H_L^\dag \ell_i) (H_L^\dag \ell_j) + \frac{c_{\bar{\nu}}^{ij}}{M_\nu}(H_R^\dag \bar{\ell}_i) (H_R^\dag \bar{\ell}_j) + \frac{c_{\nu\bar{\nu}}^{ij}}{M_\nu}(H_L^\dag \ell_i) (H_R^\dag \bar{\ell}_j) + {\rm h.c.},
\label{smyukawa}
\end{split}
\end{equation}
where $c_u^{ij}, c_d^{ij}, c_e^{ij}, c_\nu^{ij}, c_{\bar{\nu}}^{ij},c_{\nu\bar{\nu}}^{ij}  $ are dimensionless constants
and $M_u, M_d, M_\nu, M_e$ are some UV mass scales.
These dimension-5 operators can be obtained by the exchange of heavy Dirac fermions coupling to the Higgs fields and the SM fermions.
To realize the correct top quark mass,
the mass of the corresponding heavy fermion must be near the vacuum expectation value (VEV) of the $H_R$ field.%
\footnote{Alternatively, the mass term of the corresponding fermions may be below $v_R$ so that the SM top quark dominantly originate from one of the fermions rather than $\bar{q}$; see ref.~\cite{Hall:2019qwx} for details.}
The small neutrino masses can be explained by a large $M_\nu$.
If we impose the $L$ symmetry, $c_\nu,c_{\bar{\nu}} = 0$ and the neutrinos are Dirac particles.
The most general renormalizable potential of the Higgs fields $H_L$ and $H_R$ allowed
by the gauge symmetry $G_{\rm LR}$ is
\begin{align}
V(H_L,H_R)=&-\mu_L^2 |H_L|^2  -\mu_R^2|H_R|^2  +\lambda_L |H_L|^4 +\lambda_R|H_R|^4 + \lambda_{LR}|H_L|^2 |H_R|^2 . \label{eq:Higgs potential}
\end{align}
Here, $\mu_{L,R}^2$ are mass-squared parameters
and $\lambda_L, \lambda_R, \lambda_{LR}$ are dimensionless coupling constants.
They are all real parameters.

If the SM Higgs is $\Phi$, then the Yukawa interactions,
\begin{equation}
\begin{split}
    {\cal L}'_{\rm Yukawa} &=  y_{q}^{ij}\Phi q_i \bar{q}_i + \tilde{y}_{q}^{ij}\tilde{\Phi} q_i \bar{q}_i + y_{\ell}^{ij}\Phi \ell_i \bar{\ell}_i  + \tilde{y}_{\ell}^{ij}\tilde{\Phi} \ell_i \bar{\ell}_i + {\rm h.c.}, \\[1ex]
    \hvev{\Phi} &= \frac{1}{\sqrt{2}}\begin{pmatrix} v_1 & 0 \\ 0 & v_2 \end{pmatrix},
\end{split}
\end{equation}
lead to the SM fermion masses,
\begin{equation}
\begin{split}
    &m_{u}^{ij} = \frac{1}{\sqrt{2}}y_q^{ij}v_1 + \frac{1}{\sqrt{2}}\tilde{y}_{q}^{ij}v_2^*, \qquad m_{d}^{ij} = \frac{1}{\sqrt{2}}y_q^{ij}v_2 + \frac{1}{\sqrt{2}}\tilde{y}_{q}^{ij}v_1^*, \\[1ex]
    &m_{e}^{ij} = \frac{1}{\sqrt{2}}y_\ell^{ij}v_2 + \frac{1}{\sqrt{2}}\tilde{y}_{\ell}^{ij}v_1^*, 
     \qquad m_{\nu}^{ij} = \frac{1}{\sqrt{2}}y_\ell^{ij}v_1 + \frac{1}{\sqrt{2}}\tilde{y}_{\ell}^{ij}v_2^* .
\end{split}
\end{equation}
Here, $\tilde{\Phi} = \sigma_2 \Phi^* \sigma_2$ and $v \equiv \sqrt{v_1^2+v_2^2} \simeq 246\gev$.
The small SM neutrino masses require $v_1\ll v$ with $|\tilde{y}^{ij}_{\ell}| \ll 1$ or $v_2 \ll v$ with $|y^{ij}_{\ell}| \ll 1$. Instead, we may add $S H_{R}^\dag \bar{\ell} $ to give a Dirac mass term to $\bar{\nu}$, so that the SM neutrino masses are suppressed. Here $S$ is a singlet fermion. We assume that the Majorana masses of $S$ are small. 
The generic renormalizable potential of $\Phi$ and $H_R$ is
\begin{align}
    V(\Phi, H_R) = - \mu_1^2|\Phi|^2 + (\mu_2^2 \Phi^2 + {\rm h.c.})  -\mu_R^2|H_R|^2  +\lambda_R|H_R|^4 +  \cdots, \label{eq:Higgs potential2}
\end{align}
where $\cdots$ includes self quartic couplings of $\Phi$ and quartic couplings between $H_R$ and $\Phi$.
Since the model contains two $SU(2)_L$ doublet Higgses that couple to the SM fermions, tree-level flavor-changing neutral currents are in general induced by the exchange of a heavy Higgs, which sets a strong lower bound on $v_R$, see e.g., ref.~\cite{Bertolini:2019out}. Such constraints can be avoided by extending the model~\cite{Guadagnoli:2010sd,Mohapatra:2013cia,Mohapatra:2019qid} without changing our baryogenesis scenario, and hence we do not impose flavor constraints in the present paper.

If the electroweak scale is naturally explained by some dynamics such as supersymmetry and compositeness, it is plausible that the mass scale $\mu_R$ is determined by the same dynamics and not far above the electroweak scale. It is also possible that the smallness of the electroweak scale is natural as long as there are no fields which have large mass scales and couple to the SM Higgs.
The new gauge bosons, however, directly couple to the SM Higgs. The mass of the new gauge bosons should not be much above the electroweak scale. In both scenarios, the new particles in the model should not be much heavier than the electroweak scale and may be discovered at the LHC or future collider experiments.

%%%%%%%%%%%%%%%%%%%%%%%%%%%%%%%%%%%%%%%%%%%%%%%%
\subsection{New chiral leptons}
  \label{newlepton}
%%%%%%%%%%%%%%%%%%%%%%%%%%%%%%%%%%%%%%%%%%%%%%%%

To produce baryon asymmetry during the new phase transition in the similar way as electroweak baryogenesis,
we require that the global $B-L$ symmetry is anomalous under the extended gauge interactions.
This is achieved by introducing a new chiral lepton sector. We borrow the fermion content of ref.~\cite{Kim:2017qxo} discussed in the context of a dark sector model, 
which has a structure different from that of the SM fermion sector.
The charge assignments of the new fermions under $G_{\rm LR}$
and the $L$ symmetry are
\begin{equation}
\label{eq:charges}
\begin{split}
&\bar{L}_i = \Biggl( \begin{matrix}
\bar{E}_i \\ \bar{N}_i
\end{matrix}  \Biggr)
: ({\bf 1}, {\bf 1}, {\bf 2}, \frac{1}{2})_{L=-1} ,  \qquad \widetilde{L} = \Biggl( \begin{matrix}
\mathcal{E} \\ \mathcal{X}
\end{matrix}  \Biggr): ({\bf {1}}, {\bf 1}, {\bf 2}, -\frac{3}{2})_{L=1} , \\[1ex]
&{E}_i : ({\bf 1}, {\bf 1}, {\bf 1}, -1)_{L=1} ,  \quad N_i : ({\bf {1}}, {\bf 1}, {\bf 1}, 0)_{L=1} , \quad
\bar{\mathcal{E}} : ({\bf 1}, {\bf 1}, {\bf 1}, 1)_{L=-1} ,  \quad \bar{\mathcal{X}} : ({\bf {1}}, {\bf 1}, {\bf 1}, 2)_{L=-1} , \\[1ex]
\end{split}
\end{equation}
where $i = 1,2,3$ is the generation index.
There are four $SU(2)_R$ doublets $\bar{L}_i , \widetilde{L}$ and 
five singlets $E_i, \bar{\mathcal{E}}, \bar{\mathcal{X}}$ charged under the $U(1)_X$.
We have also introduced three singlets $N_i$ to make  $\bar{N}_i$ inside $\bar{L}_i$ massive by forming Dirac masses
after the symmetry breaking.
Gravitational and gauge anomalies are absent as $\Tr Q_X = 0$ and $\Tr Q_X^3 = 0$. 
The number of $SU(2)_R$ doublets is even and hence the non-perturbative $SU(2)$ anomaly \cite{Witten:1982fp} is avoided.
It should be stressed that three generations are required from the structure of the theory.

The baryon and lepton number global symmetries are anomalous under the extended gauge symmetry,
\begin{equation}
\begin{split}
&\partial^\mu j^{(B)}_\mu = - \frac{3}{32\pi^2} \left[ g_L^2 W_{L \mu\nu} \widetilde{W}_L^{\mu\nu}
- g_R^2 W_{R \mu\nu} \widetilde{W}_R^{\mu\nu}\right] , \\[1ex]
&\partial^\mu j^{(L)}_\mu = - \frac{1}{32\pi^2} \left[ 3g_L^2 W_{L \mu\nu} \widetilde{W}_L^{\mu\nu}
-5 g_R^2 W_{R \mu\nu} \widetilde{W}_R^{\mu\nu} -5 g_X^2 B_{X \mu\nu} \widetilde{B}_X^{\mu\nu} \right] , \\[1ex]
\end{split}\label{eq:anomaly}
\end{equation}
where $W_L$, $W_R$, and $B_X$ are the $SU(2)_L$, $SU(2)_R$, and $U(1)_X$ gauge field strengths respectively, and
$\widetilde{W}_{L \mu\nu} \equiv \frac{1}{2} \epsilon_{\mu\nu\rho\sigma} W_L^{\rho\sigma},
\widetilde{W}_R$, and $ \widetilde{B}_X$ are their duals.
$g_L, g_R$, and $g_X$ are the  $SU(2)_L$, $SU(2)_R$, and $U(1)_X$ gauge coupling constants respectively.
We can see that the model violates any linear combinations of the baryon and lepton number symmetries, including $B-L$, through anomaly.

All the new fermions become massive by 
the following Yukawa interactions with the Higgs field $H_R$ after the symmetry breaking,
\begin{equation}
\begin{split}
\mathcal{L}_{R} = y_E^{ij} H_R \bar{L}_i {E}_j  + y_N^{ij} H_R^\dagger  \bar{L}_i {N}_j
+ y_{\mathcal{E}} H_R \widetilde{L} \, \bar{\mathcal{E}} + y_{\mathcal{X}} H_R^\dagger \widetilde{L} \, \bar{\mathcal{X}} + \rm h.c. 
\end{split} \label{eq:Yukawa coupling}
\end{equation}
Here, $ y_E^{ij}, y_N^{ij}, y_{\mathcal{E}}, y_{\mathcal{X}}$ are dimensionless coupling constants. 
The first and second terms have the same structure as the SM quark Yukawa couplings.
The number of fermion generations involved in these couplings is also three.
Thus, as in the case of the SM quark sector~\cite{Kobayashi:1973fv},
these terms contain one physical complex phase which cannot be eliminated by field redefinition.
Gauge eigenstates are written in terms of mass eigenstates as
\begin{equation}
\begin{split}
\label{eq:mass rotation}
&\bar{E}_i = (U_{\bar{E}})_i^{\ j}(\bar{E}_{m})_j \; , \qquad E_i = (U_{E})_i^{\ j}(E_m)_j \; , \\[1ex]
&\bar{N}_i = (U_{\bar{N}})_i^{\ j}(\bar{N}_m)_j \; , \qquad N_i = (U_N)_i^{\ j}(N_m)_j \; ,
\end{split}
\end{equation}
where the subscript $m$ is for mass eigenstates and $U_{E,\bar{E}, N, \bar{N}}$ are unitary matrices. In the following, we drop the subscript $m$. Without loss of generality, we take $m_{E_3}>m_{E_2}> m_{E_1}$ and $m_{N_3}>m_{N_2}> m_{N_1}$. 
We define the unitary matrix $K \equiv U_{\bar{E}}^{\dagger}U_{\bar{N}}$ similar to the CKM matrix.
Using the same parametrization as the CKM, this matrix can be parameterized as
\begin{equation}
\begin{split}
  K =
  \begin{pmatrix}
    c_{12}c_{31} & s_{12}c_{31} & s_{31}e^{-i\delta}\\
    -s_{12}c_{23} - c_{12}s_{23}s_{31}e^{i\delta} & c_{12}c_{23}-s_{12}s_{23}s_{31}e^{i\delta} & s_{23}c_{31}\\
    s_{12}s_{23} - c_{12}c_{23}s_{31}e^{i\delta} & -c_{12}s_{23} - s_{12}c_{23}s_{31}e^{i\delta} & c_{23}c_{31}\\
  \end{pmatrix} .
  \label{eq:CKM parametrization}
\end{split}
\end{equation}
Here, $s_{ij} \equiv \sin \theta_{ij}, c_{ij} \equiv \cos \theta_{ij}$, and $\delta$ is the CP phase.
We can define a basis-independent quantity to measure the CP violating effect of the new lepton sector $J'$,
similar to the Jarlskog invariant for the CKM, as
\begin{equation}
\begin{split}
  \label{eq:Jarlskog}
  \mathrm{Im}(K_{ij}K_{kl}K^{*}_{il}K^{*}_{kj}) \equiv J^{\prime}\sum_{m,n}\epsilon_{ikm}\epsilon_{jln} \; .
\end{split}
\end{equation}
With the parameterization \eqref{eq:CKM parametrization},
$J^{\prime}$ is given by
\begin{equation}
\begin{split}
  \label{eq:Jarlskog parameter}
  J^{\prime} = \mathrm{Im}(K_{11}K_{22}K^{*}_{12}K^{*}_{21}) = s_{12}s_{23}s_{31}c_{12}c_{23}c_{31}^2\sin \delta \; .
\end{split}
\end{equation}
This is the CP-violation source we consider for baryogenesis during the new phase transition.
Complex phases in the third and fourth terms of \eqref{eq:Yukawa coupling} can be eliminated by rotating the phases of $\widetilde{L}$ and $\bar{\mathcal{X}}$.

Without any interaction with the SM leptons, this new lepton sector possesses its own lepton number. This is reduced to the common lepton number by interactions between the new lepton sector and the SM lepton sector.
The most general form of such Yukawa interactions is
\begin{equation}
\begin{split}
\mathcal{L}_{\rm int} = g^i_{\bar{\mathcal{E}}} H_L \ell_i  \, \bar{\mathcal{E}} + g_{{E}}^{ij} H_R  \bar{\ell}_i {E}_j
+ g_{{N}}^{ij} H_R^\dagger \bar{\ell}_i N_j + \rm h.c.,
\label{portal interaction}
\end{split}
\end{equation}
where $g^i_{\bar{\mathcal{E}}}, g_{{E}}^{ij},  g_{{N}}^{ij}$ are dimensionless coupling constants. The first term is absent if the SM Higgs is a bi-fundamental scalar $\Phi$.
We assume that these couplings are relatively small and their CP-violating phases do not play a role in baryogenesis.

We assume that the Majorana mass terms of $N_i$ are negligible, which may be understood by an approximate lepton number symmetry.
The $E_i$ and $\mathcal{\bar{E}}$ fields may have a Dirac mass. We also assume that the Dirac mass term is negligible, which can be understood by an approximate $Z_2$ symmetry with $(\bar{L}_i, E_i, N_i) \to - (\bar{L}_i, E_i, N_i)$ and $(\widetilde{L},\bar{\mathcal{E}},\bar{\mathcal{X}})\to (\widetilde{L},\bar{\mathcal{E}},\bar{\mathcal{X}})$.
The $Z_2$ symmetry is explicitly broken by the last two terms in \eqref{portal interaction},
which makes the lightest $Z_2$-odd particle unstable.
We comment on a possibility that the $Z_2$ symmetry is unbroken and the lightest particle among $(\bar{L}_i, E_i, N_i)$
becomes a dark matter candidate in sec.~\ref{sec:DM}.
In this case, if the SM Higgs is $\Phi$, there is no remormalizable interaction that allows the decay of the new fermions into SM particles. We need a higher dimensional interaction or a flipped charged assignment discussed in the next paragraph.

The $U(1)_X$ and $L$ charges in eq.~(\ref{eq:charges}) may be flipped. The model works as well for that case. The interactions between the new lepton sector and the SM lepton sector are modified from eq.~(\ref{portal interaction}),
\begin{align}
    {\cal L}’_{\rm int} = g^i_{\bar{\mathcal{E}}} H_R \bar{\ell}_i  \, \bar{\mathcal{E}} + g_{{E}}^{ij} H_L  \ell_i {E}_j
+ g_{{N}}^{ij} H_L^\dagger \ell_i N_j + m_L^{ij}\bar{L}_i\bar{\ell}_j + \rm h.c.,
\label{portal interaction2}
\end{align}
where the last three terms break the $Z_2$ symmetry.

%%%%%%%%%%%%%%%%%%%%%%%%%%%%%%%%%%%%%%%%%%%%%%%%
\subsection{Generation of baryon asymmetry}
%%%%%%%%%%%%%%%%%%%%%%%%%%%%%%%%%%%%%%%%%%%%%%%%

%%%%%%%%%%%%%%%%%%%%%%%%%%%%%%%%%%%%%%%%%%%%%%%%%%%%%%%%%%%%%
\begin{figure}[t]
  \centering
    \begin{minipage}[h]{0.49\linewidth}
    \hspace{0.5cm}
    \includegraphics[width=3in]{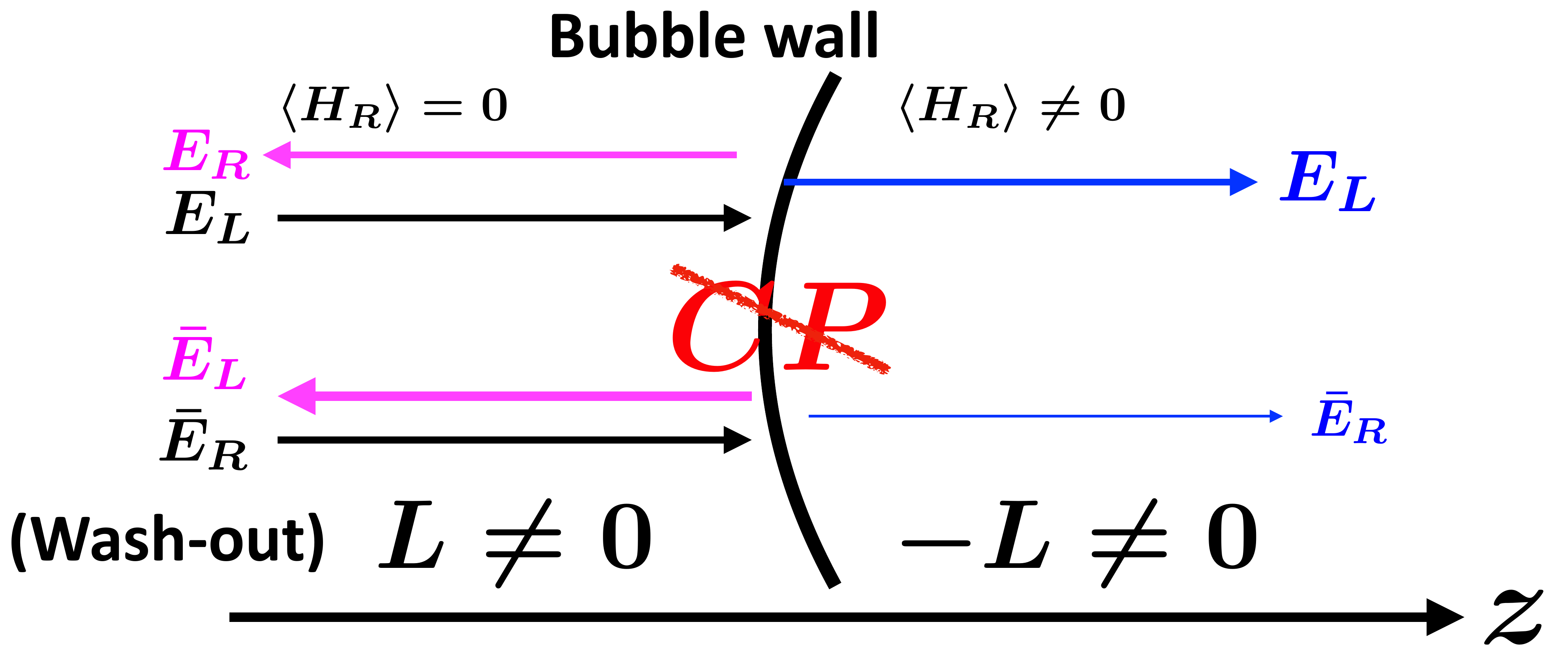}
    \end{minipage}
    \begin{minipage}[h]{0.49\linewidth}
    \hspace{1.5cm}
    \includegraphics[width=2in]{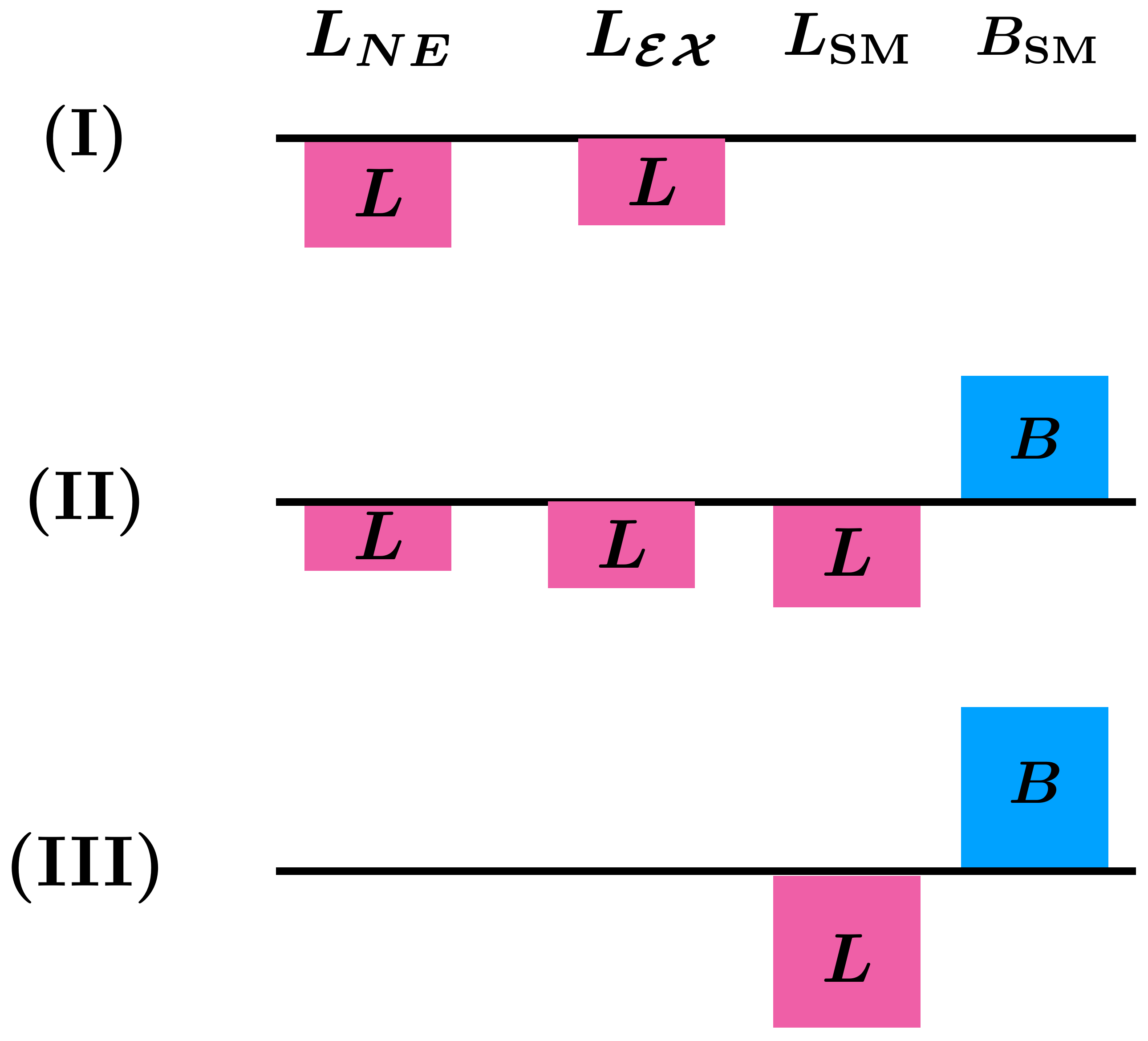}
    \end{minipage}
\vspace{0.3cm}
\caption{
An illustration of a lepton asymmetry generated by the $E$ lepton scattering with the bubble wall is given in the left panel.
Here, $E_{L (R)}$ and $\bar{E}_{R(L)}$ are the left (right)-handed $E$ lepton and its anti-particle.
For instance, $E_{L}$ particles are injected from the symmetric phase $\langle H_R \rangle = 0$.
Some are reflected by the bubble wall while others are transmitted into the broken phase $\langle H_R \rangle \neq 0$.
Thickness of the arrows describes the strength of reflections and transmissions.
Due to the CP violation, the particles and anti-particles are reflected or transmitted differently.
A nonzero lepton number in the symmetric phase is washed out. 
In the right panel, a schematic picture for the evolution of asymmetries is shown.
$L_{\rm SM}$ and $B_{\rm SM}$ denote the lepton number and baryon number asymmetries carried by the SM leptons and quarks, respectively.
The steps (I), (II) and (III) are explained in the main text.
}
\label{baryon asymmetry}
\end{figure}
%%%%%%%%%%%%%%%%%%%%%%%%%%%%%%%%%%%%%%%%%%%%%%%%%%%%%%%%%%%%%

We now describe the baryogenesis mechanism based on the first-order phase transition associated with
the breaking of $G_{\rm LR}\to G_{\rm SM}$. Hereafter, we call it the $G_{\rm LR}\to G_{\rm SM}$ phase transition.
We assume that the $G_{\rm LR} \to G_{\rm SM}$ phase transition first takes place
and then the ordinary electroweak phase transition follows.
The baryogenesis mechanism is similar to the SM electroweak baryogenesis, but the crucial difference is that nonzero lepton number rather than baryon number is generated
by a nontrivial bubble wall scattering process.

The first-order $G_{\rm LR}\to G_{\rm SM}$ phase transition proceeds via bubble nucleations.
Nucleated bubbles expand and eventually coalesce with each other until they fill the Universe.
During this process, the bubbles interact with thermal plasma.
Since the $N_i$ and $E_i$ fields coupled to $H_R$ acquire masses via the Higgs mechanism inside the bubble,
a flux of the new leptons outside the bubble see the bubble wall as a potential barrier,
and hence they are quantum mechanically scattered.
If CP-violating interactions are involved in this scattering process,
the reflection coefficients of the new leptons and their antiparticles are different from each other.
Then, a lepton number asymmetry $L_{NE}$ carried by the new leptons $N_i,E_i,\bar{N}_i, \bar{E}_i$
is generated in front of the bubble wall.
Since the lepton number is globally conserved in the wall scattering process,
the same amount of $L_{NE}$ with opposite sign is produced inside and outside the bubble.
The $SU(2)_R$ sphaleron processes convert a part of $L_{NE}$ outside the bubble into an asymmetry $L_{\mathcal{E}\mathcal{X}}$ carried by $\tilde{L}, \bar{\mathcal{E}},\bar{\mathcal{X}}$ and a $B+L$ asymmetry of the SM particles.
The produced $B+L$ asymmetry is washed out by the $SU(2)_L$ sphaleron processes.
This process is schematically shown in the left panel of fig.~\ref{baryon asymmetry} for the $E$ lepton scattering.

The generated $L$ asymmetry is transferred into a $B$ asymmetry through the following three steps.
(I) Since a $B-L$ violating process associated with the $SU(2)_R$ gauge interaction is decoupled
inside the bubble,
nonzero lepton number densities $L_{NE}$ and $L_{\mathcal{E}\mathcal{X}}$ remain after the completion of the $G_{\rm LR}\to G_{\rm SM}$ phase transition.
(II) A part of the $L$ asymmetry is transferred into a $B$ asymmetry via the ordinary electroweak sphaleron process
if the new lepton fields are in thermal equilibrium with the SM particles through
the portal interactions \eqref{portal interaction}.
(III) Finally, the new leptons decay into the SM leptons and the $L$ asymmetry carried by the new leptons
is completely transferred into the SM sector.
For simplicity, we only consider the case where the decays occur before the decoupling of the electroweak sphaleron. Even if this is not the case, the resultant baryon asymmetry changes only by an $O(1)$ factor.
Then, the injected $L$ asymmetry is further transferred into a $B$ asymmetry via the sphaleron process.
The evolution of $B$ and $L$ in these steps is schematically shown in the right panel of fig.~\ref{baryon asymmetry}.

%%%%%%%%%%%%%%%%%%%%%%%%%%%%%%%%%%%%%%%%%%%%%%%%
\subsection{A dark matter candidate}
  \label{sec:DM}
%%%%%%%%%%%%%%%%%%%%%%%%%%%%%%%%%%%%%%%%%%%%%%%%

In the previous subsection, we described the baryogenesis mechanism assuming that the $N_i,E_i$ particles are unstable. The mechanism, however, works as well even if the $Z_2$ symmetry $(\bar{L}_i, E_i, N_i) \to - (\bar{L}_i, E_i, N_i)$ is exact. The lightest particle among the $N_i,E_i$ sector is stable, and if $m_{N_1}< m_{E_1}$, $N_1$ is stable
and a dark matter candidate. Even for this case, the baryon asymmetry is still produced from $L_{\mathcal{X}\mathcal{E}}$.
We leave the discussion on phenomenology of the $N_1$ dark matter to a future work, but we briefly describe how the correct dark matter abundance can be obtained.

The freeze-out abundance of $N_1$ seems too large because the neutral $Z'$ boson
associated with the $G_{\rm LR}\to G_{\rm SM}$ breaking
must be heavy as we will see in sec.~\ref{sec:Z'andWR}, and the pair annihilation of $N_1$ is ineffective. The correct abundance can be, however, obtained by the coannihilation~\cite{Griest:1990kh} with $E_1$ if $m_{N_1} \sim m_{E_1}$. For this to occur, $N_1$ must be converted into $E_1$ via the exchange of the charged $W_R^\pm$ boson  during the freeze-out process, which is effective  for $v_R \lesssim 100$ TeV.

The asymmetry of $N_i,E_i$ leads to a too large dark matter abundance unless $m_{N_1} \sim $ GeV. For such a small mass, the coannihilation requires $m_{E_1}\sim$ GeV, which is excluded. This problem can be avoided by introducing a small Majorana mass term of $N_i$, so that the washout of $L_{NE}$ occurs inside the bubble within the Hubble time scale. The Majorana mass term may also effectively washout the asymmetry outside the bubble and erase all the asymmetry produced by the scattering with the bubble wall.
However, since this requires washout within the timescale of the diffusion (see eq.~\eqref{eq:tdiff}), which is much shorter than the Hubble time scale, the washout outside the bubble can be avoided while that inside the bubble can efficiently occur by an appropriate choice of the magnitude of the Majorana mass term.

%%%%%%%%%%%%%%%%%%%%%%%%%%%%%%%%%%%%%%%%%%%%%%%%
\subsection{A left-right symmetric theory}
  \label{sec:LR}
%%%%%%%%%%%%%%%%%%%%%%%%%%%%%%%%%%%%%%%%%%%%%%%%

The model we have described so far does not possess a left-right symmetry; the extra fermions are charged only under $SU(2)_R$. It is, however, possible to obtain the model from
a left-right symmetric theory by its spontaneous breaking.

The gauge symmetry at a UV scale is $SU(3)_c\times SU(2)_{\tilde{L}}\times SU(2)_{\tilde{R}}\times U(1)_{B-L} \times SU(2)_{XL}\times SU(2)_{XR}$ with left-right symmetry $SU(2)_{\tilde{L}}\leftrightarrow SU(2)_{\tilde{R}}$ and $SU(2)_{XL}\leftrightarrow SU(2)_{XR}$. The SM fermions are charged under $SU(3)_c\times SU(2)_{\tilde{L}}\times SU(2)_{\tilde{R}}\times U(1)_{B-L}$. The extra fermions we have introduced in eq.~\eqref{eq:charges} are charged under $SU(2)_{XR}\times U(1)_{B-L}$. The left-right symmetry requires another set of chiral fermions charged under $SU(2)_{XL}\times U(1)_{B-L}$.

$SU(2)_{\tilde{R}} \times SU(2)_{XR}$ symmetry is broken to $SU(2)_R$ by a VEV of a bi-fundamental scalar.
The left-right symmetry predicts a scalar in bi-fundamental of $SU(2)_{\tilde{L}}\times SU(2)_{XL}$, but we can choose the parameter of the left-right symmetric potential so that only the $SU(2)_{\tilde{R}} \times SU(2)_{XR}$ charged scalar obtains a VEV. The $SU(2)_{XL}\times U(1)_{B-L}$ symmetry is broken to $U(1)_{X}$ by a scalar with a charge $({\bf 2}, 1/2)$. The scalar couples to the extra chiral fermions charged under $SU(2)_{XL} \times U(1)_{B-L}$ and give masses to them. $SU(2)_{\tilde{L}}$ remains as $SU(2)_L$.

In the UV theory, the SM Higgs $H_L$ and $\Phi$ are charged under $SU(2)_{\tilde{L}}\times U(1)_{B-L}$ and $SU(2)_{\tilde{L}}\times SU(2)_{\tilde{R}}$, respectively.
The Higgs $H_R$ must be charged under $SU(2)_{XR} \times U(1)_{B-L}$ so that the Yukawa couplings in eq.~(\ref{eq:Yukawa coupling}) are obtained without suppression. If the SM Higgs is $H_L$, $H_R$ needs $SU(2)_{\tilde{R}}$ charge to obtain the SM yukawa couplings in eq.~(\ref{smyukawa}). This can be achieved by obtaining $H_R$ as a linear combination of an $SU(2)_{XR}$ doublet and an $SU(2)_{\tilde{R}}$ doublet.

In this setup, the strong CP problem is solved by combining the left-right symmetry with space-time parity in the same manner as the models in the literature~\cite{Beg:1978mt,Mohapatra:1978fy,Babu:1988mw,Babu:1989rb}. The space-time parity forbids the term $G\tilde{G}$. The correction to this term from quark masses can be also suppressed in the following way.
If the SM Higgs is $H_L$, the SM quark masses are given by eq.~(\ref{smyukawa}). The space-time parity requires that $c^{ij}$ are Hermitian, so that that $\theta$ terms remain zero~\cite{Babu:1988mw,Babu:1989rb,Hall:2018let}. Two-loop correction gives non-Hermitial quark masses~\cite{Hall:2018let}, but the correction can be smaller than the experimental upper bound on the $\theta$ term.
If the SM Higgs is $\Phi$, although the coupling $y_q^{ij}$ and $y_{\tilde{q}}^{ij}$ are Hermitian, the possible phase in $v_1$ and $v_2$ may generate non-Hermitian quark masses. The left-right symmetry with a space-time parity, $\Phi \leftrightarrow \Phi^\dag$, forbids the CP phases in the most of terms in the potential of $\Phi$, except for the quartic terms of the form,
\begin{align}
\label{eq:Phi_CP phase}
    V = e^{i \alpha}\left(\Phi^2 |\phi_1|^2 + \Phi^{\dag 2} |\phi_2|^2 \right) + {\rm h.c.}, 
\end{align}
where $\phi_{1,2}$ are scalar fields with the left-right transformation $\phi_1 \leftrightarrow \phi_2$. After the left-right symmetry is spontaneously broken by $\hvev{\phi_1} \neq \hvev{\phi_2}$, as is assumed above, this term gives complex VEVs $v_{1,2}$. To solve the strong CP problem, the potential term~(\ref{eq:Phi_CP phase}) must be forbidden by an extra symmetry.
%, such as supersymmetry~\cite{Kuchimanchi:1995rp,Mohapatra:1995xd} or discrete symmetry~\cite{Beg:1978mt,Mohapatra:1978fy}.
Supersymmetry can forbid the above quartic term while allowing for a quadratic term $\Phi^2$~\cite{Kuchimanchi:1995rp,Mohapatra:1995xd}, thereby avoiding an accidental $U(1)$ symmetry leading to a light Nambu-Goldstone boson. In non-supersymmetric theories, we may introduce a $Z_4$ symmetry that forbids the above quartic term but allows for a quartic term $\Phi^4$~\cite{Mohapatra:1978fy}, avoiding a light Nambu-Goldstone boson. The second Higgs is necessarily around the weak scale, but such a possibility is still allowed~\cite{Arkani-Hamed:2020yna}.

%%%%%%%%%%%%%%%%%%%%%%%%%%%%%%%%%%%%%%%%%%%%%%%%
\subsection{Variants of the model}
%%%%%%%%%%%%%%%%%%%%%%%%%%%%%%%%%%%%%%%%%%%%%%%%

The structure of the model with new particles in eq.~(\ref{eq:charges}) is following.
The $SU(2)_R$ sphaleron process and the scattering of three generations of chiral $\bar{L}_i, E_i,N_i$ leptons with
the bubble wall generate a lepton asymmetry.
The gauge anomaly is cancelled by $\tilde{L}$, $\bar{\cal E}$, and $\bar{\cal X}$.
Then, we can consider variants of the model.
The gauge anomaly can be instead cancelled by three generations of $L_i,\bar{E}_i,\bar{N}_i$ which have gauge charges opposite to those of $\bar{L}_i,E_i,N_i$. The fermions $L_i,\bar{E}_i,\bar{N}_i$ obtain their masses by their Yukawa coupling with $H_R$. In this case, the total $B-L$ asymmetry produced by the scattering and the $SU(2)_R$ sphaleron vanishes. However, if one of the asymmetries of the $L_i,\bar{E}_i,\bar{N}_i$ sector and the $\bar{L}_i,E_i,N_i$ sector is not transferred into the SM sector before the electroweak phase transition, the $B-L$ asymmetry that the SM sector feels is non-zero, and a non-zero baryon asymmetry remains. This can be enforced by (approximate) symmetry $(L_i,\bar{E}_i,\bar{N}_i)\rightarrow -(L_i,\bar{E}_i,\bar{N}_i)$ or $(\bar{L}_i,E_i,N_i) \rightarrow - (\bar{L}_i,E_i,N_i)$. The $Z_2$ symmetry, if it is exact, results in a dark matter candidate. The estimation of the baryon asymmetry for the model with eq.~(\ref{eq:charges}) is directly applicable to this variant model. The gauge charges of the new fermions are vector-like, but the (approximate) $Z_2$ symmetry makes the theory chiral.

A similar idea is applicable to a model with new heavy quarks $Q_i,\bar{U}_i,\bar{D}_i$ and $\bar{Q}_i,U_i,D_i$, where $Q_i,\bar{Q}_i$ are $SU(2)_R$ doublets and the others are singlets. $Q_i,\bar{U}_i,\bar{D}_i$ obtain their masses from their Yukawa couplings with $H_R$, and so do $\bar{Q}_i,U_i,D_i$. In this case, the $Z_2$ symmetry must be explicitly broken so that there are no exotic charged stable particles. The estimation of the baryon asymmetry given in the next section is applicable to this case with proper modification. We, however, do not discuss this case further.

%%%%%%%%%%%%%%%%%%%%%%%%%%%%%%%%%%%%%%%%%%%%%%%%
\section{Baryogenesis}\label{baryogenesis}
%%%%%%%%%%%%%%%%%%%%%%%%%%%%%%%%%%%%%%%%%%%%%%%%

In this section, we discuss the mechanism of baryogenesis based on the model presented in sec.~\ref{extendedgauge} and sec.~\ref{newlepton},
and calculate the amount of the baryon asymmetry.
In sec.~\ref{sec:sphaleron}, the condition of the first-order $G_{\rm LR}\to G_{\rm SM}$ phase transition is discussed
by calculating the thermal effective Higgs potential.
In sec.~\ref{sec:SFOPT}, we derive the sphaleron decoupling condition associated with the $SU(2)_R$ gauge interaction,
which is a necessary condition to avoid the washout of $B-L$ inside the bubble.
The parameter regime which satisfies the sphaleron decoupling condition is also clarified.
In sec~\ref{sec:reflection}, we show that $E$ and $N$ asymmetries generated in front of the moving bubble wall can be calculated by solving the effective Dirac equation, which is defined on the bubble wall and on the thermal background.
We then take account of the diffusion effect and the sphaleron process in the symmetric phase
and show the expression of the produced baryon asymmetry.
In sec.~\ref{sec:Delta}, we explicitly calculate thermal self-energies and damping rates for $E$ and $N$ leptons, which are necessary to solve the effective Dirac equation.
The kinematic property of quasiparticles is also discussed in the small momentum region.
In sec.~\ref{sec:decoherence effect}, we calculate the reflection probability to estimate the $E$ and $N$ asymmetries generated in front of the wall.
We utilize the calculation method originally applied to the electroweak baryogenesis within the SM framework.
Finally, the produced baryon asymmetries in the minimal model and an extended model are presented in sec.~\ref{sec:yukawa}.

%%%%%%%%%%%%%%%%%%%%%%%%%%%%%%%%%%%%%%%%%%%%%%%%
\subsection{The first-order $G_{\rm LR}\to G_{\rm SM}$ phase transition}
  \label{sec:sphaleron}
%%%%%%%%%%%%%%%%%%%%%%%%%%%%%%%%%%%%%%%%%%%%%%%%

Since we assume that the $G_{\rm LR}\to G_{\rm SM}$ phase transition takes place before the electroweak phase transition,
the potential during the $G_{\rm LR}\to G_{\rm SM}$ phase transition is obtained
by setting $H_L=0$ in \eqref{eq:Higgs potential} or $\Phi=0$ in \eqref{eq:Higgs potential2}.
Taking the background $H_R$ field as $H_R = ( \phi_R/\sqrt{2},0)^t$, the tree-level potential is
\begin{equation}
\begin{split}
V_{\rm tree} = -\frac{\mu_R^2}{2} \phi_R^2 +\frac{\lambda_R}{4}\phi_R ^4 \; . \label{eq:tree potential}
\end{split}
\end{equation}

The finite-temperature effective potential including the Coleman-Weinberg potential of $\phi_R$ is calculated
by the standard background field method~\cite{Dolan:1973qd,Quiros:1999jp}.
We take account of one-loop corrections from $SU(2)_R\times U(1)_X$ gauge bosons
and new leptons coupled to $H_R$ with the largest and the second largest Yukawa couplings.\footnote{We assume that the coupling between $H_R$, and $H_L$ or $\Phi$ is sufficiently small so that a loop correction from $H_L$ or $\Phi$ can be safely neglected.}
We adopt the renormalization condition used in ref.~\cite{Dine:1992wr}, which is given by $V'(\phi_R=v_R)|_{T=0}=0$ and $V''(\phi_R=v_R)|_{T=0}=2\lambda_R v_R^2$ where $V|_{T=0}$ denotes the sum of the tree-level potential and the Coleman-Weinberg potential.
With this renormalization condition, the VEV and mass of $H_R$ evaluated at the tree-level vacuum are
not shifted by quantum corrections.
We use the resummation prescription proposed in ref.~\cite{Arnold:1992rz}.
The total effective potential denoted by $V_{\rm eff}$ is then given by the sum of the zero temperature part $V|_{T=0}$
and the finite temperature effective potential.

We include the above-mentioned corrections in numerical analysis, but to illustrate the condition of the first-order phase transition, here we consider the effective potential at the one-loop order with the high-temperature expansion.
The effective potential is approximately given by
\begin{equation}
\begin{split}
&V_{\rm eff} \simeq  D(T^2-T_0^2) \phi_R^2 -ET\phi_R^3 +\frac{\lambda_R (T)}{4} \phi_R^4 \; , \label{eq:effective potential}
\end{split}
\end{equation}
where $D$, $E$ and $T_0$ are independent of temperature and determined by the model parameters.
The quartic coupling $\lambda_R (T)$ depends on temperature.
The potential shape at each temperature has been studied in refs.~\cite{Anderson:1991zb,Dine:1992wr}.
At very high temperature, there is a unique minimum at the origin.
As temperature decreases due to the cosmic expansion,
an extra minimum appears at $\phi_R = v_R (T)\neq 0$.
This extra minimum is separated from the minimum at the origin by a potential barrier.

The potential barrier is originated from the negative cubic term in \eqref{eq:effective potential} that is generated by bosonic Matsubara zero modes.
In the present setup, such zero modes can be provided by the $SU(2)_R \times U(1)_X$ gauge bosons and the Higgs field $H_R$.
Here, we assume that $g_X$ and $\lambda_R$ are rather small and
the $U(1)_X$ gauge boson and the $H_R$ field do not give relevant contributions to the cubic term.
A small $\lambda_R$ is required for a first order phase transition, as we will see later.
In addition, since the longitudinal modes of the $SU(2)_R$ gauge bosons obtain thermal masses at the one-loop order, their zero-mode contributions are screened by plasma effects~\cite{Arnold:1992rz}.
On the other hand, the transverse modes of the gauge bosons do not obtain thermal masses
at the one-loop order~\cite{Linde:1980ts,Gross:1980br}, and hence we focus on these modes.
Counting the number of the transverse modes of the $SU(2)_R$ gauge bosons, we obtain
\cite{Dine:1992wr}
\begin{equation}
\begin{split}
E \simeq  \frac{3g_R^3}{16\pi} \; .
\label{eq:cubic term}
\end{split}
\end{equation}
The nonzero $E$ drives a fluctuation-induced first-order phase transition
when the minimum at $\phi_R = v_R (T)$ is energetically favored.

One needs to check the validity of the perturbative analysis to determine the order of the phase transition.
In fact, it has been shown in refs.~\cite{Weinberg:1974hy,Linde:1980ts,Gross:1980br,Arnold:1994bp}
that a long-wavelength fluctuation can lead to the breakdown of perturbative expansion in finite-temperature field theories.
For example, the perturbative analysis indicates that the electroweak phase transition in the SM with the observed Higgs mass
is of the weak first order, but lattice studies~\cite{Kajantie:1995kf,Kajantie:1996qd,Rummukainen:1998as} have confirmed that it is actually a crossover transition.
In the present $SU(2)$ Higgs-Kibble model, non-perturbative effects cannot be neglected for $g_R^2/\lambda_R <1$.
Fortunately, as we will see in the next subsection, the sphaleron decoupling condition that must be satisfied to realize baryogenesis requires  $g_R^2/\lambda_R\gg1$.
Therefore, in the parameter region satisfying this condition,
the perturbative analysis is valid and the $G_{\rm LR}\to G_{\rm SM}$ phase transition is expected to be of the first order.

We now calculate $v_R (T_{\rm n})/ T_{\rm n}$, where $T_{\rm n}$ is the bubble nucleation temperature of the $G_{\rm LR}\to G_{\rm SM}$ phase transition.
This quantity is important in the discussion of the sphaleron decoupling condition.
For an illustrative purpose,
we first obtain an approximate expression of $v_R (T_{\rm n})/ T_{\rm n}$
by identifying it with $v_R (T_R)/ T_R$, where $T_R$ is the critical temperature at which the two vacua are degenerated with each other.
The effective potential at $T = T_R$ can be written as
\begin{equation}
\begin{split}
V(T_R) = \frac{\lambda_R (T_R)}{4}\phi_R^2 \left(\phi_R - v_R (T_R) \right)^2 \; .
\label{eq:critical tem effective potential}
\end{split}
\end{equation}
By comparing this form of the potential to \eqref{eq:effective potential}, we obtain
\begin{equation}
\begin{split}
\frac{v_R (T_{\rm n})}{T_{\rm n}} \simeq \frac{v_R (T_R)}{T_R}
= \frac{2E}{\lambda_R (T_R)}\simeq \frac{2E}{\lambda_R} \; .
\label{eq:the fraction}
\end{split}
\end{equation}
In the last equality, we neglect finite-temperature corrections to the quartic coupling $\lambda_R$. As we will see, $v_R(T_{\rm n})/T_{\rm n}$ is required to be large enough, which requires small $\lambda_R$.

We note that eq.~\eqref{eq:the fraction} must be regarded as a rough estimate.
As we will see in the next subsection, the sphaleron decoupling condition requires
a large $v_R (T_{\rm n} )/ T_{\rm n}$ which is realized for a tiny $\lambda_R$ and a large $g_R$.
However, for such a large $v_R (T_{\rm n} )/ T_{\rm n}$,
the high-temperature expansion is not justified.
Furthermore, the contribution from the Coleman-Weinberg potential is significant for a tiny $\lambda_R$.
In our numerical calculations, we estimate $v_R (T_{\rm n} )/ T_{\rm n}$ by using the total effective potential without resorting to the high-temperature expansion.

Although $T_{\rm n}$ is typically very close to $T_R$, it can be rigorously determined by evaluating the decay rate of the false vacuum.
The decay rate per unit volume and per unit time is expressed as
\begin{align}
\Gamma_{B} = \mathcal{A}(T) e^{-\mathcal{B}},
\end{align}
where $\mathcal{A}(T)\sim T^4$ and $\mathcal{B}$ is a bounce configuration. 
Without a significant supercooling, the bounce is given by the $O(3)$ symmetric solution, $\mathcal{B}=S_3/T$
\cite{Linde:1980tt},
which is obtained by solving the differential equation,
\begin{align}
\frac{d^2 \phi_R}{dr^2} +\frac{2}{r}\frac{d\phi_R}{dr} = \frac{\partial V_{\rm eff}}{\partial \phi_R} \, , \label{eq:bounce eq}
\end{align}
with the boundary conditions,
\begin{align}
\phi_R (r\to \infty) = \phi_R^{\rm false} \, , \qquad \left.\frac{d\phi_R}{dr}\right|_{r=0} = 0 \, , \label{eq:bounce boundary}
\end{align}
where $\phi_R^{\rm false}$ denotes the position of the false vacuum.
The action is then obtained as
\begin{align}
S_3= \int^\infty_0 dr 4\pi r^2 \left(\frac{1}{2}\left( \frac{d\phi_R}{dr} \right)^2 + V_{\rm eff} (\phi_R)  \right).
\end{align}
Here, $\phi_R$ is the solution of eqs.~\eqref{eq:bounce eq} and \eqref{eq:bounce boundary}.
The nucleation temperature $T_{\rm n}$ is defined through $\Gamma_{B}(T_{\rm n}) = H^4(T_{\rm n})$ where $H(T)= 1.66 \sqrt{g_*} \, T^2/ m_{\rm Pl}$ is the Hubble parameter
with $m_{\rm Pl} \simeq 1.22\times10^{19} \gev$ and $g_*$ being the Planck mass and the number of the effective degrees of freedom
for the radiation energy density.
We find
\begin{align}
\left. \frac{S_3}{T} \right|_{T=T_{\rm n}}\simeq 130 \, . \label{eq:nucleation temperature}
\end{align}
Here we take $T_{\rm n} = 5 \, \rm TeV$ and $g_\ast = 143$ as reference values. 

\subsection{The sphaleron decoupling condition}\label{sec:SFOPT}

The model has not only the ordinary electroweak sphaleron process~\cite{Klinkhamer:1984di} but also the sphaleron process associated with the $SU(2)_R$ gauge interaction.
In the $G_{\rm LR}$ symmetric phase at $T>T_{\rm n}$,
the baryon and lepton number violations by the $SU(2)_R$ sphaleron processes are not suppressed.
We can rewrite \eqref{eq:anomaly} as 
\begin{equation}
\begin{split}
&\partial^\mu j^{(B-L)}_\mu =  -\frac{2}{32\pi^2} \, g_R^2 W_{R \mu\nu} \widetilde{W}_R^{\mu\nu} \; , \\[1ex]
&\partial^\mu j^{(B+L)}_\mu =  \frac{6}{32\pi^2} \, g_L^2 W_{L \mu\nu} \widetilde{W}_L^{\mu\nu} + \frac{8}{32\pi^2} \, g_R^2 W_{R \mu\nu} \widetilde{W}_R^{\mu\nu} \; ,
\label{eq:anomaly1}
\end{split}
\end{equation}
where we neglect the contribution from the $U(1)_X$ gauge field.
We can see that $B-L$ is violated by the $SU(2)_R$ sphaleron process and $B+L$ is violated by both the $SU(2)_L$ and $SU(2)_R$ sphaleron processes.

As the Universe cools down, the $G_{\rm LR}\to G_{\rm SM}$ phase transition occurs
at the nucleation temperature $T_{\rm n}$.
In the broken phase, the $SU(2)_R$ sphaleron transition rate is 
\begin{equation}
\begin{split}
\Gamma_{\rm sph}^R (T)= A(T) \, e^{-\frac{E^R_{\rm sph}(T)}{T}} .
\label{reaction rate}
\end{split}
\end{equation}
Here $E^R_{\rm sph} (T)$ is the sphaleron energy,
which is given by
\cite{Klinkhamer:1984di}
\begin{equation}
\begin{split}
E_{{\rm sph}}^R (T) = \frac{4\pi v_R(T)}{g_R} B(\lambda_R/g_R^2),
\label{sphaleron energy}
\end{split}
\end{equation}
where $B(x)$ is a function with $B(0) = 1.6$ and $B(\infty ) = 2.7$.
The expression of $A(T)$ can be found in refs.~\cite{Funakubo:2009eg, Fuyuto:2014yia}.
For successful baryogenesis, the $B-L$ violating process by the $SU(2)_R$ gauge interaction must be decoupled
in the broken phase.
Such a sphaleron decoupling condition is conservatively given by $\Gamma^R_{\rm sph} (T_{\rm n}) < H(T_{\rm n})$.
Using the reaction rate \eqref{reaction rate} and the form of the sphaleron energy \eqref{sphaleron energy},
the condition is given by
\begin{equation}
\begin{split}
\frac{v_R(T_{\rm n})}{T_{\rm n}} \gtrsim 40\times \frac{g_R}{4\pi B(\lambda_R/g_R^2)} \; .
\label{eq:sphaleron decoupling condition}
\end{split}
\end{equation}
Here we take $T_{\rm n} = 5 \, \rm TeV$ and $g_\ast = 143$ as reference values.
%%%%%%%%%%%%%%%%%%%%%%%%%%%%%%%%%%%%%%%%%%%%%%%%
\begin{figure}[t]
\centering
\begin{minipage}[h]{0.49\linewidth}
        \includegraphics[width=2.5in]{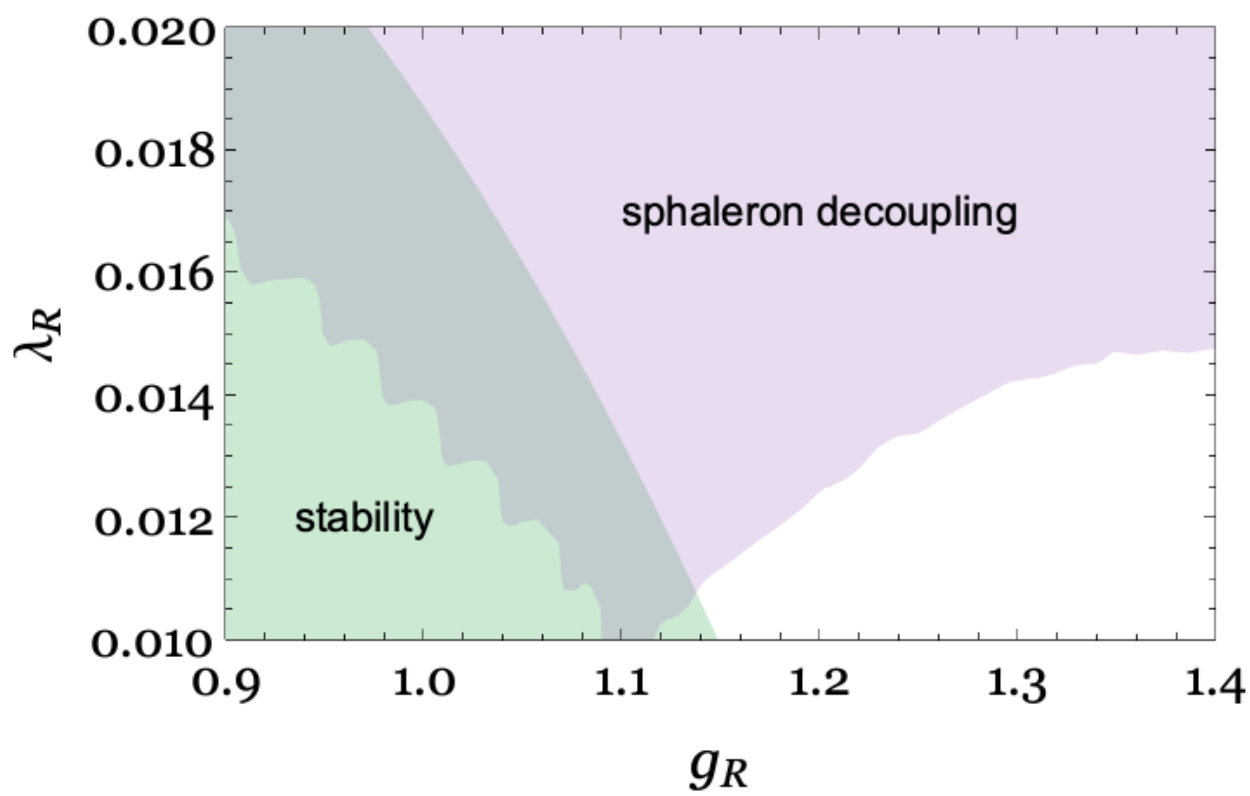}
    \end{minipage}
    \begin{minipage}[h]{0.49\linewidth}
        \includegraphics[width=2.5in]{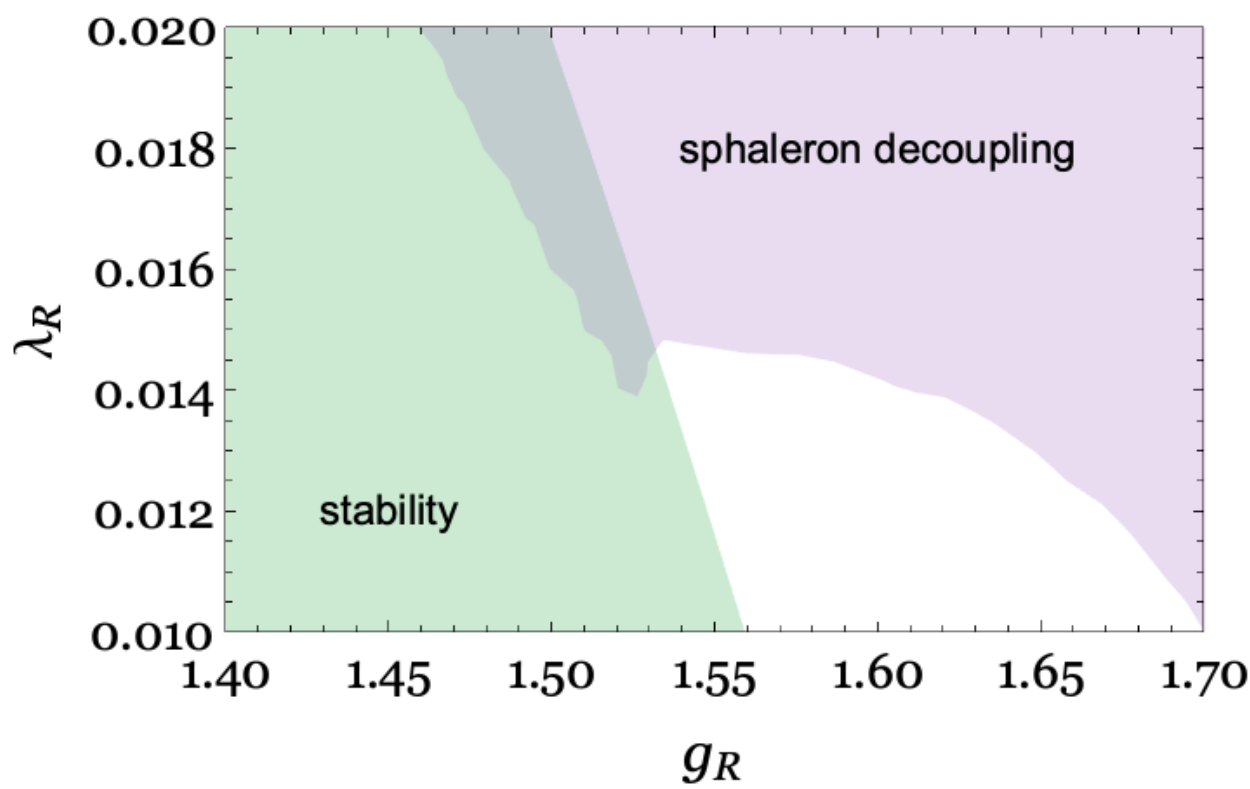}
    \end{minipage}
\caption{
The region of $g_R$ and $\lambda_R$ excluded by the sphaleron decoupling condition (purple)
and the stability of the zero-temperature effective potential below $10 v_R = 150\, {\rm TeV}$ (green).
We take $y_{E_3}=0.8$, $y_{E_2}=0.7$, $y_{N_2}=0.7$ and $\ynth=0.8$ in the left panel
and $\yntwo=0.89$ and $\ynth=1.35$ in the right panel.
}
\label{fig:decoupling}
\end{figure}
%%%%%%%%%%%%%%%%%%%%%%%%%%%%%%%%%%%%%%%%%%%%%%%%
Fig.~\ref{fig:decoupling} shows the region of $g_R$ and $\lambda_R$ which satisfies
the sphaleron decoupling condition \eqref{eq:sphaleron decoupling condition} with large Yukawa couplings $y_{E_3} = 0.8,~y_{E_2}=0.7,~\ynth=0.8,~\yntwo = 0.7$ (left) and $\ynth=1.35,~\yntwo=0.89$ (right).
In the right panel of the figure, $y_{E_3}$ is assumed to be much smaller than $y_{N_{2,3}}$
and ignored.
The VEV is set to be $v_R = 15 \tev$ in both panels.
The thermal effective potential is evaluated by using the method outlined in the previous subsection.
Here we approximate $v_R(T_{\rm n}) / T_{\rm n}$ with $v_R(T_R) / T_R$, while the latter is less than
the former.
Since the quantum correction from $y_{N_3,N_2,E_3,E_2}$ negatively contributes to the running of the quartic coupling towards high energy scales,
$V(\phi_R)|_{T=0}$ is not bounded from below for a large field value.
We require that the zero-temperature effective potential is stable until $\phi_R=10 v_R$,
$V(\phi_R=0)|_{T=0}<V (10v_R)|_{T=0}$, so that a UV completion that stabilizes the potential does not affect the dynamics around the scale $v_R$.
From the figure, we can easily see that a tiny $\lambda_R$ and a large $g_R$ are required
to satisfy both the sphaleron decoupling condition and the stability of the effective potential.

In the parameter range shown in fig.~\ref{fig:decoupling}, the approximation
$v_R(T_R)/ T_R\simeq v_R(T_{\rm n}) / T_{\rm n}$ is justified,
while the approximation becomes unreliable
for a larger $g_R$ and a smaller $\lambda_R$.
For example, in the right panel of fig.~\ref{fig:decoupling}, the ratio $v_R(T_R)/ T_R$ differs from the true value of $v_R(T_{\rm n}) / T_{\rm n}$ only by $20\%$ for $g_R =1.6$ and $\lambda_R=0.01$.

%%%%%%%%%%%%%%%%%%%%%%%%%%%%%%%%%%%%%%%%%%%%%%%%
\subsection{Asymmetry from quantum reflections}
  \label{sec:reflection}
%%%%%%%%%%%%%%%%%%%%%%%%%%%%%%%%%%%%%%%%%%%%%%%%

So far, we clarified the condition of the first order phase transition as well as the sphaleron decoupling condition.
We shall now discuss the generation of the baryon asymmetry by the moving bubble wall.

During the bubble expansion, a flux of the new leptons in the symmetric phase see the bubble wall as a potential barrier and are quantum mechanically scattered.
The reflection coefficients of the new leptons and their antiparticles can be calculated
by solving their Dirac equations on the non-trivial bubble wall background configuration and on the thermal background.
We can safely assume a planer bubble wall perpendicular to a Cartesian coordinate $z$~\cite{Carrington:1993ng}.
In the rest frame of the bubble wall, the wall configuration $\phi (T,z)$ can be generally parameterized as $\phi (T,z)=v_R (T)\xi (z)$, where the function $\xi (z)$ is a monotonically increasing function and satisfies $\xi(z= -\infty)=0$ and $\xi(z= +\infty)=1$.
In this parameterization, $z\to - \infty$ ($z\to +\infty$) corresponds to the symmetric (broken) phase.
After integrating out the interaction of a fermion with the thermal bath, the fermion can be described as a collective excitation called a {\it quasiparticle}.
The propagation and scattering of the new leptons with the wall are governed by the following effective Dirac equation,
\begin{equation}
\begin{split}
\begin{pmatrix}
\left( \omega +{\boldsymbol \sigma}\cdot \bm{p} \right) \mathbb{1} + \Sigma_L (\omega, \bm{p}) & M(T,z)\\
M^\dag (T,z)& \left( \omega -{\boldsymbol \sigma}\cdot \bm{p} \right) \mathbb{1} +\Sigma_R (\omega,\bm{p})
\end{pmatrix}
\begin{pmatrix}
\Psi_L\\
\Psi_R
\end{pmatrix}
=0 \; . \label{eq:Dirac eq}
\end{split}
\end{equation}
Here, $\psi^i_{L(R)} \equiv (\Psi_{L(R)})_i \,\, (i = 1,2,3)$ denotes the left (right)-handed new lepton $\bar{E}_i$
($E_i^*$) or $\bar{N}_i$ ($N_i^*$).
The thermal self-energy for the left (right)-handed new lepton $\Sigma_{L(R)}(\omega,\bm{p})$
and the $z$-dependent new lepton mass $M(T,z)\equiv M(T) \xi (z)$ are
$3 \times 3$ matrices. We take the mass eigenstate basis, $M_{N(E)}={\rm diag}(m_{N_1 (E_1)}(T),m_{N_2 (E_2)}(T),m_{N_3 (E_3)}(T))$ for the $N (E)$ particles.
$\mathbb{1}$ is the $3 \times 3$ unit matrix.

The lepton number density generated in front of the bubble wall $n^r_L$ is expressed
in terms of the reflection coefficient matrix $R^{ij}_{LR}~(R^{ij}_{RL})$
for the left (right)-handed new lepton $\psi^i_{L(R)}$ reflected into
the right (left)-handed $\psi^j_{R(L)}$.\footnote{
In sec.~\ref{sec:large momentum regime}, we will see that the left-handed new lepton can be reflected into the left-handed new lepton in the presence of abnormal modes with a large momentum. However, such a process gives a negligible contribution to the resulting baryon asymmetry, and hence, we can ignore it.
}
Generically, the reflection coefficient can be obtained by solving the Dirac equation with an appropriate boundary condition corresponding to the injection of $\psi_{L(R)}$ from the symmetric phase and then extracting the reflected component of $\psi_{R(L)}$. 
The calculation of the reflection coefficient including the definition of the boundary condition of eq.~\eqref{eq:Dirac eq} will be discussed in sec.~\ref{sec:decoherence effect}.

A moving wall reflects the left-handed new leptons in the symmetric phase with a probability ${\rm Tr}[R_{LR}^\dag R_{LR}]$ and the left-handed anti-new leptons with a probability ${\rm Tr}[\bar{R}_{LR}^\dag \bar{R}_{LR}]$
where $\bar{R}_{LR}$ is the CP conjugation of $R_{LR}$.
For the right-handed (anti-)new leptons, $L$ and $R$ are exchanged.
Then, the lepton number density $n^r_L$ is given by
\begin{equation}
\begin{split}
n^r_L =\int \frac{d\omega}{2\pi} \left\{n_L (\omega) {\rm Tr} [R^\dag_{LR}R_{LR}-\bar{R}^\dag_{LR} \bar{R}_{LR}] + (L \leftrightarrow R) \right\} \; ,
\label{eq: lepton number density}
\end{split}
\end{equation}
where $n_{L(R)}$ is the distribution function of the left (right)-handed new lepton in the symmetric phase.
We assume that the bubble wall velocity is small, $v_w \equiv |\bm{v}_w| \ll 1$, and work in the wall rest frame.
To the first order in $v_w$, the Lorentz-boosted distribution function is given by
$n(\omega) = n_0(\omega+\bm{p}\cdot \bm{v}_w)$ where $\bm{p}$ is the momentum carried by the new lepton
and $n_0 (\omega) \equiv 1/(e^{\omega/T}+1)$ is the Fermi-Dirac distribution function.
The CPT invariance requires $R_{LR} =\bar{R}_{RL}$ which
simplifies \eqref{eq: lepton number density} as
\begin{equation}
\begin{split}
n^r_L &= -\int \frac{d\omega}{2\pi}\left(n_L (\omega)- n_R (\omega) \right)\Delta (\omega) \\[1ex]
&\simeq -\int \frac{\rmd \omega}{2\pi}n_0(\omega)[1 - n_0(\omega)]\frac{(\bm{p}_R - \bm{p}_L) \cdot \bm{v}_w}{T} \Delta(\omega) + \mathcal{O}(v_w^2) \; , \\[1.5ex]
\Delta (\omega) &\equiv {\rm Tr}[\bar{R}^\dag_{LR} \bar{R}_{LR}-R^\dag_{LR}R_{LR}] \; .
\label{eq:reflection number} 
\end{split}
\end{equation}
Here, $\omega$ and $\bm{p}_{L(R)}$ are the energy and momentum of the injected left (right)-handed new lepton.
We can see that a non-zero wall velocity and a CP-violating $\Delta (\omega)$ are needed to realize a non-zero $n^r_L$. 

Since the lepton number is globally conserved in the wall scattering process,
the asymmetry in the symmetric phase ($n^r_L$) and that in the broken phase ($-n^r_L$)
add up to zero.
However, the generated lepton number in front of the bubble wall diffused into the symmetric phase
is partly washed-out by the $SU(2)_R$ sphaleron process until it is captured by the moving bubble wall,
otherwise there is no lepton number asymmetry after the completion of the phase transition.
In the wall rest frame, the diffusion length of the lepton number is given by $\sqrt{D_L t}$,
where $D_L$ is the diffusion constant,
while the bubble wall travels a distance $v_w t$.
Then, the typical time scale in which the moving bubble wall captures the diffused lepton number is
given by
\begin{align}
\label{eq:tdiff}
    \tau_{\rm diff} \equiv D_L/ v_w^2.
\end{align}

To see the development of the lepton number density,
we first consider only the diffusion effect, assuming that
the time scale of the $SU(2)_R$ sphaleron process is longer than the diffusion time scale $\tau_{\rm diff}$.
The lepton number density is then governed by a diffusion equation
\cite{Farrar:1993hn,Cline:1995dg}
\begin{align}
\frac{\partial}{\partial t}n_L (t,z) = D_L \frac{\partial^2}{\partial z^2} n_L (t,z) \, .
\label{eq:diffusion equation}
\end{align}
The boundary condition is given by the number density at the wall, $n_L(t,z = v_w t) = n_L^r$.
In our model, the diffusion constant $D_L$ is dominated by
the contribution of the $W_R$ gauge boson
\cite{Joyce:1994zn},
\begin{align}
  \label{eq:Diffusion constant}
  D^{-1}_{L} = \frac{45}{7\pi} \alpha^{2}_R T \log \left(\frac{32 T^2}{M^{2}_{\rm Debye}}\right),
\end{align}
where $M^{2}_{\rm Debye} = 13 g_R^2 T^2/6$ is the Debye mass of the $W_R$ gauge boson~\cite{Weldon:1982aq,Weldon:1982bn}. 
To solve \eqref{eq:diffusion equation}, we take the ansatz that the lepton number density
is a function of $z'\equiv z+v_w t$ and rewrite the equation in terms of $z'$, 
\begin{align}
D_L \frac{\partial^2}{\partial z'^2} n_L(z') = v_w \frac{\partial}{\partial z'}n_L(z') \, .
\end{align}
The solution of this equation is
\begin{align}
  n_L(z')=n_L^r e^{v_w z'/D_L} \, . \label{eq:lepton number density distribution}
\end{align}

We now turn on the $SU(2)_R$ sphaleron process which converts $L$ into $B-L$.
The rate of the $SU(2)_R$ sphaleron process in the symmetric phase per unit time and per unit volume is
given by
\begin{align}
\Gamma^R_{\rm sph} = \kappa \alpha_R^5 T^4, \label{reactionrate}
\end{align}
where $\alpha_R \equiv g_R^2 / 4 \pi$ and $\kappa\simeq 20$ is a numerical coefficient \cite{Moore:1997sn,Bodeker:1999gx,Moore:2000ara,DOnofrio:2014rug}.
In the broken phase, the $SU(2)_R$ sphaleron process is decoupled and $B-L$ is conserved.
Then, the time evolution of the $B-L$ number density $n_{B-L}(z,t)$ outside ($z'<0$) and inside ($z'>0$) the bubble wall is governed by the following equation:
\begin{align}
    \partial_t n_{B-L}(z,t)  = 
    \begin{cases}
    \frac{6 \Gamma^R_{\rm sph}}{T^3}n_L(z^\prime)~~&(z'<0)\\
    0~~&(z'>0) \, .
    \end{cases}\label{eq:sphaleron wash-out}
\end{align}
Here, we used the formulation in~\cite{Cline:1995dg,Joyce:1994zn} assuming that the transfer of $L_{NE}$ by the Yukawa interactions of $\bar{L}$, $E$, and $N$ with other leptons within the diffusion time scale is not efficient and the only interaction that reduces $L_{NE}$ is the $SU(2)_R$ sphaleron. If the Yukawa interactions are efficient, the wash-out rate is reduced by an $O(1)$ factor.
At a given position $z$, we integrate eq.~\eqref{eq:sphaleron wash-out} with respect to $t$ from $t=-\infty$ ($z'=-\infty$) to $t=-z/v_w$ ($z'=0$) where the $SU(2)_R$ sphaleron process is decoupled.
By changing the integration variable from $t$ to $z'$, we obtain
\begin{align}
\label{eq:final lepton asymmetry}
n_{B-L}(z'=0)= \frac{6\Gamma^R_{\rm sph}}{T^3 v_w}\int^{0}_{-\infty} dz^\prime n_L^r e^{v_w z^\prime /D_L}=\frac{6\Gamma^R_{\rm sph}}{T^3} \frac{D_L}{v_w^2} n_L^r \, .
\end{align}
In this calculation, we have used $n_{B-L}(z'=-\infty)=0$.
Since we have assumed that the $SU(2)_R$ sphaleron time scale $\tau_{\rm sph}=( 6\Gamma_{\rm sph}/T^3)^{-1}$ is longer than $\tau_{\rm diff}$, the resultant $B-L$ number density is suppressed by $\tau_{\rm diff}/\tau_{\rm sph}<1$.
On the other hand, when $\tau_{\rm diff} > \tau_{\rm sph}$, we expect that the lepton number density diffused into the symmetric phase is almost washed-out by the $SU(2)_R$ sphaleron process.
In this case, there would be no suppression factor from the diffusion, $n_L\simeq n^r_L$.
We thus finally obtain
\begin{align}
    n_{B-L}\simeq {\rm min}\left\{1,~\frac{\tau_{\rm diff}}{\tau_{\rm sph}}\right\}n^r_L \, . \label{diffusion suppression}
\end{align}

The asymmetry of the new leptons is converted into that of the SM leptons by the scatterings or decays of the new leptons through the portal couplings in eq.~(\ref{portal interaction}) or (\ref{portal interaction2}). As long as this conversion efficiently occurs before the electroweak phase transition, the produced lepton asymmetry is further converted into the baryon asymmetry via the ordinary electroweak sphaleron process.
This condition is easily satisfied. For example, the last two terms in eq.~(\ref{portal interaction}) mix the new leptons with the SM leptons after $H_R$ gets a VEV. The mixing converts the asymmetry of $E, N$ into that of the SM leptons with the rate,
\begin{align}
    \label{eq:decay rate}
    \Gamma_{E,N} \sim \frac{1}{32 \pi}\frac{g_L^2g_{E,N}^2 v_R^2}{m_{E,N}} \, ,
\end{align}
where $g_L$ is the $SU(2)_L$ gauge coupling.
This rate is larger than the Hubble expansion rate around the electroweak phase transition if $g_{E,N} > 10^{-9}$. We assume this case in the following discussions.
As discussed in sec.~\ref{sec:DM}, if the $Z_2$ symmetry to obtain a dark matter candidate is introduced, the lepton asymmetry of the SM fermions is generated by the decay of $\bar{\mathcal{X}}$ and $\bar{\mathcal{E}}$ leptons, whose asymmetry is produced from that of $E$ and $N$ by the $SU(2)_{R}$ sphaleron processes, through the first term of eq.~(\ref{portal interaction}) or (\ref{portal interaction2}).

The amount of the baryon asymmetry is estimated by imposing the chemical equilibrium condition including the electroweak sphaleron process.
Assuming a crossover electroweak phase transition, the baryon asymmetry is given by~\cite{Harvey:1990qw}
\begin{equation}
\begin{split}
\frac{n_B}{s} \simeq \frac{12}{37}\frac{n_{B-L}}{s} = \frac{12}{37} \mathrm{min}\left \{1,\frac{6 \Gamma_{\rm sph}^R}{T^3}\frac{D_L}{v_W^2}\right \}\frac{n_L^r}{s}. \label{eq:baryon asymmetry}
\end{split}
\end{equation}
Even if the electroweak phase transition is of the strong first order, the result is qualitatively unchanged.
In the above equation, we find that a large $SU(2)_R$ gauge coupling makes the produced baryon asymmetry large because the reaction rate of the $SU(2)_R$ sphaleron process in the symmetric phase \eqref{reactionrate} is very sensitive to the coupling.

%%%%%%%%%%%%%%%%%%%%%%%%%%%%%%%%%%%%%%%%%%%%%%%%
\subsection{Thermal self-energy}
  \label{sec:Delta}
%%%%%%%%%%%%%%%%%%%%%%%%%%%%%%%%%%%%%%%%%%%%%%%%

We here estimate the thermal self-energy in the Dirac equation \eqref{eq:Dirac eq}.
The real part of the thermal self-energy $\Sigma_{L(R)}$ with high-temperature expansion, $\omega,~ p\equiv |\bm{p}| \ll T$, is expressed as~\cite{Farrar:1993sp}
\begin{align}
    {\rm Re} \, \Sigma_{L,R}(\omega,\bm{p})=\Omega^2_{L,R}\left[\pm \frac{\bm{\sigma}\cdot\bm{p}}{p^2}\left(1-F\left(\frac{\omega}{p}\right)\right)-\frac{1}{\omega}F\left(\frac{\omega }{p}\right)\right]. \label{eq:thermal self-energy}
\end{align}
Here, the function $F$ is given by
\begin{align}
~F(x)\equiv \frac{x}{2}\log\left[\frac{x+1}{x-1}\right], \label{eq:thermal self-energy2}
\end{align}
and $\Omega_{L(R)}$ is the thermal mass matrix for the left (right)-handed new lepton.
Since the new lepton sector is chiral under the $SU(2)_R\times U(1)_X$ symmetry,
the thermal masses of the left and right-handed new leptons are different.
We define $\Omega_{R}\equiv \Omega_{\bar{L}}$ for the right-handed new leptons and $\Omega_{L}\equiv \Omega_{E(N)}$ for the left-handed $E~(N)$ leptons.
At a high temperature, the leading order thermal masses of the $\overline{L},~E$, and $N$ leptons
are~\cite{Weldon:1982bn}
\begin{equation}
\begin{split}
  \Omega_{\bar{L}}^2 &\equiv (\Omega_{\bar{L},0} + \Delta \Omega_{\bar{L}})^2 = \Omega_{\bar{L},0}^2 + 2 \Omega_{\bar{L},0} \Delta \Omega_{\bar{L}}\\[1ex]
  &= \frac{\pi \alpha_R T^2}{2}\left(\frac{3}{4} + \frac{\tan^2 \theta_X}{4} \right) \mathbb{1}
  + \frac{\pi \alpha_R T^2}{2} \frac{M_E^\dagger M_E
  + K M_N^\dagger M_N K^{\dagger}}{4 m_{W_R}^2} \; , \\[1ex]
  \Omega^{2}_E &\equiv (\Omega_{E,0} + \Delta \Omega_{E})^2 = \Omega_{E,0}^2 + 2 \Omega_{E,0} \Delta \Omega_{E}
  = \frac{\pi \alpha_R T^2}{2} \tan^{2 }\theta_X \mathbb{1}
  + \frac{\pi \alpha_R T^2}{2} \frac{M_E^\dagger M_E}{2m_{W_R}^2} \; ,\\[1ex]
  \Omega^{2}_N &\equiv \Delta \Omega_{N}^2
  =  \frac{\pi \alpha_R T^2}{2} \frac{M_N^\dagger M_N}{2m_{W_R}^2},
  \label{eq:OmegaL}
\end{split}
\end{equation}
where $\tan \theta_X \equiv g_X/g_R$ is defined just like the weak mixing angle in the SM.
The matrix $K$ is defined in \eqref{eq:CKM parametrization}.
$\Omega_{\bar{L}, 0}^2$ and $\Omega_{E, 0}^2$ are generation independent parts proportional to the unit $3 \times 3$ matrix.
For later use, we define
\begin{equation}
\begin{split}
\Omega_+ \equiv \frac{\Omega_{{L}, 0}+\Omega_{R,0}}{2}  \; ,
\qquad \Omega_- \equiv \frac{\Omega_{R,0} -\Omega_{L,0}}{2} \; .
\end{split}
\end{equation}
Generation dependent parts are given by $\Delta \Omega_{\bar{L}}^2$, $\Delta \Omega_{E}^2$ and $\Delta \Omega_{N}^2$. 

Let us now consider the dispersion relations of the quasiparticles.
The presence of the thermal masses $\Omega_{L(R)}$ changes the dispersion relations from those at zero temperature.
For simplicity, we here neglect the flavor dependent interactions.
The dispersion relations for the left (right)-handed new leptons in the symmetric phase
$\omega^s_{L(R), \pm}$ and in the broken phase $\omega^b_{\pm, \pm}$ are obtained
by setting the determinant of the effective Dirac equation \eqref{eq:Dirac eq} to zero.
For a small momentum, since $\omega \simeq \Omega_{L(R)}$ when $p\simeq 0$, the thermal self-energy can be approximated as
\begin{align}
    \mathrm{Re}\ \Sigma_{L(R)} = (\omega - 2\Omega_{L(R)})\mathbb{1}_{2\times 2} \mp \frac{\bm \sigma \cdot \bm p}{3}, \label{eq:small momentum approximation}
\end{align}
where $-$ and $+$ are for $L$ and $R$, respectively.
With this approximation, we obtain
\begin{align}
\omega^s_{L(R), \pm} &= \Omega_{L(R), 0} \pm \frac{p}{3} \; , \qquad z<0 \; , \label{eq:dispersion relation in s}\\[1ex]
\omega^b_{\pm, \pm} &= \Omega_+ \pm \sqrt{\frac{m^2}{4}+\left(\Omega_- \pm \frac{p}{3}\right)^2} \; , \qquad z>0 \; ,
\label{eq:dispersion relation in b}
\end{align}
where $p \equiv |\bm{p}|$ and $m$ is the Dirac mass of the new lepton.
We have approximated a thermal self-energy in the broken phase as the one in the symmetric phase.
This approximation is valid for small Yukawa couplings.
The dispersion relations show that a quasiparticle has two modes;
one is a dressed fermion with a positive group velocity $d\omega / dp > 0$ called a normal mode while 
the other is a hole state of an anti-fermion with the same chirality
but a negative group velocity $d\omega / dp < 0$ called an abnormal mode~\cite{Weldon:1989ys}.

%%%%%%%%%%%%%%%%%%%%%%%%%%%%%%%%%%%%%%%%%%%%%%%%%%%%%%%%%%%%%
\begin{figure}[t]
\centering\includegraphics[width=8cm]{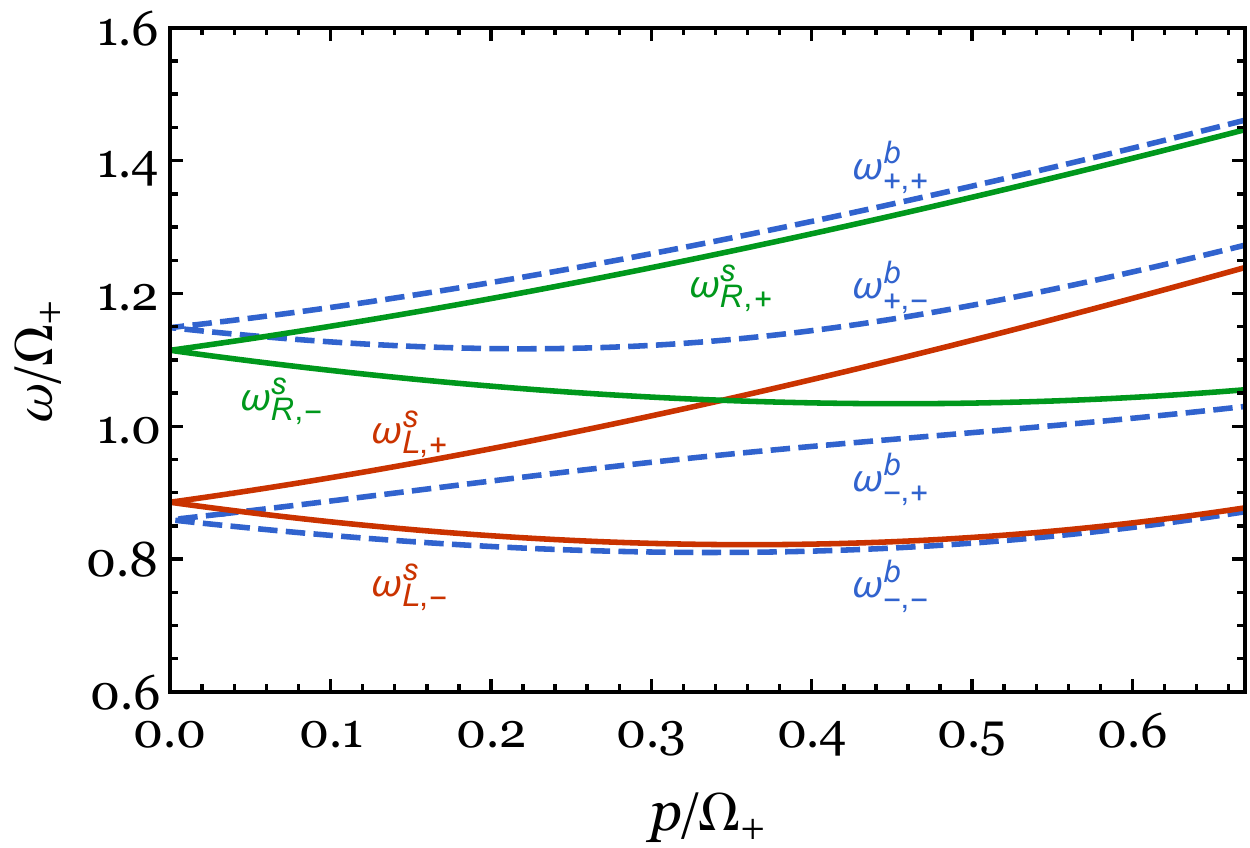}
\caption{
The dispersion curves for an $E$ lepton in the symmetric phase $\omega^s_{R+}$, $\omega^s_{R-}$ (green solid), $\omega^s_{L+}$, $\omega^s_{L-}$ (red solid) and in the broken phase $\omega^b_{+,+}$, $\omega^b_{+,-}$,
$\omega^b_{-,+}$, $\omega^b_{-,-}$ (blue dotted).
The definitions of $\omega$ are given in \eqref{eq:dispersion relation in s} and \eqref{eq:dispersion relation in b}.
We here take $g_R=0.6$, $y_{E}=0.015$ and $v_R(T_R)/T_R=3$.}
\label{fig:dispersion curves}
\end{figure}
%%%%%%%%%%%%%%%%%%%%%%%%%%%%%%%%%%%%%%%%%%%%%%%%%%%%%%%%%%%%%

The dispersion curves in the symmetric and in the broken phase are depicted in Fig.~\ref{fig:dispersion curves}
for an $E$ lepton.
In the symmetric phase, the positive branch in \eqref{eq:dispersion relation in s} corresponds to
a normal mode for each of the left and right-handed
while the negative branch is an abnormal mode.
In the broken phase, the $(+,+)$ and $(-,-)$ branches in eq.~\eqref{eq:dispersion relation in b}
are normal modes for $p<3\Omega_-$, and
in the absence of the Dirac mass they are right and left-handed respectively.
The $(+,-)$ and $(-,+)$ branches are right and left-handed abnormal modes for $p < 3\Omega_-$.
Since the Dirac mass mixes the left and right-handed modes in the broken phase, the $(+,-)$ and $(-,-)$ branches become a left-handed normal mode and a right-handed abnormal mode for $p>3 \Omega_-$, respectively, at $p=3 \Omega_-$ where the level crossing takes place in the symmetric phase.
For the $(+,+)$ and $(-,+)$ branches, such a level crossing does not take place.

We next discuss the imaginary part of the thermal self-energy and the resultant decoherence effects.
As originally pointed out in refs.~\cite{Gavela:1993ts,Huet:1994jb,Gavela:1994ds,Gavela:1994dt}, incoherent scatterings of particles with medium destroy the quantum mechanical CP-violating process, which suppresses $\Delta(\omega)$.
It has been shown in ref.~\cite{Gavela:1993ts} that this decoherence is captured by the imaginary part of the thermal self-energy.
At zero momentum, the dominant contribution to the imaginary part of the thermal self-energy for each of $\bar{L},E$ and $N$, ${\rm Im} \Sigma_{\bar{L},E,N} \equiv 2 \gamma_{\bar{L},E,N}$, is given by~\cite{Braaten:1992gd,Joyce:1994zn}
\begin{align}
  &\gamma_{\bar{L}} \simeq \alpha_RT\left(1 + \frac{3}{8}\tan^2 \theta_X\right), \nonumber \\[1ex] 
  &\gamma_{E} \simeq \frac{3}{2}\alpha_R T  \tan^2 \theta_X, \label{eq:decay E}\\[1ex] 
  &\gamma_{N}\simeq \frac{T}{64\pi}y_N^2, \nonumber 
\end{align}
where $y_N={\rm diag}(y_{N_1},y_{N_2},y_{N_3})$ is the diagonalized Yukawa coupling of the $N$ lepton.
Note that the dominant contribution to the damping rate $\gamma_{\bar{L},E}$ comes from the $SU(2)_R$ and $U(1)_X$ gauge bosons
and is flavor independent while the damping rate $\gamma_{N}$ arises from the flavor-dependent Yukawa coupling.

%%%%%%%%%%%%%%%%%%%%%%%%%%%%%%%%%%%%%%%%%%%%%%%%
\subsection{Estimation of $\Delta(\omega)$}
  \label{sec:decoherence effect}
%%%%%%%%%%%%%%%%%%%%%%%%%%%%%%%%%%%%%%%%%%%%%%%%

We now estimate $\Delta(\omega)$ by considering the reflection problem by the wall.
$\Delta (\omega)$ can be decomposed into two contributions $\Delta (\omega)=\Delta_E (\omega) +\Delta_N (\omega)$, where $\Delta_{E,N} (\omega)$ is the contribution from the $E$, $N$ lepton scattering, respectively.
As we will see later in this subsection, the estimation of $\Delta_E (\omega)$ depends on the damping rates $\gamma_{\bar{L},E}$,
the thermal masses $\Omega_{\pm}$, and the new lepton mass $m_{E_i}$ because they determine the propagation of the quasiparticles.
We assume that $m_{E_2}$ and $m_{E_3}$ are not hierarchical for simplicity, but $m_{E_1}$ can be much smaller than $m_{E_{2,3}}$.
For $m_{E_{2,3}}<4\gamma_{\bar{L},E}<\Omega_+$ (parameter regime (i)),
the effects of the damping rates and the thermal masses are significant.
For $4\gamma_{\bar{L},E}<m_{E_{2,3}}<\Omega_+$ (parameter regime (ii)), the effect of the damping rates is negligible
while the thermal masses play an important role.
For $\Omega_+<m_{E_{2,3}}$ (parameter regime (iii)), the effects of the damping rates
as well as the thermal masses are subdominant.
Furthermore, it will turn out that the reflection problem for a small $\Omega_-$ corresponding to $\tan \theta_X\simeq 1$ (parameter regime (A))
is qualitatively different from that for a large $\Omega_-$ corresponding to $\tan \theta_X\lesssim 1$ (parameter regime (B)).
We therefore estimate $\Delta_E (\omega)$ for different parameter regimes listed in table.~\ref{table:parameter}.
In each parameter regime, we consider contributions from small momentum $p \, (< l_w^{-1})$
and large $p \, (> l_w^{-1})$, where $l_w$ is the wall thickness.
The contribution of $\Delta_N (\omega)$ for the $N$ lepton will be also discussed in each parameter regime.

One should note that a nonzero $\Delta (\omega)$ arises from interference between CP-even $(\mathcal{A})$ and CP-odd $(\mathcal{B})$ contributions 
in the reflection coefficient,
\begin{align}
 R_{\rm LR}=\mathcal{A}+\mathcal{B}, \qquad \bar{R}_{\rm LR}=\mathcal{A}+\mathcal{B}^*,
\end{align}
where $\mathcal{A}$ is invariant while $\mathcal{B}$ is complex-conjugated under the CP transformation.
$\Delta (\omega)$ is given by
\begin{align}
    \Delta (\omega) = -4 \, {\rm Im}\mathcal{A} \, {\rm Im}\mathcal{B}.
\end{align}
The nonzero CP-odd phase is provided by the phase $\delta$ in \eqref{eq:CKM parametrization}, ${\rm Im}\mathcal{B}\propto\delta\neq 0$. 
The CP-even phase in $\mathcal{A}$ comes from the damping rate or total reflection as we will clarify in the following subsections.

%%%%%%%%%%%%%%%%%%%%%%%%%%%%%%%%%%%%%%%%%%%%%%%%
\begin{table}[!t]
\centering
\scalebox{.75}{
\centering
  \begin{tabular}{|l|c|c|} \hline
   &
   \begin{tabular}{c}
   (A) $\tan \theta_X \simeq1$
   \end{tabular}
   & \begin{tabular}{c}
   (B) $\tan \theta_X \lesssim 1$
   \end{tabular}
   \\ \hline
    (i) $m_{E_{2,3}} <  4\gamma_{\bar{L},E} < \Omega_+$ & 
    \begin{tabular}{l}
    \hspace{-0.7cm} Suppression from the damping rates,\\[-0.5ex]
    \hspace{-0.7cm} refs.~\cite{Gavela:1993ts,Gavela:1994ds,Gavela:1994dt,Huet:1994jb},\\[-0.5ex]
    \hspace{-0.7cm} section \ref{sec:perturbation}.
    \end{tabular}& 
    \begin{tabular}{l}
    Suppression from the finite wall size,\\[-0.5ex] 
    section \ref{sec:large momentum regime}.
    \end{tabular}\\ \hline
    (ii) $4\gamma_{\bar{L},E} < m_{E_{2,3}} < \Omega_+$ &
    \begin{tabular}{l}
    Suppression from transition amplitudes,\\[-0.5ex]
    refs.~\cite{Gavela:1993ts,Gavela:1994ds,Gavela:1994dt,Farrar:1993sp,Farrar:1993hn},\\[-0.5ex]
    section \ref{sec:heavy masses}.
    \end{tabular}&
    \begin{tabular}{l}
    Suppression from the finite wall size,\\[-0.5ex]
    section \ref{sec:large momentum regime}.
    \end{tabular}
    \\  \hline
    (iii) $ \Omega_+ < m_{E_{2,3}}$ & \multicolumn{2}{|c|}{
    No significant suppression, 
    section \ref{sec:(iii)}.} \\\hline
  \end{tabular}}
  \vspace{3mm}
  \caption{
  A dominant factor of suppression on $\Delta_E (\omega)$ in each parameter regime
  discussed in the referred section. 
  The corresponding references for the SM electroweak baryogenesis
  are also summarized.
  }
  \label{table:parameter}
\end{table}
%%%%%%%%%%%%%%%%%%%%%%%%%%%%%%%%%%%%%%%%%%%%%%%%

%%%%%%%%%%%%%%%%%%%%%%%%%%%%%%%%%%%%%%%%%%%%%%%%
\subsubsection{Perturbative regime ((i)-(A))}
  \label{sec:perturbation}
%%%%%%%%%%%%%%%%%%%%%%%%%%%%%%%%%%%%%%%%%%%%%%%%

In the parameter regime of (i)-(A) in table~\ref{table:parameter},
we will see that a dominant contribution to $\Delta_E (\omega)$ comes from the small momentum region
and the reflection coefficient is strongly suppressed by the damping rates~\cite{Huet:1994jb,Gavela:1994dt}.
Let us first identify the part of the new lepton thermal masses which contains CP violation.
The generation-dependent part of the $\bar{L}$ lepton thermal mass
arising from the exchange of the charged Higgs boson can be extracted from \eqref{eq:OmegaL},
\begin{align}
&\delta \Omega_{\bar{L}}\equiv \Delta \Omega_{\bar{L}} \simeq \frac{\pi \alpha_R T^2}{4 \Omega_{+}}\frac{M_E^2 + KM_N^2K^{\dagger}}{4 M_{W_R}^2} \; , \label{eq:right flavor thermal mass}
\end{align}
in the basis where $M_{E,N}^2$ are diagonalized.
In this calculation, we have used $\Omega_{\bar{L},0}\simeq \Omega_+$, which is valid in the parameter region (A).
We can see from eq.~\eqref{eq:OmegaL} that $\Omega_{E,N}^2$ have no $K$ dependence at the order of $T^2$.
However, in the broken phase, $\Omega_{E,N}^2$ receive contributions depending on the $K$ matrix
at the order of $\log(T)$~\cite{Gavela:1993ts},
\begin{align}
\delta \Omega_E = \frac{\alpha_R}{32\pi\Omega_+} \frac{M_E K M^{2}_{N} \ln \left(M^{2}_{N} /T^2\right) K^\dag M_E}{M_{W_R}^2} \; , \label{eq:left flavor thermal mass}
\end{align}
and $\delta \Omega_N$ with replacement of $K\leftrightarrow K^\dag$ and $M_E \leftrightarrow M_N$.
These are smaller than $\delta \Omega_{\bar{L}}$ when $M_E$ or $M_N$ are smaller than
the temperature during the phase transition.

In the following discussion, we use the small momentum approximation given by eq.~\eqref{eq:small momentum approximation} and the thin-wall approximation $\xi (z)\simeq\Theta(z)$, where $\Theta(z)$ is the Heaviside step function 
The validity of such approximations will be discussed below.
Under these approximations, the effective Dirac equation \eqref{eq:Dirac eq}
of the $E$ lepton takes the following form:
\begin{equation}
\begin{split}
\begin{pmatrix}
2\left( (\omega + i\gamma_{E}  - \widetilde{\Omega}_{L} )\mathbb{1}_{2\times 2}+ \frac{1}{3}{\boldsymbol \sigma} \cdot \bm{p}\right) & \mathcal{M}(z) \mathbb{1}_{2\times 2}\\
\mathcal{M}^\dag (z) \mathbb{1}_{2\times 2}& 2\left( (\omega +i \gamma_{\bar{L}} -\widetilde{\Omega}_{R})\mathbb{1}_{2\times 2} - \frac{1}{3}{\boldsymbol \sigma}\cdot \bm{p}\right)
\end{pmatrix}
\begin{pmatrix}
E_L\\
E_R
\end{pmatrix}
=0 \; .
\label{DiraceqE}
\end{split}
\end{equation}
In this expression, $E_{L(R)}$ denotes the left (right)-handed $E$ lepton, and $\widetilde{\Omega}_{L}\equiv \Omega_{E,0} + \Delta \Omega_E +\delta \Omega_E,~\widetilde{\Omega}_{R} \equiv \Omega_{\bar{L},0}+ \delta \Omega_{\bar{L}}$ and  $\mathcal{M}(z)\equiv 3M(z)/2$.
We take $E_L$ and $E_R$ as the eigenstates of the $z$-component of the angular momentum operator,
$j_z E_L = -1/2 E_L$ and $j_z E_R = - 1/2E_R$.
The above equation depends on momentum parallel to the wall, ${\bm p}_{\parallel}$.
Following ref.~\cite{Cline:1995dg},
we set ${\bm p}_{\parallel}=0$ and reduce the Dirac equation to the 1D problem.
This approximation is valid as long as ${\bm p}_{\parallel}$ is smaller than $\Omega_+$ so that the small momentum approximation \eqref{eq:small momentum approximation} can be justified and ${\bm p}_{\parallel}$ does not affect the dispersion relation.
Hence we impose a cutoff at $\Omega_+$ on the phase-space integration with respect to ${\bm p}_{\parallel}$,
\begin{align}
    \int_0^{\Omega_+} \frac{\rmd^2 {\bm p}_{\parallel}}{(2\pi)^2} = \frac{\Omega_+^2}{4\pi}, \label{eq:tangential contribution}
\end{align}
which gives a suppression factor for the produced lepton asymmetry. There may be a non-zero contribution from $|{\bm p}_{\parallel}| > \Omega_+$~\cite{Farrar:1993hn}, but as we will see in sec.~\ref{sec:minimal model}, the parameter region (i)-(A) produces a baryon asymmetry much smaller than the observed one, and a possible contribution from $|{\bm p}_{\parallel}| > \Omega_+$ will not help.

Let us now specify the boundary condition of eq.~\eqref{DiraceqE}.
In the presence of the non-zero damping rate, the momentum and the energy of the quasiparticle spread with a width of the order of the damping rate~\cite{Braaten:1992gd}.
This implies that no plane wave solution of eq.~\eqref{DiraceqE} describing the reflection by the wall exists, and hence, it cannot be an initial nor a final state. In particular, one cannot use the boundary condition of the wave function of the initial particle $\sim e^{i p_iz}$ at $z\rightarrow - \infty$ and read off the reflection coefficient from the coefficient of $e^{- i p_fz} $.
Instead, following ref.~\cite{Huet:1994jb}, we inject an initial particle at the symmetric phase near the wall $z=0$ with a delta-function source $\delta (z)$ and read off the reflection coefficient also at $z=0$. This prescription can correctly treat the reflection problem in the thin-wall limit since the propagation in the symmetric phase is trivial.%
\footnote{
One may instead use a wave packet with a finite size that allows a spread of the momentum and the energy due to the damping~\cite{Gavela:1994dt}. The delta-function source approximation corresponds to the limit where the size of the wave packet is smaller than the size of the spreading by the damping rate. In the present case, both are $O(\gamma^{-1})$, so the approximation gives a good order of magnitude estimation.
}
Going to the position space, the Dirac equation \eqref{DiraceqE} is written as
\begin{equation}
\begin{split}
    &(-i\partial_z\mathbb{1} -p_L)E_L (z)= -i\delta (z)E_L(0)+ \mathcal{M}(z) E_R (z) \; , \\
    &(-i\partial_z\mathbb{1} -p_R)E_R (z)= -\mathcal{M}^{\dag}(z) E_L(z) \; ,
    \label{DiracPosition}
\end{split}
\end{equation}
where $p_L \equiv 3((\omega+i\gamma_E)\mathbb{1}-\widetilde{\Omega}_{L})$ and $p_R\equiv -3((\omega+i\gamma_{\bar{L}})\mathbb{1}-\widetilde{\Omega}_{R})$.
We decompose $p_{L(R)}$ into the flavor independent part $p_{L(R)}^0$ and the flavor dependent part,
$\delta p_L = -3 (\Delta \Omega_E + \delta \Omega_E)$ and $\delta p_R = 3 \delta \Omega_{\bar{L}}$.

\begin{figure}[t]
\centering\includegraphics[width=15cm]{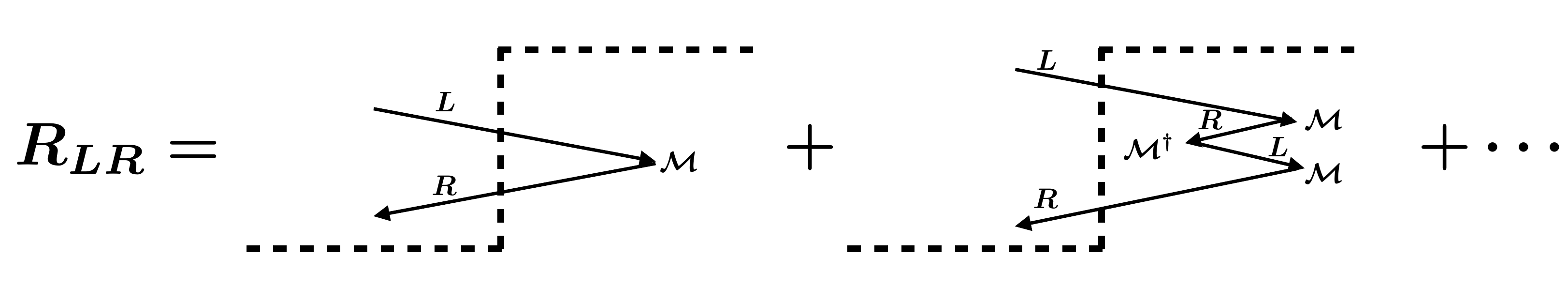}
\caption{
The schematic description of the perturbative calculation for the reflection coefficient $R_{LR}$.
The dotted line shows the bubble wall configuration.
$\mathcal{M}$ and $\mathcal{M}^\dag$ represent the mass matrix operator and its Hermitian conjugate.
The first and second terms on the right hand side correspond to $E^{(1)}_R (0)$ and $E^{(3)}_R (0)$, respectively.
}
\label{fig:scatter}
\end{figure}

Now that we have found the form of the effective Dirac equation describing the propagation of quasiparticles on the bubble wall background under the thin-wall and small momentum approximations,
let us solve the equation and calculate $\Delta_E (\omega)$.
Following the discussion of ref.~\cite{Huet:1994jb},
we iteratively solve the Dirac equation \eqref{DiracPosition}
when the Dirac mass term $\mathcal{M}$ is smaller than the damping rate.
Then, the reflection coefficient can be calculated order by order in the expansion of $\mathcal{M}$.
Let us define the propagator of the left (right)-handed new lepton satisfying
\begin{align}
    (-i\partial_z - p_{L(R)} )G_{L(R)}(z-z_0) = \mathbb{1}\delta (z-z_0) \; ,
\end{align}
where $z_0>0$ is a constant.
The solution of this equation with boundary conditions $G_L(z\to -\infty)=0$ and $G_R (z\to \infty)=0$ is given by 
\begin{equation}
\begin{split}
&G_{L}(z-z_0) =  i\Theta(z-z_0) e^{ip_L (z-z_0)}=i\Theta(z-z_0)e^{3i(\omega-\widetilde{\Omega}_L) z}e^{-3\gamma_{E} (z-z_0)} \; ,\\[1ex]
&G_{R}(z-z_0) = -i\Theta(z_0-z) e^{ip_R (z_0 - z)}=-i\Theta(z_0-z)e^{-3i(\omega-\widetilde{\Omega}_R)(z-z_0)}e^{-3\gamma_{\bar{L}} (z_0-z)} \; .
\end{split}
\end{equation}
Note that the left-handed quasiparticle propagates to the {\it positive} $z$ direction
while the right-handed quasiparticle propagates to the {\it negative} $z$ direction.
We also find that the quasiparticles are damped with the damping rate $\gamma_{E(\bar{L})}$.
The solution to the Dirac equation \eqref{DiracPosition} is then expressed by the power series with respect to $\mathcal{M}$,
\begin{equation}
\begin{split}
    &E^{\rm sol}_L(z) = E^{(0)}_L (z) + E_L^{(2)} (z)+\cdots,\\[1ex]
    &E^{\rm sol}_R (z) = E^{(1)}_R (z) +E_R^{(3)} (z)+\cdots .
\end{split}
\end{equation}
Here, $E^{(i)}_{L(R)}$ represents the solution at the $i$-th order of $\mathcal{M}$.
$E_{L(R)}^{\rm sol}$ only contains even (odd) powers of $\mathcal{M}$ because of chirality flip by the mass insertion.
With the wavefunction normalization $E_L(0) =1$, the reflection coefficient is explicitly given by
\begin{align}
    R_{\rm LR} =E^{(1)}_R (0) + E_R^{(3)} (0)+\cdots, \label{eq:reflection coefficients}
\end{align}
where
\begin{equation}
\begin{split}
    &E^{(1)}_R (0) = i\int^\infty_0 dz_1 \, e^{-ip_R z_1}\mathcal{M}^\dag e^{ip_L z_1},\\[1ex]
    &E^{(3)}_R (0) = i\int^\infty_0 dz_1 \int^0_{z_1}dz_2 \int^\infty_{z_2} dz_3 \, e^{-ip_R z_3} \mathcal{M}^\dag e^{ip_L(z_3-z_2)}\mathcal{M}e^{ip_R(z_2-z_1)} \mathcal{M}^\dag e^{ip_L z_1}.
\end{split}\label{eq:reflections}
\end{equation}
Higher order contributions such as $E^{(5)}_R (0)$ can be also calculated in the similar manner.
Fig.~\ref{fig:scatter} shows the schematic description of the perturbative calculation for the reflection coefficient.
A left-handed particle $E_L$ injected from the unbroken phase is reflected by an insertion of the mass matrix operator $\mathcal{M}$ inside the bubble and hence becomes a right-handed particle $E_R$.
The reflected right-handed particle $E_R$ escapes to the outside of the bubble and
contributes to $R_{LR}$, which corresponds to $E^{(1)}_R (0)$, or scatters again via the operator $\mathcal{M}^\dag$.

%%%%%%%%%%%%%%%%%%%%%%%%%%%%%%%%%%%%%%%%%%%%%%%%%%%%%%%%%%%%%
%%%%%%%%%%%%%%%%%%%%%%%%%%%%%%%%%%%%%%%%%%%%%%%%%%%%%%%%%%%%%

To capture the effect of the damping rate, we compute the reflection coefficient to the leading order in $m_E$, neglecting the flavor-dependent part of the thermal mass: 
\begin{align}
    &E^{s(1)}_{R} = \frac{m_E}{4\overline{\gamma}_E}\dfrac{e^{i\phi(\omega)}}{\sqrt{1+\left(\frac{\omega-\Omega_+}{\overline{\gamma}_E}\right)^2}},\label{eq:single flavor reflection}\\
    &\overline{\gamma}_E\equiv \frac{\gamma_{\bar{L}}+\gamma_E}{2}, \quad \phi(\omega)= \arctan\left(-\frac{\overline{\gamma}_E}{\omega-\Omega_+}\right).\nonumber 
\end{align}
Note that the CP even phase of the reflection coefficient $\phi(\omega)$
is provided by the damping rate $\overline{\gamma}_E$.
For $\omega \simeq \Omega_+$, a quantum mechanical resonance with a width $\overline{\gamma}_E$ takes place.
This is because the incoming left-handed particle and the reflected right-handed particle
have the similar energy there.
Since the magnitude of the reflection coefficient $|E^{s(1)}_{R} |$
cannot exceed unity, $m_E<4\overline{\gamma}_E$ is required as the perturbativity condition.
For $|\omega-\Omega_+| \gg \bar{\gamma}_E$, the amplitude as well as the CP even phase of $E_R^{(1)}$
are suppressed.
The suppression by the damping rate $\overline{\gamma}_E^{-1}$ can be understood as follows.
It is apparent from fig.~\ref{fig:dispersion curves} that there is no momentum satisfying the dispersion relations in the broken phase for $\Omega_+-m_E/2 <\omega<\Omega_++m_E/2$ due to the level crossing.
Thus, a new lepton entering from the symmetric phase whose energy is within this narrow region
appears to be totally reflected by the bubble wall and the reflection coefficient is unity at the tree-level.
As pointed out in ref.~\cite{Gavela:1994dt}, however,
a non-zero damping rate allows for a change of the energy as large as the size of the damping rate,
which leads to a suppression of the reflection coefficient. 

We now describe how to perturbatively calculate $\Delta_E (\omega) = {\rm Tr}[\bar{R}^\dag_{LR} \bar{R}_{LR}-R^\dag_{LR}R_{LR}]$ from the reflection coefficient \eqref{eq:reflection coefficients}
taking account of the flavor dependent thermal mass.
To perform the $z$ integration in \eqref{eq:reflections},
we expand \eqref{eq:reflections} in terms of $\delta p_L / \gamma_E$ and $\delta p_R / \gamma_{\bar{L}}$.
A nonzero $\Delta (\omega)$ arises from noncommutativity of $\delta p_L$ and $\delta p_R$ with $\mathcal{M}$ which gives rise to a basis-independent CP violation analogous to the Jarlskog determinant in the SM.
The first non-vanishing contribution appears at $\mathcal{O}((\delta p_L) (\delta p_R))$
and $\mathcal{O}(\mathcal{M}^4)$ as the invariant \eqref{eq:Jarlskog parameter} can be formed.
The contribution is at the seventh order of $\gamma_{\bar{L},E}^{-1}$,
\begin{align}
&\Delta_E^7 (\omega) \simeq - \frac{2}{3} \left(\frac{27 \alpha_{R} T}{32 \Omega_{+} M^{2}_{W_R}}\right)^{2} \mathcal{F} \left(\frac{\omega - \Omega_{+}}{\overline{\gamma}_E}\right) \mathcal{D}_E (6\overline{\gamma}_E)^{-6}, \label{eq:Delta7}
\end{align}
where $\mathcal{F} (x) \equiv \frac{x}{(1+x^2)^4}$.
The effect of CP violation is encapsulated in
\begin{align}
  \mathcal{D}_E &\equiv \Im \Tr [M_N^2 \log (M_N^2) K M_E^4 K^{\dagger}M_N^2 K M_E^2 K^{\dagger}] \nonumber\\
  &= J' \left[m_{N_3}^2m^2_{N_2}\log \frac{m^2_{N_3}}{m^2_{N_2}} + m^2_{N_3}m^2_{N_1}\log \frac{m^2_{N_1}}{m^2_{N_3}} + m^2_{N_2}m^2_{N_1}\log \frac{m^2_{N_2}}{m^2_{N_1}}\right]\nonumber \\
 &\quad \times (m^2_{E_3} - m^2_{E_2}) (m^2_{E_3} - m^2_{E_1}) (m^2_{E_2} - m^2_{E_1}).
  \label{eq:determinant7}
\end{align}
Note that $\mathcal{D}_E$ vanishes when there is a mass degeneracy among the $E$ or $N$ leptons.
The seventh order contribution \eqref{eq:Delta7} is proportional to the flavor non-diagonal thermal mass $\delta p_L$,
which becomes subdominant at a high temperature.

The next non-vanishing contribution appears at $\mathcal{O}(\delta p_R^3)$ and $\mathcal{O}(\mathcal{M}^6)$.
The contribution is at the ninth order of $\gamma_{\bar{L},E}^{-1}$ and given by
\begin{align}
  &\Delta^{9}_E(\omega) \simeq 4 \left( \frac{27 \pi \alpha_R T^2}{64 \Omega_+ M_{W_{R}}^2}\right)^3  \mathcal{G} \left(\dfrac{\omega-\Omega_+}{\overline{\gamma}_E}\right) \mathrm{det} \mathcal{C}  (6\overline{\gamma}_E)^{-9}, \label{eq:Delta9}
\end{align}
where $\mathcal{G} (x) = \left(\frac{1}{1+x^2}\right)^6$.
The effect of CP violation is encapsulated in
\begin{equation}
\begin{split}
\mathrm{det} \mathcal{C} &\equiv i \mathrm{det}[M_E^2, KM_N^2K^{\dagger}] \\
&= -2J' (m_{E_1}^2 - m_{E_2}^2)(m_{E_2}^2 - m_{E_3}^2)(m_{E_3}^2-m_{E_1}^2) \\
&\qquad \times(m_{N_1}^2 -m_{N_2}^2)(m_{N_2}^2- m_{N_3}^2)(m_{N_3}^2 -m_{N_{1}}^2) \label{eq:determinant}.
\end{split}
\end{equation}
As in the case of \eqref{eq:determinant7}, $\mathrm{det} \mathcal{C}$ vanishes
when two masses among the $E$ or $N$ leptons are degenerated.
This $\Delta^{9}_E(\omega)$ gives the leading contribution at a high temperature. 

The lepton number density $n_L^r$ in \eqref{eq:reflection number} is obtained by performing the integration over $\omega$.
Both $\Delta^7_E (\omega)$ and $\Delta^9_E (\omega)$ are centered at $\omega =\Omega_+$.
Since $\mathcal{F}(x)$ in $\Delta^7_E (\omega)$ is an odd function and suppressed outside
the small region of $\Omega_+ - \overline{\gamma}_{E}< \omega< \Omega_+ +\overline{\gamma}_{E}$
where $n_0 (\omega)$ is approximately constant,
the contribution from $\omega < \Omega_+$ almost cancels out that from $\omega > \Omega_+$.
The integrated lepton number density from $\Delta^7_E (\omega)$ is thus vanishingly small
compared to that from $\Delta^9_E (\omega)$.
We neglect the $\Delta_E^7 (\omega)$ contribution in the following discussion
and focus on the $\Delta^9_E (\omega)$ contribution.
Since the CP-violating effect in eq.~\eqref{eq:determinant} is proportional to the product of the mass-squared differences among the $E$ leptons and among the $N$ leptons, sufficiently large Yukawa couplings of the $N$ leptons as well as the $E$ leptons are needed to avoid a suppression from the damping rate although the Yukawa couplings of the $E$ leptons are bounded by the perturbative requirement, $m_{E}<4\overline{\gamma}_E$.

We have estimated the $\Delta_E (\omega)$ contribution under the small momentum approximation
\eqref{eq:dispersion relation in s} and \eqref{eq:dispersion relation in b} and the thin-wall approximation.
The thin-wall approximation is justified for $p < l_w^{-1}$.
As can be seen from \eqref{eq:Delta9}, the dominant contribution to
$\Delta_E^9 (\omega)$ comes from $\omega \simeq \Omega_+$ corresponding to the momentum $p\simeq 3\Omega_-$.
Thus, $p\simeq 3\Omega_- < l_w^{-1} $ is required to maintain the thin-wall approximation.
Under the assumption that the wall thickness during a bubble expansion does not significantly
differ from the one at $T_{\rm n}$,
we can evaluate $l_w$ from a bounce solution.
The wall thickness realized in our model is typically $l_w = \mathcal{O}(10\sim 100) T_{\rm n}^{-1}$,
and it becomes small when $v_R (T_{\rm n}) / T_{\rm n}$ is large for fixed $g_R$.
A small $\Omega_-$ requires $\Omega_{\bar{L}}\simeq \Omega_E$ which is realized
by taking $\tan \theta_X\simeq 1$ for the $E$ lepton scattering.
We confirm that the thin-wall approximation is actually valid for $\tan\theta_X\simeq 1$
and a large $v_R (T_{\rm n})/ T_{\rm n}$.
Since the gauge couplings $g_R$ and $g_X$ are related through the $U(1)_Y$ gauge coupling (see eq.~\eqref{eq:gauge couplings}), $\tan \theta_X \simeq 1$ corresponds to $g_R\simeq 0.5$.
While we have discussed the contribution from the $E$ lepton scattering,
the contribution from the $N$ lepton scattering
$\Delta_N (\omega) \simeq \Delta^7_N(\omega)+\Delta^9_N (\omega)$ can be also obtained
by interchanging $K$ with $K^\dag$ and $M_E$ with $M_N$ in \eqref{eq:Delta7} and \eqref{eq:Delta9}.
However, for the $N$ lepton scattering,
since $\Omega_N$ is determined by the Yukawa coupling,
the thin-wall approximation is not justified unless $\Omega_{\bar{L}}\simeq \Omega_N$ is realized.
We will argue in sec.~\ref{sec:large momentum regime} that a contribution from the parameter regime
where the thin-wall approximation is not justified is significantly suppressed by the finite wall size
unless the new lepton masses are larger than the thermal masses.
We thus mainly focus on the sufficiently small momentum regime so that the thin-wall approximation is maintained.
Moreover, we can verify that the small momentum approximation~\eqref{eq:dispersion relation in s} and \eqref{eq:dispersion relation in b}
are always satisfied when the thin-wall approximation is justified.

Using eq.~\eqref{eq:dispersion relation in s} with $\omega^s_{L, \pm} = \omega^s_{R, \pm}$,
we find $(\bm{p}_R-\bm{p}_L)\cdot \bm{v}_w /T \simeq 3(\Omega_R -\Omega_L) v_w / T$
in eq.~\eqref{eq:reflection number}.
Note that the momentum direction of an injected new lepton is opposite to the wall velocity $\bm{v}_w$
for the normal branch, while it is the same as $\bm{v}_w$ for the abnormal branch
because the direction of the group velocity is opposite to that of the momentum in the small momentum regime.

%%%%%%%%%%%%%%%%%%%%%%%%%%%%%%%%%%%%%%%%%%%%%%%%
\subsubsection{Scattering without damping ((ii)-(A))}
  \label{sec:heavy masses}
%%%%%%%%%%%%%%%%%%%%%%%%%%%%%%%%%%%%%%%%%%%%%%%%

We next consider the parameter regime of (ii)-(A) in table~\ref{table:parameter}.
As in section~\ref{sec:perturbation},
we focus on the small momentum region,
and the effective Dirac equation to be solved is given by eq.~\eqref{DiracPosition},
while the perturbative expansion in terms of $m_{E} / 4\overline{\gamma}_E$ is not valid.
To estimate $\Delta_E (\omega)$ of \eqref{eq:reflection number} in this regime,
we follow the discussion of the SM electroweak baryogenesis
in which the damping rate is neglected~\cite{Farrar:1993sp,Farrar:1993hn}.

The reflection coefficient may be computed in the following way.
Since the effect of the damping rate is negligible in this parameter regime, a plane wave can be the solution of the effective Dirac equation.
We may consider the wavefunction of $E^i_L$ at $z=-\infty$ proportional to $e^{-i\omega t + p_z z}$ where $p_z$ is the momentum of $z$ component of the $E$ lepton.
Here, we neglect the momentum tangential to the wall for simplicity.
Since this wavefunction must travel $+z$ direction, the group velocity must be positive, $\partial  \omega / \partial p_z >0$.
Under this boundary condition, the reflected wavefunction of $E^j_R$ at $z=-\infty$ is also proportional to $e^{-i\omega t + ip_z z}$ with negative group velocity, $\partial \omega / \partial p_z <0$.
Then the reflection coefficient $R^{ij}_{LR}$ is defined by the ratio of the injected wavefunction to the reflected one, which is common for the reflection problem in the ordinary quantum mechanics.

A total reflection of the $E_i$ lepton occurs for $\Omega_+-m_{E_i}/2<\omega<\Omega_++m_{E_i}/2$ since there is no suppression from the damping rate as we discussed in section~\ref{sec:perturbation}.
Furthermore, this total reflection provides a CP-even phase even in the absence of the damping rate because the reflection coefficient becomes a complex value, as explicitly confirmed in ref~\cite{Farrar:1993hn}.
Since $m_{E_3}> m_{E_2} > m_{E_1}$, all the $E$ leptons are totally reflected for $\Omega_+-m_{E_1}/2<\omega<\Omega_++m_{E_1}/2$.
In this energy region, no $E$ lepton is transmitted into the bubble wall and hence $\Delta_E (\omega) = 0$.
The dominant contribution comes from the energy region,
$\Omega_+-m_{E_2}/2<\omega<\Omega_+ -m_{E_1}/2$ and $\Omega_++m_{E_1}/2<\omega<\Omega_{+}+m_{E_2}/2$,
where the $E_2$ lepton is totally reflected and the $E_1$ lepton is partially transmitted.
For simplicity, we take $m_{E_2}\gg m_{E_1}$ in the following discussion, 
which does not affect our qualitative results.

In the case of the SM electroweak baryogenesis,
the dominant contribution comes from the path in which a strange quark is injected
from the symmetric phase and totally reflected by the bubble wall~\cite{Farrar:1993hn}.
Correspondingly, in the present case,
the $E_2$ lepton is injected from the symmetric phase and is reflected by the wall.
The total reflection of the $E_2$ lepton makes the reflection coefficient large, 
leading to an enhancement of $\Delta_E (\omega)$ at around $\omega \simeq \Omega_+$.
The peak height can be estimated by perturbation theory with respect to $\delta p_{L,R}$
in the case of $\gamma_{\bar{L},E,N} = 0$.
We consider the parameter region where $\delta p_R$ provides the dominant source of CP violation.
This is indeed the case because we mainly focus on the parameter regime $m_{E_i} < T$ during the phase transition.
The CP-even contribution with a nonzero phase $\delta_{\rm even}$ appears at the zeroth order of $\delta p_R$,
$\mathcal{A}\sim e^{i\delta_{\rm even}}$, corresponding to the total reflection.
The CP-odd contribution with the phase $\delta$ arises at the third order of $\delta p_R$.
Here, all the three generations are involved, that is,
the injected $E_2$ changes into $E_3$ which is reflected by the wall and projected back onto $E_2$
via the $E_1$ lepton.
All the new lepton flavor changing processes are mediated by $\delta p_R$.
In this path, the CP-odd contribution is estimated as
\begin{align}
\mathcal{B}\sim A (E_2 \rightarrow E_3) \, R^{E_3}_{LR} \, A (E_3 \rightarrow E_1) \, A (E_1 \rightarrow E_2),
\end{align}
where $R^{E_3}_{LR} \sim \mathcal{O}(1)$ is the tree-level reflection coefficient of the $E_3$ lepton
and $A (E_i \rightarrow E_j)$ is the transition amplitude from $E_i$ to $E_j$ through $\delta p_R$.
The path of $E_2$ changing into $E_1$ reflected by the wall only gives a negligible contribution
because the reflection coefficient of the $E_1$ lepton is suppressed by
the tiny Yukawa coupling, $y_{E_1}\ll y_{E_3}$.

The transition amplitude $A ({E_i} \rightarrow {E_j})$ is proportional to $\delta p_R$ and, 
following the standard quantum mechanical perturbation theory, divided by the energy separation between the initial state and the final state $\sim 1/|m_{E_i}-m_{E_j}|$ around the level-crossing point (see eq.~\eqref{eq:dispersion relation in b}).
Then, the amplitude is roughly given by $A (E_{i} \rightarrow E_{j}) \sim \delta p_R/|m_{E_i}-m_{E_j}|$.
More explicitly, we find
\begin{align}
&A (E_2 \rightarrow E_3) \sim \frac{\alpha_R T^2m_{N_3}^2}{\Omega_+ M_{W_R}^2|m_{E_3}-m_{E_2}|}\sim \frac{\alpha_R T^2 m_{N_3}^2}{\Omega_+ M_{W_R}^2 m_{E_3}}, \nonumber\\[1ex]
&A (E_3 \rightarrow E_1) \sim \frac{\alpha_R T^2 m_{N_3}^2}{\Omega_+ M_{W_R}^2|m_{E_3}-m_{E_1}|}\sim \frac{\alpha_R T^2m_{N_3}^2}{\Omega_+ M_{W_R}^2 m_{E_3}},\\[1ex]
&A (E_1 \rightarrow E_2) \sim \frac{\alpha_R T^2 m_{N_2}^2}{\Omega_+ M_{W_R}^2|m_{E_1}-m_{E_2}|}\sim \frac{\alpha_R T^2m_{N_2}^2}{\Omega_+ M_{W_R}^2 m_{E_2}},\nonumber
\end{align}
assuming $m_{E_3}\gg m_{E_2}\gg m_{E_1}$.
Here, we have omitted the matrix $K$.
As a conservative estimate, we approximate $\Delta_E (\omega)$ as a step function
in the range of $\Omega_+ - m_{E_2}/2<\omega<\Omega_++m_{E_2}/2$.
Taking account of the matrix $K$, $\Delta_E (\omega)$ is then given by
\begin{align}
    \Delta_E (\omega)\sim J'\left(\frac{\alpha_RT^2}{\Omega_+ M^2_{W_R}}\right)^3\frac{m_{N_3}^4m_{N_2}^2}{m_{E_3}^2m_{E_2}},
    \qquad \Omega_+-\frac{m_{E_2}}{2} <\omega<\Omega_+ +\frac{m_{E_2}}{2}. \label{eq:non-perturbative result}
\end{align}
The $\Delta_N (\omega)$ contribution is also obtained
by the replacement of $m_{E_i}\leftrightarrow m_{N_j}$.

Let us compare the above estimate of $\Delta_E (\omega)$ with $\Delta_E^9 (\omega)$ obtained by the perturbation with respect to $m_{E}/4\overline{\gamma}_E$.
A nonzero $\Delta_E (\omega)$ contribution appears at the order of $\delta p_R^3$, which is in agreement with the perturbative analysis.
$ \Delta_E (\omega)$ is not suppressed by the damping rate for $m_{E_{2,3}}/4\overline{\gamma}_E>1$, but
instead has the factor of $m_{N_3}^4m^2_{N_2}/m_{E_3}^2m_{E_2}$ coming from the transition amplitudes of $E_2 \rightarrow E_3$, $E_3 \rightarrow E_1$ and $E_{1} \rightarrow E_2$.

\subsubsection{Reflection with a large momentum exchange ((i),(ii)-(B))}\label{sec:large momentum regime}

\begin{figure}
    \centering
    \begin{minipage}[h]{0.49\linewidth}
        \includegraphics[width=2.5in]{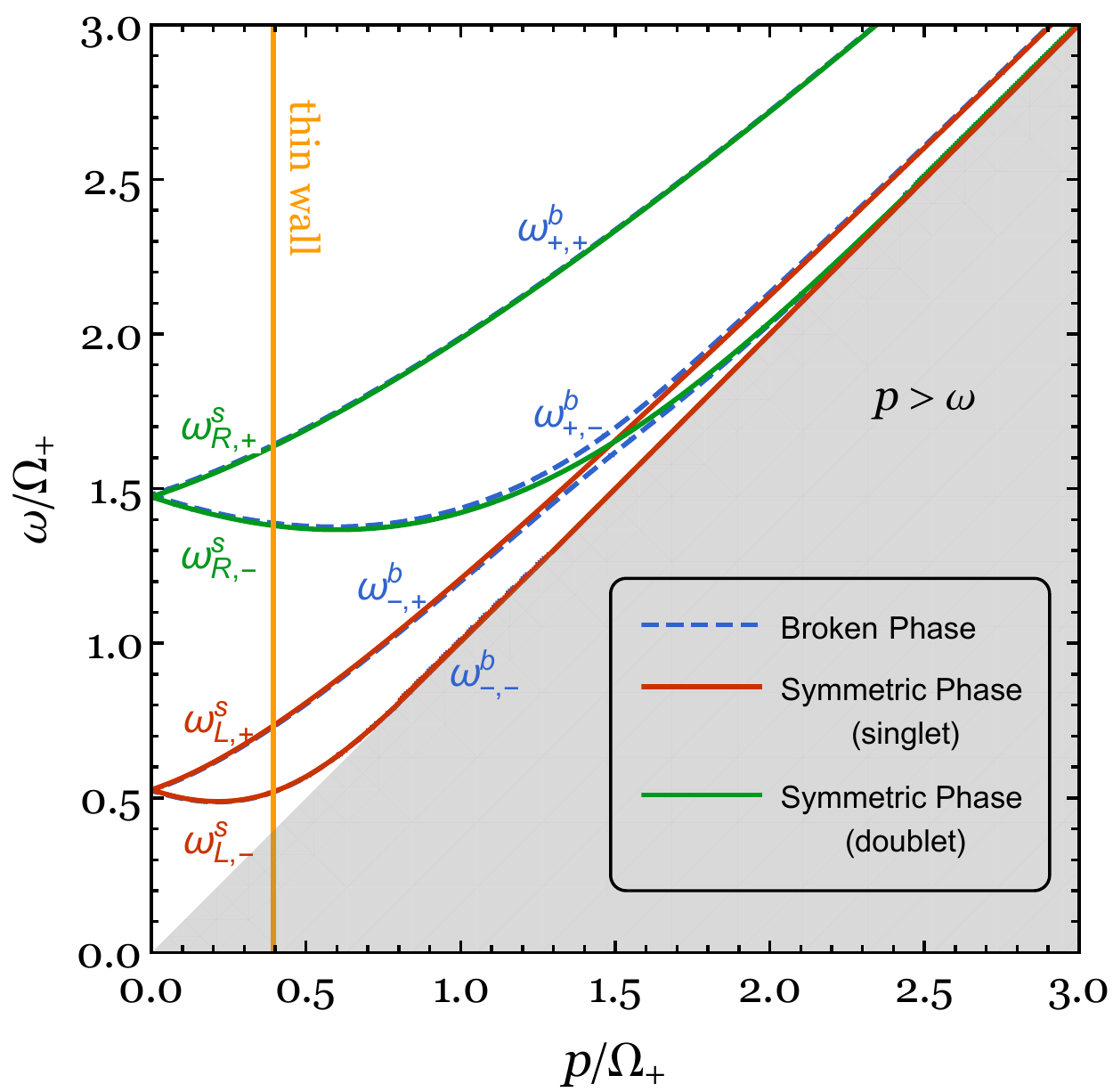}
    \end{minipage}
    \begin{minipage}[h]{0.49\linewidth}
        \includegraphics[width=2.5in]{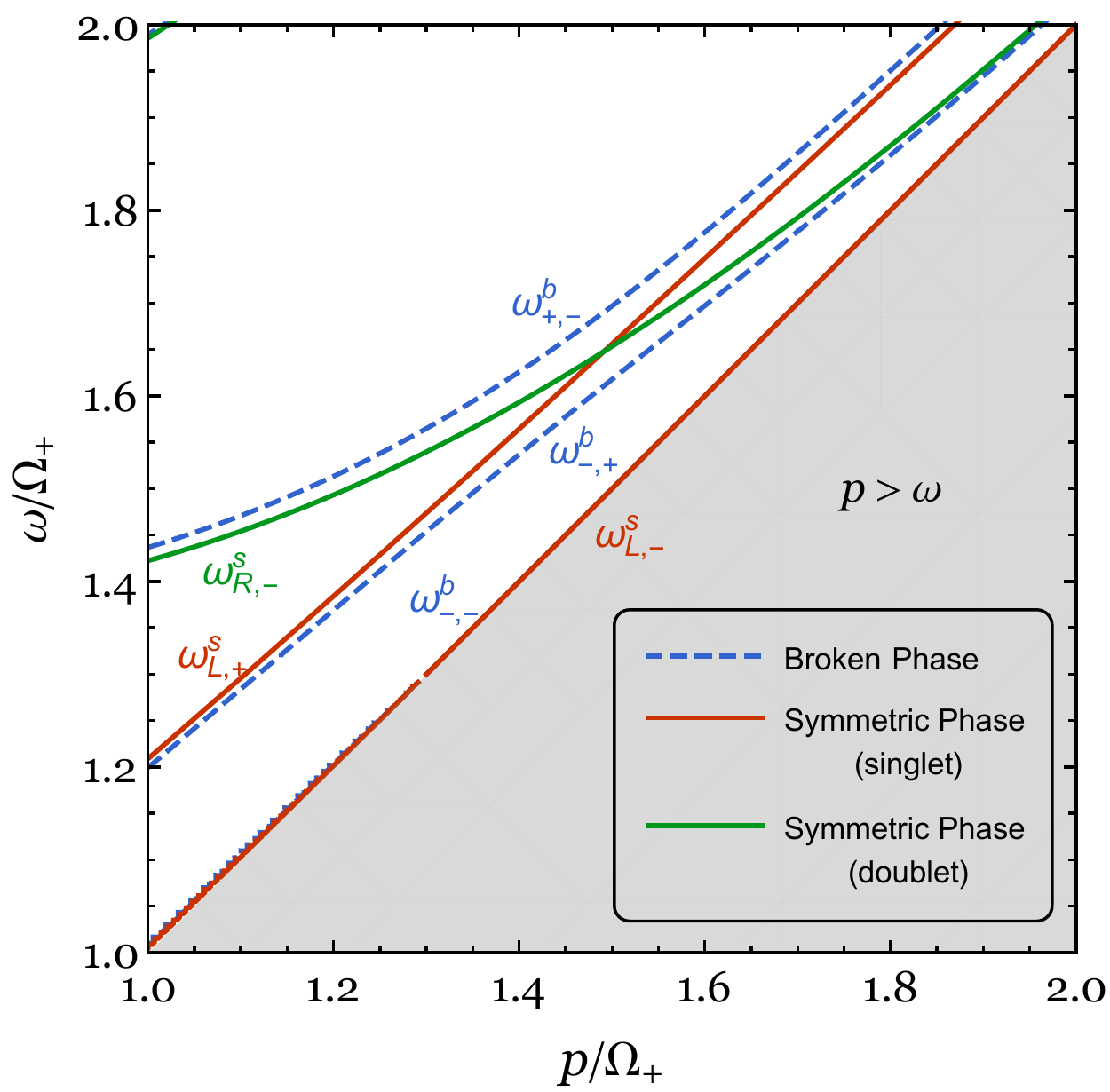}
    \end{minipage}
    \caption{Dispersion curves of the left (green solid) and right-handed (red solid) $E$ lepton in the symmetric phase and those in the broken phase (blue dashed) for $y_{E}=0.015,~g_R = 1.2$, and $v_R(T_c)/T_c=3$.
    In the left panel, the momenta to the left of the orange solid line satisfy the condition of the thin-wall approximation,
    $p < l_w^{-1}=T_{\rm n}/10$.
    The right panel focuses on the level-crossing point.
    }
    \label{fig:dispersion large gauge coupling}
\end{figure}

To discuss the reflection problem for the parameter regimes (i)-(B) and (ii)-(B) in table~\ref{table:parameter},
let us consider the dispersion curves of the new leptons.
Fig.~\ref{fig:dispersion large gauge coupling} shows the dispersion curves of the $E_1$ lepton obtained from
eq.~\eqref{eq:thermal self-energy} for $y_{E_1}=0.015$, $g_R=1.2$ and $v_R(T_R)/T_R = 3$.
Here, the thermal self-energy in the broken phase is approximated by that in the symmetric phase,
which is justified for the parameter regimes (i) and (ii).
We can see that the level crossing of the $(+,-)$ and $(-,-)$ branches
takes place at a large momentum region.
As discussed in sections~\ref{sec:perturbation} and \ref{sec:heavy masses},
$\Delta (\omega)$ is non-zero only at around the level crossing point.
Then, we need to consider the reflection problem without relying on the small momentum approximation.

%%%%%%%%%%%%%%%%%%%%%%%%%%%%%%%%%%%%%%%%%%%%%
%%%%%%%%%%%%%%%%%%%%%%%%%%%%%%%%%%%%%%%%%%%%%

We first clarify possible reflection processes for the left-handed new lepton injected from the symmetric phase at $z=-\infty$.
In fig.~\ref{fig:dispersion large gauge coupling},
the frequencies for the normal branches are always increasing functions of $p$, while those for
the abnormal branches are decreasing functions for a small $p$ but increasing functions for a large $p$.
This implies that the group velocity of the abnormal branch has a direction opposite to the momentum for a small $p$
but the same direction for a large $p$.
In the reflection process, the direction of the group velocity must be flipped,
while that of the momentum can be arbitrary.
The energy conservation and the angular momentum conservation must be satisfied.
In addition, note that helicity is always opposite to chirality for the abnormal branch.
Then, we can see that reflections in the small momentum region are forbidden by the energy conservation.
In the large momentum region,
allowed processes are ${\rm LN}\to {\rm RN}$, ${\rm LA}\to {\rm RA}$, ${\rm LN}\to{\rm LA}$ and ${\rm LA} \to{\rm LN}$ where LN (LA) and RN (RA) denote the left-handed normal (abnormal) and the right-handed normal (abnormal) modes, respectively.

All the allowed reflection processes require large momentum exchanges,
which are generally larger than the inverse of the wall thickness realized in our model, $l_w^{-1}\sim T_{\rm n}/10$.
It has been argued in ref.~\cite{Farrar:1993hn} that
reflection coefficients of such processes are significantly suppressed by the semi-classical factor $e^{-\Delta p \, l_w}$
where $\Delta p$ denotes the change of the momentum during the reflection process.
On the other hand, transmission processes require only tiny momentum exchanges.
Moreover, refs.~\cite{Weldon:1989ys,Gavela:1994dt} have pointed out that
the abnormal mode in the large momentum region is exponentially decoupled, that is, the reflection process
involving an abnormal mode with a large momentum is exponentially suppressed.
The similar discussion can be applied to the reflection problem
for the right-handed new lepton injected from the symmetric phase.
We therefore conclude that the produced lepton asymmetry in the cases (i)-(B) and (ii)-(B) is significantly suppressed by the finite wall thickness and negligible.

%%%%%%%%%%%%%%%%%%%%%%%%%%%%%%%%%%%%%%%%%%%%%%%%
\subsubsection{The zero-temperature contribution ((iii)-(A),(B))}
  \label{sec:(iii)}
%%%%%%%%%%%%%%%%%%%%%%%%%%%%%%%%%%%%%%%%%%%%%%%%

We here consider the parameter region where the new lepton mass is larger than the thermal mass,
$m_{E_{i}}>\Omega_+$.
In the broken phase, the dispersion relation of the new lepton is then determined
by the mass obtained via the Higgs condensation rather than the thermal mass.
The abnormal modes are decoupled in the large momentum and large mass region~\cite{Weldon:1989ys,Petitgirard:1991mf},
and thus, the dispersion curve in the broken phase is approximately given by
$\omega \simeq \sqrt{p_b^2+m_{E_i}^2}$ where $p_b$ is the momentum of the $E$ lepton.
Since $m_{E_i}>\Omega_+$, the total reflection is easily realized for a small injected momentum
$ p_s\sim \omega \lesssim m_{E_i}$.
For $\omega<m_{E_1}$, no $E$ lepton can be transmited into the broken phase and hence $\Delta_{E} (\omega)=0$.
In addition, for $\omega >m_{E_3}$, the reflection coefficient is significantly suppressed by the wall thickness
because of a large momentum in the symmetric phase, $\omega \sim p_s \gtrsim m_{E_3}\gg l_w^{-1}$.
Then, $\Delta (\omega)$ is non-negligible only for $m_{E_1}< \omega <m_{E_3}$.
Within this energy region, the total reflection of $E_{2,3}$ or only $E_3$ can take place in the thick-wall regime.
We have explicitly calculated the reflection coefficient by using the formula shown in ref.~\cite{Farrar:1993sp}
and confirmed that the CP-even phase produced by the total reflection is not suppressed in the thick-wall regime.
Therefore, no strong suppression on $\Delta_E (\omega)$ from the wall thickness is expected for $m_{E_1}< \omega < m_{E_3}$.

To find a rough estimation for $\Delta_{E} (\omega)$,
we first note that $\Delta_E (\omega)$ vanishes when there is a mass degeneracy among the $E$ or $N$ leptons.
This suppression is evaded for $y_{E_i,N_i}-y_{E_j,N_j} = \mathcal{O}(1)$ $(i>j)$.
Secondly, $\Delta_E (\omega)$ vanishes when one of the mixing angles defined in eq.~\eqref{eq:CKM parametrization} is zero,
but to assume $c_{ij}\sim s_{ij} = \mathcal{O}(1)$ can avoid the suppression.
Thirdly, as noted at the beginning of this subsection,
a nonzero $\Delta_E (\omega)$ is obtained by the interference between the CP-even phase and the CP-odd phase.
The CP-even phase is provided by the total reflection, while the CP-odd phase comes from loop processes
involving the $SU(2)_R$ gauge boson and the $N$ leptons.
These loop processes are at the order of $\alpha_R^2$ since all the new lepton generations must be involved.
This statement was confirmed by the authors of ref.~\cite{Gavela:1994ds} within the SM framework at zero-temperature.%
\footnote{However, they assume the thin-wall approximation for the calculation of $\Delta (\omega)$, which is not justified in our case, and thus, one cannot utilize the results for a quantitative estimation.}
Finally, there is a suppression from the gauge boson or $N$ lepton masses if the energy of the $E$ lepton
injected from the symmetric phase is smaller than those masses. If $m_{N_i}$ is also small, this process is further suppressed by the smallness of $m_{N_i}$.
As mentioned, the typical energy of the $E$ lepton relevant for the calculation of $\Delta_E (\omega)$
is $\omega\sim m_{E_{2,3}}$.
Assuming $m_{W_R} \sim m_{N_{2,3}}$,
$\Delta_{E}(\omega)$ can be roughly estimated as
\begin{align}
    \Delta_{E} (\omega) \sim & \left( \frac{\alpha_R}{4\pi}\right)^2 \times {\rm min}\left\{1,~\left(\frac{m^2_{E_{2,3}}}{m^2_{W_R}}\right)^2\right\} ,
    \qquad m_{E_1}<\omega<m_{E_3} \, ,\label{eq:(iii)}
\end{align}
where $1/(4\pi)^2$ comes from a loop factor. 
We do not expect a suppression from the transition amplitude presented in eq.~\eqref{eq:non-perturbative result}
as long as the momentum of an injected particle is comparable to the mass.
$\Delta_N (\omega)$ can be also obtained by replacement $m_{E_i}\leftrightarrow m_{N_i}$ in eq.~\eqref{eq:(iii)}, assuming $m_{E_{2,3}}\sim m_{W_R}$.
For a more quantitative estimate of $\Delta (\omega)$,
we need to exactly solve the effective Dirac equation defined on a non-trivial wall background,
which is beyond the scope of the present paper. 

Although for simplicity $m_{E_{2,3}}\sim m_{E_1}>\Omega_+$ has been assumed in the above discussion,
$m_{E_1} < \Omega_+$ is also possible because the reflection of the $E_1$ lepton is not needed
for a non-zero $\Delta_E (\omega)$ as long as the $E_{2,3}$ leptons are totally reflected.
In this case, we expect an additional suppression factor from the transition amplitude
similar to eq.~\eqref{eq:non-perturbative result}
when the injected energy is smaller than $m_{E_{2,3}}$.

It is worthwhile to note that, in the parameter regimes (i),(ii)-(A), to maintain the small momentum approximation,
we only consider the tangential momentum smaller than the thermal mass.
On the other hand, in the parameter regime (iii), the effective Dirac equation is approximately given by
the zero-temperature form.
Since any tangential momentum can be set to zero by the Lorentz transformation,
it can be as large as the temperature.
Then, there is no suppression from a small tangential momentum given by eq.~\eqref{eq:tangential contribution},
unlike the parameter regimes (i),(ii)-(A).
We thus expect that the parameter regime (iii) gives the maximal baryon asymmetry realized in our model.

%%%%%%%%%%%%%%%%%%%%%%%%%%%%%%%%%%%%%%%%%%%%%%%%
\subsection{Results}
  \label{sec:yukawa}
%%%%%%%%%%%%%%%%%%%%%%%%%%%%%%%%%%%%%%%%%%%%%%%%

We now estimate the baryon asymmetry generated in our model.
In sec.~\ref{sec:minimal model}, it is found that the produced asymmetry is smaller than the observed value
for the parameter regions (i) and (ii) in table~\ref{table:parameter} even though the model parameters are highly optimized.
For the parameter region (iii), on the other hand, the qualitative estimation shows that
the observed baryon asymmetry can be produced.
In sec.~\ref{sec:extra interaction}, we discuss an extended model to realize a sufficient baryon asymmetry
even for the parameter region (i).

\subsubsection{The minimal model}\label{sec:minimal model}

Let us find the largest baryon asymmetry produced by the $E$ lepton scattering for the parameter regions (i) and (ii).
As discussed in section~\ref{sec:large momentum regime}, 
the baryon asymmetry is exponentially suppressed for the parameter regions (i)-(B) and (ii)-(B),
and hence we focus on the parameter regions (i)-(A) and (ii)-(A).
For the parameter regime (i)-(A), as mentioned in section~\ref{sec:decoherence effect},
the $\Delta^{7}_E (\omega)$ contribution is negligibly small compared to $\Delta^{9}_E(\omega)$.
Thus, we use $\Delta_E (\omega)\simeq \Delta^9_E (\omega)$ in this regime.
The expressions of $\Delta_E (\omega)$ for the parameter regions (i)-(A) and (ii)-(A)
are then given by eq.~\eqref{eq:Delta9} and eq.~\eqref{eq:non-perturbative result}, respectively.
The expressions show that $\Delta_E (\omega)$ is suppressed by the damping rate for $m_{E_i}<4\overline{\gamma}_E$
corresponding to the region (i)-(A) while it is suppressed by the transition amplitudes for
$(\Omega_+ \!>) \, m_{E_i}>4\overline{\gamma}_E$ of the region (ii)-(A).
At the boundary of these parameter regimes $m_{E_i}\simeq 4\overline{\gamma}_{E}$, there is no suppression from both of the damping rate and the transition amplitudes, and hence the largest baryon asymmetry is realized.
We then focus on the region near the boundary.
Note that $\Delta_E (\omega)$ calculated by eq.~\eqref{eq:Delta9} and eq.~\eqref{eq:non-perturbative result}
at this boundary give the same order results, and
we can use both expressions in the following estimation.

The model parameters relevant to the amount of the baryon asymmetry are the new lepton Yukawa couplings
$y_{E_i}$ and $y_{N_i}$, the gauge coupling $g_R$ of $SU(2)_R$, the size of CP violation in the new lepton sector $J'$
and the quartic coupling $\lambda_R$.
Note that the $U(1)_X$ gauge coupling $g_X$ is determined by the condition \eqref{eq:gauge couplings}.
The wall velocity can be estimated by computing the friction acting on the wall~\cite{Moore:1995si,Moore:1995ua,Konstandin:2014zta,Kozaczuk:2015owa}.
However, the determination of the wall velocity requires an out-of-equilibrium computation,
which is beyond the scope of the present paper.
Instead, we treat the wall velocity as a free parameter.
We numerically find that $\lambda_R \simeq 0.001$ and $g_R \simeq 0.55$ produce the largest baryon asymmetry
while maintaining the small momentum and thin-wall approximations.
The sphaleron decoupling condition is also satisfied.
For these values of $\lambda_R$ and $g_R$, the largest Yukawa coupling of the $N_3$ lepton
allowed by the stability of the effective potential at zero temperature for $\phi_R < 10 \, v_R$
is $y_{N_3} \simeq 0.55$.
In addition, we have confirmed that $m_{E_3}=4\overline{\gamma}_E$ is satisfied for $y_{E_3} \simeq 0.022$.
The other Yukawa couplings $y_{E_{1,2},{N_{1,2}}}$ are also determined to maximize the baryon asymmetry,
and we have found $y_{E_1,N_1}/y_{E_3,N_3} \simeq 0.1$, $y_{E_2,N_2}/y_{E_3,N_3} \simeq 0.7$.
We set $J'$ to the maximal value, $J'=0.08$.
For the wall velocity, we take the representative value $v_w =0.1$ as a benchmark point.
For this parameter set, we find $v_R (T_{\rm n})/ T_{\rm n} \simeq 7.6$, $v_R(T_R)/T_R \simeq 5.5$, $l_wT_{\rm n} \simeq 40$ and 
$n_B/s\sim 10^{-3}\times {\rm BAU}$ where ${\rm BAU} \equiv 8.7 \times 10^{-11}$
is the observed amount of the baryon asymmetry~\cite{Tanabashi:2018oca}.
The most significant suppression comes from the small gauge coupling $g_R=0.55$
which suppresses the reaction rate of the $SU(2)_R$ sphaleron process in the symmetric phase
as discussed in section~\ref{sec:Delta}.
Note also that our estimation only includes a contribution from the small tangential momentum; see eq.~\eqref{eq:tangential contribution}.
Even if the cut-off of the tangential momentum is of the order of the temperature rather than the thermal mass, one cannot realize the observed baryon asymmetry.
We therefore conclude that it is impossible to realize the observed baryon asymmetry from $\Delta_E (\omega)$ for the parameter regions (i)-(A) and (ii)-(A).

For the parameter region (iii),
the naive estimation of $\Delta_E (\omega)$ in eq.~\eqref{eq:(iii)} leads to the baryon asymmetry,
\begin{align}
    \frac{n_B}{s}&\sim 1.1 \times 10^{-10} \times \left(\frac{g_R}{1.2}\right)^4\times 
    \left(\mathrm{min}\left\{1,~4\left(\frac{\yeth}{0.8}\right)^2 \left(\frac{1.2}{g_R}\right)^2\right\}\right)^2. \label{eq:(iii) result}
\end{align}
In this calculation, we take $v_w=0.3,~\yeone=0.1$ and $v_R(T_{\rm n})/T_{\rm n} = 2.5$ as a benchmark point.
The quartic coupling $\lambda_R$ is taken to realize $v_R (T_{\rm n})/T_{\rm n}=2.5$ for a given $g_R$.
The region of $g_R$ and $\lambda_R$ satisfying the sphaleron decoupling condition is shown in the left panel of fig.~\ref{fig:decoupling}.
We see that the observed baryon asymmetry can be produced by the $\Delta_{E}(\omega)$ contribution
for the parameter region (iii).

We here comment on the contribution from the $N$ lepton scattering $\Delta_N(\omega)$.
Since the left-handed $N$ leptons are gauge singlet under $G_{\rm LR}$,
they do not receive thermal masses and damping rates from gauge bosons
(see eqs.~\eqref{eq:OmegaL} and~\eqref{eq:decay E}).
Hence, the thermal effect is subdominant for the $N$ lepton scattering, unlike the case of the $E$ lepton scattering.
In fact, for the optimized parameter set described above,
we have confirmed that the $\Delta_N (\omega)$ contribution comes from the parameter regime (iii)
which gives the baryon asymmetry comparable to eq.~\eqref{eq:(iii) result}.
We therefore conclude that only the parameter region (iii) can realize the observed baryon asymmetry.

%%%%%%%%%%%%%%%%%%%%%%%%%%%%%%%%%%%%%%%%%%%%%%%%%%%
\begin{figure}
    \centering
    \begin{minipage}[h]{0.47\linewidth}
        \includegraphics[width=2.5in]{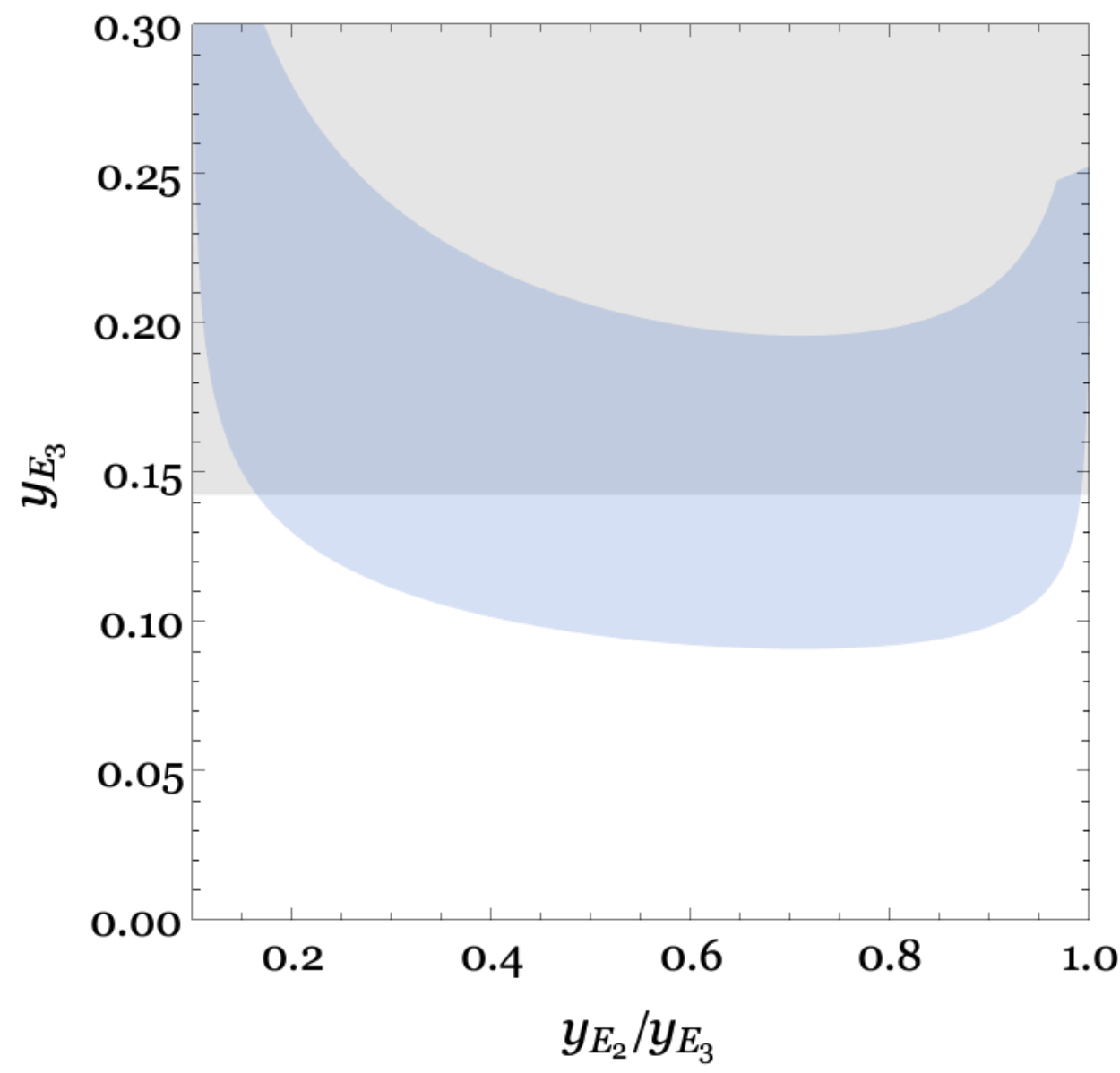}
    \end{minipage}
    \begin{minipage}[h]{0.47\linewidth}
        \includegraphics[width=2.5in]{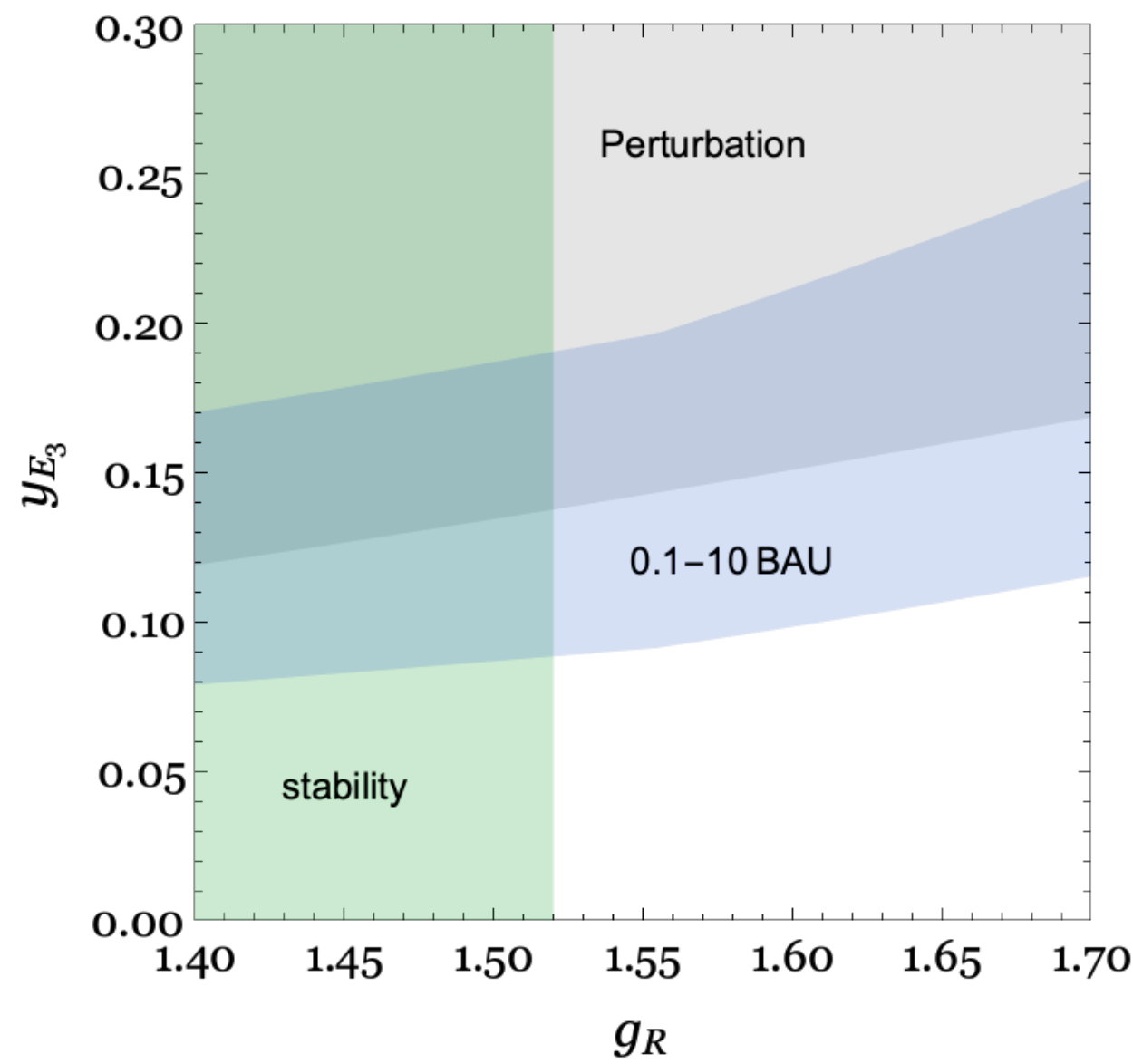}
    \end{minipage}
    \caption{The parameter region that produces $0.1-10$ BAU for the extended model.
    In the left panel for the $(y_{E_2}/y_{E_3},y_{E_3})$-plane, the model parameters are taken as
    $g_R=1.55,~\lambda_R=0.014,~\tilde{g}_E = 1.17,~J'=0.08,~y_{N_1}=0.03,~y_{N_2}=0.89$ and $y_{N_3}=1.35$.
    In the right panel for the $(g_R,y_{E_3})$-plane, we take $\yeone=0.1\yeth$, $\lambda_R =0.014,~J^\prime=0.08,~\ynone=0.03,~\yntwo=0.89,~\ynth=1.35$ and $v_w=0.48$.
    The coupling $\tilde{g}_E$ is chosen in such a way that $3\Omega_-=l_w^{-1}$ is satisfied.
    The gray colored region does not satisfy the perturbative requirement, $m_{E_i}< 4\overline{\gamma}_E$.
    The zero-temperature effective potential is not stable for $\phi_R < 10 \, v_R$ in the green colored region.
    }
    \label{fig:yg}
\end{figure}
%%%%%%%%%%%%%%%%%%%%%%%%%%%%%%%%%%%%%%%%%%%%%%%%%%

%%%%%%%%%%%%%%%%%%%%%%%%%%%%%%%%%%%%%%%%%%%%%%%%
\subsubsection{A model with an extra interaction}
  \label{sec:extra interaction}
%%%%%%%%%%%%%%%%%%%%%%%%%%%%%%%%%%%%%%%%%%%%%%%%

If the $E$ lepton obtains an additional thermal mass from an interaction other than the $U(1)_X$ gauge interaction,
the level crossing can occur for a small momentum even for a larger $g_R$.
Then, we can avoid the most significant suppression from the reaction rate from the sphaleron process,
while the perturbative calculation outlined in sec.~\ref{sec:perturbation} can be applied.
This situation can be achieved by an $O(1)$ portal coupling $g_E$ in eq.~(\ref{portal interaction}).
In this case, however, the SM left-handed charged leptons become mixtures of $\bar{E}$ and $e$,
and hence the lepton universality is in general violated.
For example, the branching ratio of the $Z$ boson into $\tau$ and $e$ are observed to be the same with an accuracy of $0.1\%$,
requiring $v_R > 10^6$ GeV.
Also, $g_E$ breaks the $Z_2$ symmetry $(\bar{L}_i, E_i, N_i) \to - (\bar{L}_i, E_i, N_i)$,
and thus, $N_1$ can no longer be a dark matter candidate.

We can instead introduce interactions with additional particles.
For instance, if there exist Dirac $SU(2)_L$ doublets $\tilde{\ell}_i = (\bm{1}, \bm{2}, \bm{1}, -1/2)_1$ and $\tilde{\bar{\ell}}_i = (\bm{1}, \bm{2}, \bm{1}, 1/2)_{-1}$,
the following Yukawa coupling can be introduced:
\begin{align}
    {\cal L} = \tilde{g}_E H_L^{(\dag)} \tilde{\bar{\ell}}_{i} E_i + {\rm h.c.}, \label{eq:additional thermal mass}
\end{align}
where $H_L$ or $H_L^{\dag}$ depends on the $U(1)_X$ charge of $\tilde{\bar{\ell}}_i$.
To straightforwardly apply the formulation in sec.~\ref{sec:perturbation}, we assume the universal $\tilde{g}_E$,
giving a universal contribution $\Delta \Omega_E^2 = \tilde{g}_E^2 T^2/8$ to $\Omega_E$ defined in eq.~\eqref{eq:OmegaL} for simplicity.
We take the value of $\tilde{g}_E$ so that the level crossing occurs in the small momentum and thin-wall regions where the perturbative computation for $\Delta_E(\omega)$ can be applied.

The left panel of fig.~\ref{fig:yg} shows the region with $n_B/s=0.1\times {\rm BAU} \sim 10\times {\rm BAU}$
including the additional thermal mass $\Delta \Omega^2_E$ for $g_R=1.55,~\lambda_R =0.014,~\tilde{g}_E=1.17,~J'=0.08,~y_{N_1}=0.03,~y_{N_2}=0.89,~y_{N_3}=1.35$, and $v_w=0.48$.
In this benchmark point, we obtain $v_R (T_{\rm n}) / T_{\rm n}=3.96$ and $l_w T_{\rm {n}}=15$.
To show that the observed baryon asymmetry is realized for the parameter region (i),
we have included only $\Delta_{E}^9 (\omega)$
without the possible $N$ lepton contribution $\Delta_N(\omega)$.
The right panel shows the viable region
on the $(g_R,~y_{E_3})$-plane for $\lambda_R=0.014,~J'=0.08,~y_{E_1}=0.1y_{E_2},~y_{N_1}=0.03,~y_{N_2}=0.89,~y_{N_3}=1.35$
and $v_w=0.48$.
The value of $\tilde{g}_E$ is determined in such a way that
$3\Omega_- = l_w^{-1}$ is satisfied to maintain the thin-wall approximation.
We have confirmed that $v_R(T_{\rm n}) /T_{\rm n} \simeq 4.2$ does not change significantly for $1.4<g_R<1.7$.
The kink of the blue band at $g_R \simeq1.55$ appears because the suppression factor from the diffusion
discussed around eq.~\eqref{diffusion suppression} disappears above this point.
Note that a large gauge coupling $g_R$ not only evades the suppression from the reaction rate of the sphaleron process
in the symmetric phase but also allows for a large Yukawa coupling without making the effective potential unstable
just above the symmetry breaking scale.
Thus, $\Delta_E (\omega)$ can be significantly enhanced compared to the case of the minimal model
for the parameter regions (i)-(A) and (ii)-(A).

%%%%%%%%%%%%%%%%%%%%%%%%%%%%%%%%%%%%%%%%%%%%%%%%
\section{Phenomenology}
\label{pheno}
%%%%%%%%%%%%%%%%%%%%%%%%%%%%%%%%%%%%%%%%%%%%%%%%

In the present framework of baryogenesis, it is plausible that the scale of the new gauge symmetry breaking
is not far above the electroweak scale.
Some of the new particles are charged under the SM gauge symmetry and may be within the LHC reach. 
We here discuss direct and indirect searches for such new particles and their implications 
for the model parameter space.
The amount of stochastic GWs generated by the first-order $G_{\rm LR} \to G_{\rm SM}$ phase transition
is also estimated.

%%%%%%%%%%%%%%%%%%%%%%%%%%%%%%%%%%%%%%%%%%%%%%%%
\subsection{Direct searches}
%%%%%%%%%%%%%%%%%%%%%%%%%%%%%%%%%%%%%%%%%%%%%%%%

The model predicts new gauge bosons associated with the extended gauge symmetry
and new matter fermions charged under the gauge interaction.
They can be searched at the LHC or future colliders.

%%%%%%%%%%%%%%%%%%%%%%%%%%%%%%%%%%%%%%%%%%%%%%%%
\subsubsection{New gauge bosons}
  \label{sec:Z'andWR}
%%%%%%%%%%%%%%%%%%%%%%%%%%%%%%%%%%%%%%%%%%%%%%%%

The neutral $Z^\prime$ boson can be produced at colliders and decay into a pair of SM fermions as well as new leptons.
Using the parameterization in refs.~\cite{Salvioni:2009mt,Aaboud:2017buh},
the $Z'$ coupling with a SM fermion $\psi$ is given by
\begin{equation}
\begin{split}
 \label{eq:Z coupling}
 &{\cal L}_{Z'} =   \sqrt{g^2 + {g'}^2} \gamma'Z'_\mu \left( {\rm sin}\theta_M Y  + {\rm cos} \theta_M \left(B-L\right) \right)\psi^\dag \bar{\sigma}^{\mu} \psi, \\[1ex]
  &\gamma' {\rm sin}\theta_M \equiv  \frac{g_R^{2}}{\sqrt{g^2 + {g'}^2} \sqrt{g_{R}^2 + g_X^2}},
  \qquad {\rm tan}\theta_M = - 2 \frac{g_R^2}{g_R^2 + g_X^2},
\end{split}
\end{equation}
where $\bar{\sigma}^{\mu} \equiv (1, -\vec{\sigma})$ with $\vec{\sigma}$ being Pauli matrices.
A SM fermion has a $U(1)_X$ charge $X = (B - L)/2$. Here, $B$ and $L$ are the baryon and lepton numbers, respectively. $Y$ is the ordinary hypercharge. The parameter $\gamma^{\prime}$ in the second line
measures the relative strength of the coupling of $Z^{\prime}$
to SM fermions compared to that of the SM $Z$ boson.
Such a new $Z'$ boson has been searched for at the LHC, which gives constraints on the mass $m_{Z'}$
and the relative strength $\gamma^{\prime}$.
Currently, the most stringent bound comes from the ATLAS dilepton search
with a luminosity of 139 $\rm fb^{-1}$~\cite{Aad:2019fac}.
The bound on $m_{Z'}$ and $\gamma'$ depends on the parameter $\theta_M$ defined above.
In our model, for $g_R$ within $[0.5,1.7]$,
$\theta_M$ varies within an approximate range of $[0.8,1.1]$.
According to ref.~\cite{Aaboud:2017buh},
the mass limit is almost the same for $\theta_M$ in this range.
We thus ignore the difference of $\theta_M$ and adopt the constraint on the $Z'_{\chi}$ model in ref.~\cite{Aad:2019fac}.
The $Z^\prime$ boson can also decay into new leptons.
To be conservative, we assume that all the new leptons except for $N_3$
are lighter than $m_{Z^\prime}/2$.
Under this assumption, the branching ratios of $Z'$ into electrons and muons are reduced
by a factor of two in comparison with the $Z'_{\chi}$ model.

The charged $W_R^\pm$ bosons can be also produced and decay into right-handed fermion pairs,
including SM fermions and new leptons.
We rescale the bound in ref.~\cite{Aad:2019wvl} by the square of $g_R/g$ and the branching ratios.

%%%%%%%%%%%%%%%%%%%%%%%%%%%%%%%%%%%%%%%%%%%%%%%%%
\begin{figure}[t]
  \centering
\includegraphics[width=0.49\linewidth]{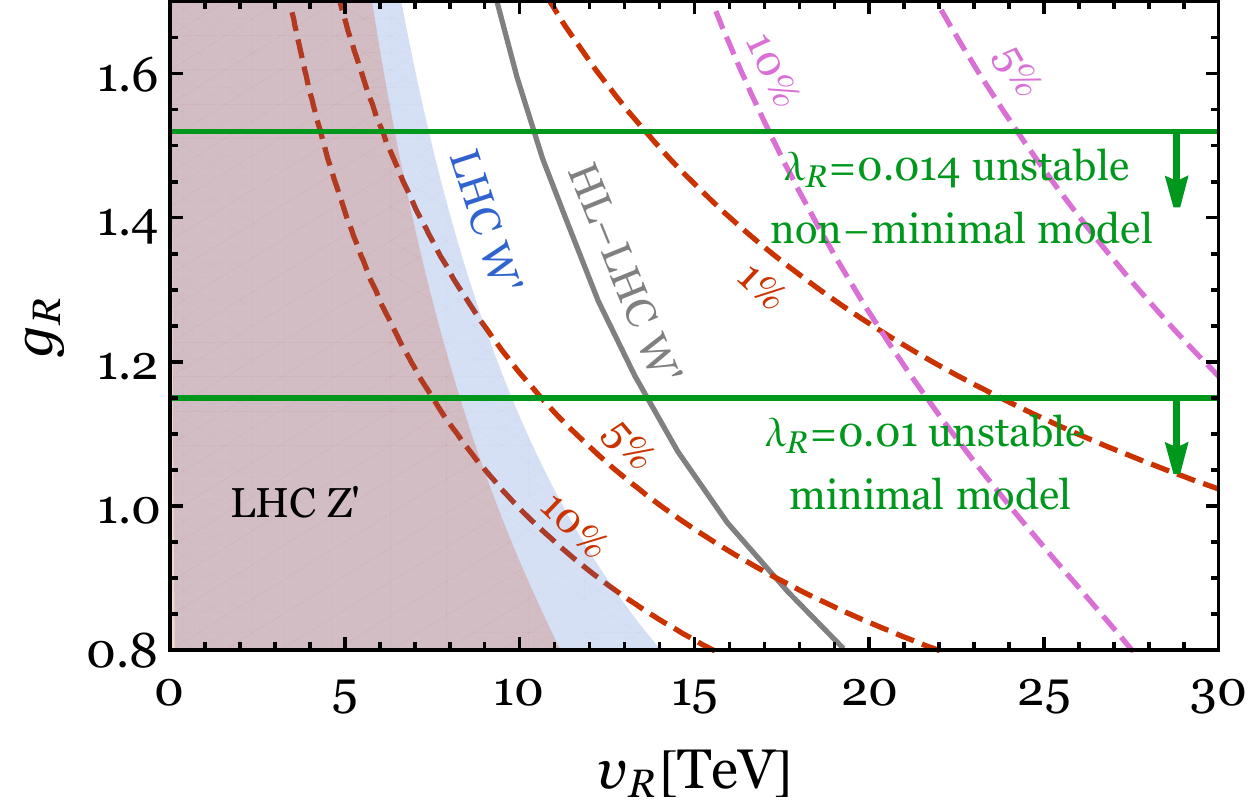}
\caption{The constraints on $g_R$ and $v_R$ at the LHC.
The red and blue shaded regions are excluded by $Z^\prime$ and $W^\prime$ searches, respectively.
The gray solid line denotes a prospective constraint from the $W^\prime$ search at the HL-LHC.
Below the green solid lines, the potential of $H_R$ becomes unstable at an energy scale below $10v_R$
for the minimal and extended models.
The red and magenta dashed lines show 10\%, 5\% and 1\% fine-tuning in the SM Higgs mass-squared parameter
against quantum corrections from the heavy gauge bosons,
with the SM Higgs being a bi-doublet and an $SU(2)_L$ doublet, respectively.
}
  \label{fig:gv}
\end{figure}
%%%%%%%%%%%%%%%%%%%%%%%%%%%%%%%%%%%%%%%%%%%%%%%%%

The current constraints on $g_R$ and $v_R$ from the LHC are shown in Fig.~\ref{fig:gv}.
We can see that $v_R \gtrsim 10$ TeV is required for $g_R\sim 1$.
Lower bounds on $g_R$ from the stability of the effective potential are also shown in the figure.
Here, the Yukawa couplings are the same as those in Fig.~\ref{fig:decoupling}. 
We choose $\lambda_R = 0.01$ and $0.014$ for the minimal model and the model with an extra interaction, respectively. 
For these values of $\lambda_R$, the lower bound on $g_R$ from the sphaleron decoupling condition is weaker than that from the stability condition.
The prospective constraint from the High-Luminosity LHC (HL-LHC) experiment is also plotted in Fig.~\ref{fig:gv}.
The $W_R^\pm$ boson mass and coupling are expected to be further probed by the HL-LHC~\cite{ATL-PHYS-PUB-2018-044}.

The SM Higgs mass-squared parameter receives a quantum correction from the coupling with the $W_R^\pm$ boson
in the case that the SM Higgs is a bi-doublet,
$\delta m_H^2 \simeq \frac{g_R^2 m_{W_R}^2}{16\pi^2}$.
When the SM Higgs is an $SU(2)_L$ doublet, the $Z^\prime$ boson gives the dominating correction,
$\delta m_H^2 \simeq \frac{g_X^2 m_{Z^\prime}^2}{16 \pi^2}$.
The required fine-tuning in the Higgs mass-squared parameter is plotted in Fig.~\ref{fig:gv}.
A milder fine-tuning favors a smaller $v_R$.

%%%%%%%%%%%%%%%%%%%%%%%%%%%%%%%%%%%%%%%%%%%%%%%%
\subsubsection{New leptons}
  \label{sec:new leptons}
%%%%%%%%%%%%%%%%%%%%%%%%%%%%%%%%%%%%%%%%%%%%%%%%

Next, let us discuss collider searches for the new leptons.
We here assume that the $N$ and $E$ particles decay and do not provide a dark matter candidate.
The discussion of the case that $N_1$ becomes a dark matter candidate is left for a future study.
Since the $SU(2)_L\times U(1)_Y$ neutral leptons $N$ are produced only through the $W_R^\pm$ or $Z'$ exchange,
their production cross sections are highly suppressed.
In addition, the masses of new charged leptons ${\cal E}$ and ${\cal X}$ in our models
are irrelevant for the baryon asymmetry.
We thus only discuss collider phenomenology of the hypercharged $E$ leptons, focusing on the lightest one, $E_1$.
The $E_1$ particle may be light, say $O(100)$ GeV, while the correct amount of the baryon asymmetry is produced.

The $E_1$ lepton is produced through the $Z$ or $\gamma$ exchange.
For $m_{E_1} < m_{N_1}$, it decays into $\nu + W$ or $e/\mu/\tau + Z/h$ via the mixing with SM leptons
given by the portal coupling of eq.~(\ref{portal interaction}).
The LEP experiment puts a lower bound of around 100 GeV on masses of new charged particles~\cite{Achard:2001qw}. The latest LHC search, which focuses on a new lepton with a mass higher than 100 GeV, gives an upper limit on the production cross section for a given mass~\cite{Sirunyan:2019ofn}. The production cross section of $E_1$, however, satisfies this limit for any masses.
Furthermore, if the $E_1$ lepton dominantly decays into $\tau$, it is even difficult to probe this particle
at the future HL-LHC \cite{Bhattiprolu:2019vdu}.
If $E_1$ dominantly decays into $e$ or $\mu$, the HL-LHC is expected to be sensitive to $E_1$ 
with a mass below 300 GeV~\cite{Abdullah:2016avr}.
For $m_{E_1} > m_{N_1}$, $E_1$ may decay into $N_1$ and jets or $e/\mu/\tau + \nu$ via the off-shell $W_R^\pm$ exchange.
The $N_1$ particle further decays into $e/\mu/\tau + W$ or $\nu + Z/h$.
Since the resultant decay products contain the same objects
as the case of $m_{E_1}$ directly decaying to SM particles,
the LHC searches provide similar constraints and sensitivities,
although more dedicated searches may be more efficient.

So far we have implicitly assumed that $E_1$ decays promptly.
If the portal coupling is small, $E_1$ may be observed as a long-lived particle.
Searches for anomalous tracks give a lower bound of $m_{E_1} > $ 800 GeV for $E_1$
which reaches the muon spectrometer~\cite{Aaboud:2019trc}, corresponding to $\tau>O(10)$ ns.

%%%%%%%%%%%%%%%%%%%%%%%%%%%%%%%%%%%%%%%%%%%%%%%%
\subsection{Indirect searches}
%%%%%%%%%%%%%%%%%%%%%%%%%%%%%%%%%%%%%%%%%%%%%%%%

Some precision measurements may also constrain the model parameter space indirectly.
Here, we consider the electroweak precision test and EDM measurements.
The former is important when the SM gauge group is extended,
and the latter gives a stringent constraint on the usual electroweak baryogenesis scenarios.
We will see that both measurements do not give strong constraints on our model parameter space.

%%%%%%%%%%%%%%%%%%%%%%%%%%%%%%%%%%%%%%%%%%%%%%%%
\subsubsection{The electroweak precision test}
\label{sec:precision}
%%%%%%%%%%%%%%%%%%%%%%%%%%%%%%%%%%%%%%%%%%%%%%%%

The SM Higgs doublet couples not only to the $SU(2)_L \times U(1)_Y$ gauge bosons
but also to the $Z^{\prime}$ boson which is a linear combination of the $W_R^3$ and $B_X$ bosons.
This contributes to a tree-level correction to the $Z$ boson mass, which is expressed by the $T$ parameter in the $S$, $T$, $U$ framework~\cite{Peskin:1991sw}.
The leading effect can be computed by the effective operator after integrating out the $Z^{\prime}$ boson,
\begin{align}
  \label{eq:effective operator}
  \mathcal{L}_{\rm eff} \supset  \frac{\sin^2 \theta_X g_X^2}{8m_{Z^{\prime}}^2}(H_{L}^{\dagger}D_{\mu}H_{L} - (D_{\mu}H_L)^{\dagger}H_{L})^{2},
\end{align}
where we have defined $\tan \theta_X \equiv g_X/g_R$. Then, following the step described in ref.~\cite{Skiba:2010xn},
we obtain a new contribution to the $T$ parameter, 
\begin{align}
  \label{eq:T}
 \Delta T = - \frac{\sin^2 \theta_X g_X^2v^{2}}{4 {\alpha_{\rm em}} m_{Z^{\prime}}^2}=
 -\frac{\sin^4 \theta_X v^2}{\alpha_{\rm em} v_R^2}.
\end{align}
Here, $\alpha_{\rm em}$ is the fine structure constant and $v \simeq 246 \, \rm GeV$ is the SM Higgs VEV.
When $S=0$, the lower limit on the $T$ parameter is $-0.042$ at 95\% confidence level~\cite{Tanabashi:2018oca}. The lower bound on the symmetry breaking scale $v_R$ is thus given by
\begin{align}
  v_R > 6.4 \, \mathrm{TeV} \left( \frac{\sin \theta_X}{0.45} \right).
\end{align}
The bound from the electroweak precision test is generally weaker than that from direct collider searches in our framework.

%%%%%%%%%%%%%%%%%%%%%%%%%%%%%%%%%%%%%%%%%%%%%%%%
\subsubsection{The electron electric dipole moment}
%%%%%%%%%%%%%%%%%%%%%%%%%%%%%%%%%%%%%%%%%%%%%%%%

%%%%%%%%%%%%%%%%%%%%%%%%%%%%%%%%%%%%%%%%%
\begin{figure}[t]
  \centering
  \includegraphics[width=0.4\linewidth]{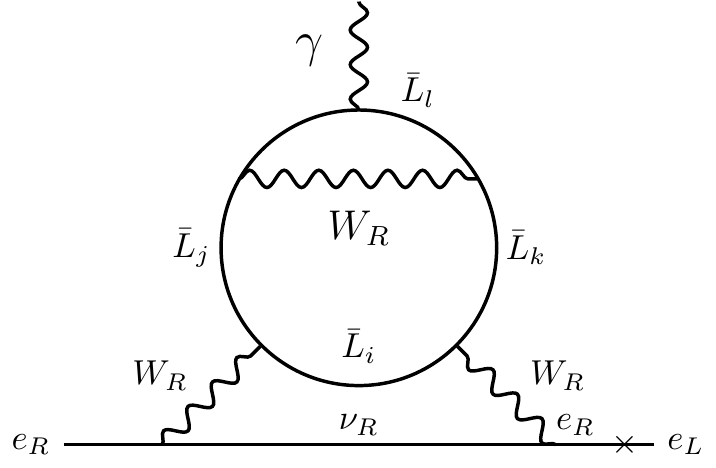}
  \caption{A three-loop contribution to the electron EDM in our model.}
  \label{fig:edm}
\end{figure}
%%%%%%%%%%%%%%%%%%%%%%%%%%%%%%%%%%%%%%%%%%

A CP violating new physics may generate observable electric dipole moments (EDMs).
We focus on the electron EDM, which is now the most sensitive probe of general CP violating new physics.
Our model predicts a new CP violating phase in the matrix $K$ of \eqref{eq:CKM parametrization}.
The lowest order contribution to the electron EDM involving this complex phase is shown in Fig.~\ref{fig:edm}.
Note that the diagram must contain all the three generations of new leptons $\bar{L}_i \,\, (i =1,2,3)$
because three generations of new leptons are essential for the appearance of the physical complex phase
as in the case of the ordinary CKM phase.
The diagram must also contain a chirality flip of the electron, which is provided by an electron mass insertion.
Then, an order of magnitude estimation of the electron EDM is given by
\begin{align}
  \label{eq:edm result}
  \frac{d_e}{e} \sim \left(\frac{1}{16 \pi^2}\right)^3 \frac{m_e}{m_{W_R}^2} J^\prime,
\end{align}
where $J^\prime$ has been defined in eq.~\eqref{eq:Jarlskog parameter}
and $m_{W_R}$ is the $W_R^\pm$ boson mass.
This estimation gives a negligibly small value of the electron EDM.
In fact, for a benchmark point of $g_R=1.5$, $v_R = 15 \tev$ and $J'=0.08$,
we obtain $d_e \simeq  10^{-33} \, e\rm cm$
which is much smaller than the current experimental limit, $d_e < 1.1 \times 10^{-29} \, e\rm cm$
\cite{an2018improved}.
Thus, our model parameter space is not constrained by EDM measurements
in contrast to the usual electroweak baryogenesis scenarios. 

There may be also one-loop corrections to the EDM involving the portal couplings in eqs.~\eqref{portal interaction} or \eqref{portal interaction2}. Similarly, lepton flavor-violating processes could be introduced by one-loop corrections.
However, the portal coupling does not have to be large for successful baryogenesis (see eq.~\eqref{eq:decay rate} and the discussion below) and
hence the EDM and lepton-flavor violation do not serve as model-independent probes of the scenario.
%the magnitude of the EDM or lepton-flavor violation is model-dependent.

%%%%%%%%%%%%%%%%%%%%%%%%%%%%%%%%%%%%%%%%%%%%%%%%  
\subsection{Gravitational wave signals}
%%%%%%%%%%%%%%%%%%%%%%%%%%%%%%%%%%%%%%%%%%%%%%%%

The first-order $G_{\rm LR} \to G_{\rm SM}$ phase transition may generate a stochastic gravitational wave (GW) background that can be observed by future space-based interferometers.\footnote{GWs from the spontaneous breaking of
$SU(2)_R\times U(1)_{B-L}$ have been discussed in ref.~\cite{Brdar:2019fur},
although the matter content considered there differs from our model.
For general reviews on GWs from first-order phase transitions, see e.g. refs.~\cite{Caprini:2015zlo,Fujikura:2018duw,Caprini:2019egz}.}
A GW signal from a phase transition is characterized by two important parameters:
the ratio of the vacuum energy released by the phase transition
to the radiation energy of the Universe $\alpha$,
and the duration of the phase transition $\beta$.
The former is related to the change of the trace of the energy-momentum tensor across the bubble wall,
\begin{align}
\epsilon =  \Delta V_{\rm eff} - \frac{T}{4}\frac{\partial \Delta V_{\rm eff}}{\partial T} \, . \label{eq:alpha}
\end{align}
Here, $\Delta V_{\rm eff}\equiv V(\phi_R^{\rm false},T_{\rm n}) -V (\phi_R^{\rm true},T_{\rm n})$ is the free energy difference between the false vacuum and the true vacuum at the nucleation temperature, $T=T_{\rm n}$,
and $\phi_R^{\rm false \, (true)}$ is the position of the false (true) vacuum.
Using this $\epsilon$, the parameter $\alpha$ is defined as
\begin{align}
\alpha =\left. \frac{\epsilon}{\frac{\pi^2}{30}g_* T^4}\right|_{T=T_{\rm n}}.
\end{align}
The duration of a phase transition is a characteristic time (length) scale of the phase transition given by
\cite{Caprini:2015zlo}
\begin{align}
\beta \equiv \left[ H(T) T \frac{d}{dT} \left.\left(\frac{S_3}{T}\right) \right]\right|_{T=T_{\rm n}}\label{eq:beta}.
\end{align}
The bounce action $S_3$ has been defined in sec.~\ref{sec:sphaleron}.

The total GW signal can be divided into the following two parts,
\begin{align}
\Omega_{\rm tot}h^2 \simeq \Omega_{\rm sw}h^2 + \Omega_{\rm tur}h^2,
\end{align}
where $\Omega_{\rm sw}h^2$ and $\Omega_{\rm tur}h^2$ are sourced by the sound wave and turbulence of plasma,
respectively.
We neglect a contribution from bubble collisions
\cite{Turner:1990rc,Kosowsky:1991ua,Turner:1992tz,Kosowsky:1992vn,Jinno:2016vai,Jinno:2017fby}
since the most of the energy released by the phase transition is expected to be converted into the thermal plasma in our model.
The sound wave contribution is given by~\cite{Hindmarsh:2015qta}
\begin{equation}
\begin{split}
&\Omega_{\rm sw}h^2 = 2.65\times 10^{-6} \\
&\qquad \qquad \times H(T_{\rm n})\tau_{\rm shock} \left(\frac{H(T_{\rm n})}{\beta} \right) \left(\frac{\kappa_{\rm sw} \alpha}{1+\alpha} \right)^2 \left(\frac{100}{g_*}\right)^{\frac{1}{3}}v_w \left( \frac{f}{f_{\rm sw}}\right)^3 \left( \dfrac{7}{4+3(f/f_{\rm sw})^2 }\right)^{\frac{7}{2}},\\[2ex]
&f_{\rm sw} = 1.9\times 10^{-2} \, {\rm mHz} \times \frac{1}{v_w} \left(\frac{\beta}{H(T_{\rm n})}\right) \left( \frac{T_{\rm n}}{100 \, {\rm GeV}} \right)\left( \frac{g_*}{100}\right)^{\frac{1}{6}} .
\end{split}
\end{equation}
Here, $\kappa_{\rm sw}$ is the efficiency factor of the sound wave and $H\tau_{\rm shock}$ represents the reduction factor due to the short-lasting sound waves.
We assume that the wall velocity is smaller than the sound velocity of the thermal plasma,
$v_w\lesssim c_s\equiv 1/\sqrt{3}$ (i.e., the deflagration regime), which would be realized
by the friction of the thermal plasma
\cite{Bodeker:2009qy,Bodeker:2017cim,Ellis:2019oqb, Mancha:2020fzw,Hoeche:2020rsg,Vanvlasselaer:2020niz}.
Under this assumption, $\kappa_{\rm sw}$ can be fitted by the following formula as found in
ref.~\cite{Espinosa:2010hh}:
\begin{align}
\kappa_{\rm sw}\simeq \frac{c_s^{11/5} \kappa_A \kappa_B}{(c_s^{11/5}-v_w^{11/5})\kappa_B+v_w c_s^{6/5}\kappa_A} \, ,
\end{align}
where 
\begin{equation}
\begin{split}
    &\kappa_A \simeq v_w^{6/5}\frac{6.9\alpha}{1.36-0.037\sqrt{\alpha}+\alpha} \, , \\
    &\kappa_B \simeq \frac{\alpha^{2/5}}{0.017+(0.997+\alpha)^{2/5}} \, .
\end{split}
\end{equation}
The reduction factor is given by~\cite{Ellis:2018mja,Cutting:2019zws,Ellis:2020awk}
\begin{align}
    H(T_{\rm n}) \tau_{\rm shock} =
    {\rm min}\left[1,~(8\pi)^\frac{1}{3}\left(\frac{{\rm max}\{v_w,c_s\}}{\beta/H(T_{\rm n})} \right)\left(\frac{4}{3}\frac{1+\alpha}{\kappa_{\rm sw}\alpha}\right)^\frac{1}{2} \right].
\end{align}
On the other hand, the turbulence contribution is expressed as
\cite{Caprini:2009yp,Binetruy:2012ze}
\begin{equation}
\begin{split}
&\Omega_{\rm tur}h^2 = 3.35 \times 10^{-4} \\
&\qquad \qquad \times \left(\frac{H(T_{\rm n})}{\beta}\right) \left(\frac{\kappa_{\rm tur}\alpha}{1+\alpha}\right)^{\frac{3}{2}} \left(\frac{100}{g_*}\right)^{\frac{1}{3}}v_w \left( \dfrac{(f/f_{\rm tur})^3}{(1+(f/f_{\rm tur}))^\frac{11}{3}(1+8\pi f/h_*)}\right) ,\\[2ex]
&f_{\rm tur}= 2.7 \times 10^{-2} \, {\rm mHz} \times \frac{1}{v_w} \left(\frac{\beta}{H(T_{\rm n})}\right) \left( \frac{T_{\rm n}}{100 \, {\rm GeV}}\right) \left( \frac{g_*}{100}\right)^\frac{1}{6}, \\[1ex]
&h_* = 16.5 \times 10^{-3} \, {\rm mHz} \times \left( \frac{T_{\rm n}}{100 \, {\rm GeV}}  \right)\left( \frac{g_*}{100}\right)^\frac{1}{6},
\end{split}
\end{equation}
where $\kappa_{\rm tur}$ is the efficiency factor of the plasma turbulence.
We assume $\kappa_{\rm tur} = 0.05 \kappa_{\rm sw}$, following ref.~\cite{Caprini:2015zlo}.

%%%%%%%%%%%%%%%%%%%%%%%%%%%%%%%%%%%%%%%%%%%%%%%%%%%
\begin{figure}
    \centering
    \includegraphics[width=0.6\linewidth]{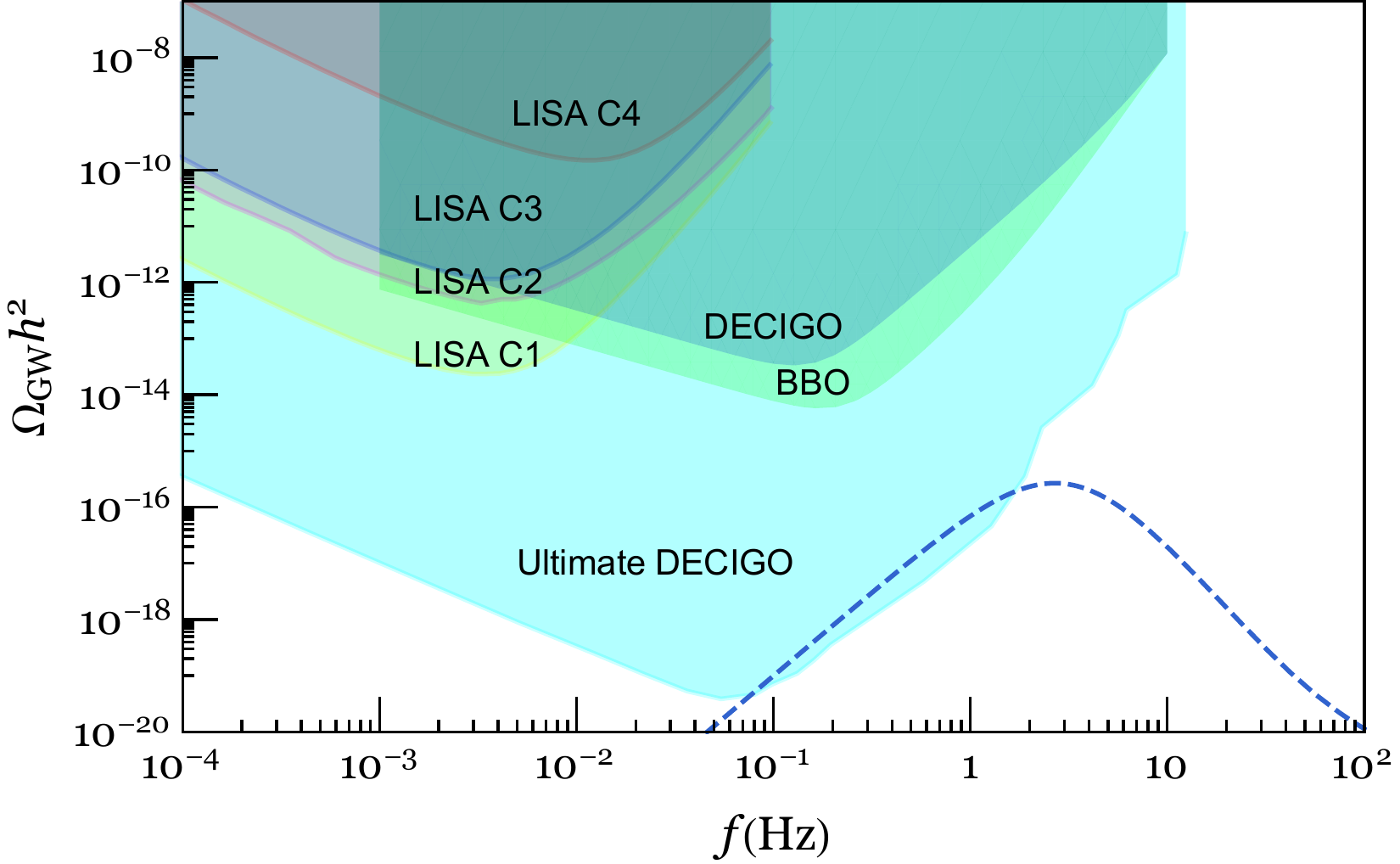}
    \caption{
    The generated GW signal for a benchmark point of $\ynth=1.35$, $\yntwo=0.89$ $g_R=1.55$,  $\lambda_R=0.014$,
    $v_w=0.48$ and $v_R=15 \tev$ (blue dashed). 
    Colored regions are probed by LISA, DECIGO, BBO and U-DECIGO.
    }
    \label{fig:gw}
\end{figure}
%%%%%%%%%%%%%%%%%%%%%%%%%%%%%%%%%%%%%%%%%%%%%%%%%%%

The generated GW signal for a benchmark point is shown in fig.~\ref{fig:gw}.
We take $\ynth=1.35$, $\yntwo=0.89$, $g_R=1.55$,  $\lambda_R=0.014$, $v_w=0.48$
and $v_R=15 \tev$, which reproduces the observed amount of baryon asymmetry.
In this benchmark point, we obtain $T_{\rm n}=4.34 \tev,~\beta/H (T_{\rm n})=1471$ and $\alpha = 0.03$.
Sensitivity curves of future space-based interferometers LISA
\cite{Caprini:2015zlo,Klein:2015hvg}, DECIGO, BBO~\cite{Yagi:2011wg}, and ultimate-DECIGO~\cite{Kudoh:2005as,Kuroyanagi:2014qaa} (U-DECIGO)
are also plotted.
The GW signal is within the reach of U-DECIGO.

%%%%%%%%%%%%%%%%%%%%%%%%%%%%%%%%%%%%%%%%%%%%%%%%
\section{Conclusion}
\label{conclusions}
%%%%%%%%%%%%%%%%%%%%%%%%%%%%%%%%%%%%%%%%%%%%%%%%

In this paper, we have proposed a new baryogenesis scenario based on a first-order phase transition
associated with the spontaneous breaking of an extended gauge symmetry, $SU(2)_R\times U(1)_X\to U(1)_Y$.
Our model contains three generations of new chiral leptons that make the $B-L$ symmetry anomalous.
A moving bubble wall from the first-order phase transition
and the $SU(2)_R$ sphaleron process play essential roles in generation of the $B-L$ asymmetry
in a similar way to the SM electroweak baryogenesis.
To avoid a $B-L$ wash-out inside the bubble wall, the strength of the phase transition
is required to satisfy the sphaleron decoupling condition.
The $B-L$ asymmetry carried by new leptons is generated
by quantum mechanical reflections of the new leptons by the wall. 
In our model, a new source of CP violation is a complex phase in the mixing matrix of the new lepton sector similar to the CKM phase.
The reflection coefficient is computed by solving the effective Dirac equation on the thermal and bubble wall backgrounds.
We have taken account of damping effects on quasiparticles in the effective Dirac equation,
and in sec.~\ref{sec:perturbation}, obtained an analytical expression for the produced baryon asymmetry
by using a perturbative expansion with thin-wall and small-momentum approximations
when new leptons are sufficiently light.
We also qualitatively estimated the amount of the baryon asymmetry for heavy new leptons in a small momentum regime,
for light new leptons in a large momentum regime, and for heavy new leptons in a large momentum regime
in secs.~\ref{sec:heavy masses}, \ref{sec:large momentum regime}, and~\ref{sec:(iii)},  respectively.
By the detailed analysis, it turned out that the observed baryon asymmetry can be explained
for very heavy new leptons in a large momentum regime.
In the model with the extra interaction \eqref{eq:additional thermal mass},
we found that the observed baryon asymmetry may be produced
for light new leptons in a small momentum regime where the perturbation theory is reliable.

There are some ambiguities in our estimation of the baryon asymmetry.
In the small new lepton mass regime, we reduced the computation of the scattering coefficient
to a one-dimensional problem and simply introduced a cut-off to the momentum parallel to the wall.
We took the cut-off at the thermal mass because the kinematics of the new leptons is dominated by the thermal effect. 
Our estimation of the baryon asymmetry given by  eq.~\eqref{eq:(iii)} in the large mass regime
is based on a simple dimensional analysis.
Such ambiguities will be ameliorated by exactly solving the corresponding Dirac equation.
It is worthwhile to note that even if there exists a large enhancement from these ambiguities
on the produced baryon asymmetry,
we can realize the observed value by taking smaller Yukawa couplings of the new leptons.
However, if there is a suppression more than a factor of $0.1$, it is difficult to realize the observed baryon asymmetry.

To avoid a severe fine-tuning, the scale of the new gauge symmetry breaking
is favored to be near the electroweak scale.
New gauge bosons associated with the extended gauge symmetry
and new matter fermions predicted by the model may be within the reach of the LHC or future experimental programs.
The LHC searches for the new gauge bosons have already put a lower bound on the symmetry breaking scale
at around $10 \, \rm TeV$.
We have also discussed implications of the electroweak precision test and the electron EDM measurement for our model.
It has been shown that both measurements do not give strong constraints on the model parameter space,
in contrast to the usual electroweak baryogenesis scenarios.
The new first-order phase transition can produce a stochastic GW background.
We showed that a benchmark point providing the observed baryon asymmetry predicts a GW signal
that can be detected by the U-DECIGO.
As discussed in sec.~\ref{sec:DM}, when the model respects a $Z_2$ symmetry,
the lightest neutral new fermion is a dark matter candidate.
Such a possibility will be further explored in a future study.  

%#######################

%%%%%%%%%%%%%%%%%%%%%%%%%%%%%%%%%%%%%%%%%%%%%%%%
\section*{Acknowledgements}
We would like to thank Shintaro Eijima, Hiroyuki Ishida, Sho Iwamoto, Teppei Kitahara and Koji Tsumura for discussions.
This work was supported in part by JSPS Grants-in-Aid for Research Fellows No.~20J12415
(K.F.), DOE grant DOE-SC0010008 (I.R.W.), and Friends of the Institute for Advanced Study (K.H.).
%%%%%%%%%%%%%%%%%%%%%%%%%%%%%%%%%%%%%%%%%%%%%%%%

%%%%%%%%%%%%%%%%%%%%%%%%%%%%%%%%%%%%%%%%%%%%%%%%
\bibliographystyle{JHEP}
\bibliography{EWBG_with_Chiral_Matter}

\providecommand{\href}[2]{#2}\begingroup\raggedright\begin{thebibliography}{100}

\bibitem{Sakharov:1967dj}
A.~D. Sakharov, \emph{{Violation of CP Invariance, C asymmetry, and baryon
  asymmetry of the universe}},
  \href{https://doi.org/10.1070/PU1991v034n05ABEH002497}{\emph{Pisma Zh. Eksp.
  Teor. Fiz.} {\bfseries 5} (1967) 32}.

\bibitem{tHooft:1976rip}
G.~'t~Hooft, \emph{{Symmetry Breaking Through Bell-Jackiw Anomalies}},
  \href{https://doi.org/10.1103/PhysRevLett.37.8}{\emph{Phys. Rev. Lett.}
  {\bfseries 37} (1976) 8}.

\bibitem{tHooft:1976snw}
G.~'t~Hooft, \emph{{Computation of the Quantum Effects Due to a
  Four-Dimensional Pseudoparticle}},
  \href{https://doi.org/10.1103/PhysRevD.18.2199.3,
  10.1103/PhysRevD.14.3432}{\emph{Phys. Rev.} {\bfseries D14} (1976) 3432}.

\bibitem{Manton:1983nd}
N.~S. Manton, \emph{{Topology in the Weinberg-Salam Theory}},
  \href{https://doi.org/10.1103/PhysRevD.28.2019}{\emph{Phys. Rev.} {\bfseries
  D28} (1983) 2019}.

\bibitem{Klinkhamer:1984di}
F.~R. Klinkhamer and N.~S. Manton, \emph{{A Saddle Point Solution in the
  Weinberg-Salam Theory}},
  \href{https://doi.org/10.1103/PhysRevD.30.2212}{\emph{Phys. Rev.} {\bfseries
  D30} (1984) 2212}.

\bibitem{Kuzmin:1985mm}
V.~A. Kuzmin, V.~A. Rubakov and M.~E. Shaposhnikov, \emph{{On the Anomalous
  Electroweak Baryon Number Nonconservation in the Early Universe}},
  \href{https://doi.org/10.1016/0370-2693(85)91028-7}{\emph{Phys. Lett.}
  {\bfseries 155B} (1985) 36}.

\bibitem{Cabibbo:1963yz}
N.~Cabibbo, \emph{{Unitary Symmetry and Leptonic Decays}},
  \href{https://doi.org/10.1103/PhysRevLett.10.531}{\emph{Phys. Rev. Lett.}
  {\bfseries 10} (1963) 531}.

\bibitem{Kobayashi:1973fv}
M.~Kobayashi and T.~Maskawa, \emph{{CP Violation in the Renormalizable Theory
  of Weak Interaction}}, \href{https://doi.org/10.1143/PTP.49.652}{\emph{Prog.
  Theor. Phys.} {\bfseries 49} (1973) 652}.

\bibitem{Farrar:1993sp}
G.~R. Farrar and M.~E. Shaposhnikov, \emph{{Baryon asymmetry of the universe in
  the minimal Standard Model}},
  \href{https://doi.org/10.1103/PhysRevLett.71.210.2,
  10.1103/PhysRevLett.70.2833}{\emph{Phys. Rev. Lett.} {\bfseries 70} (1993)
  2833} [\href{https://arxiv.org/abs/hep-ph/9305274}{{\ttfamily
  hep-ph/9305274}}].

\bibitem{Farrar:1993hn}
G.~R. Farrar and M.~E. Shaposhnikov, \emph{{Baryon asymmetry of the universe in
  the standard electroweak theory}},
  \href{https://doi.org/10.1103/PhysRevD.50.774}{\emph{Phys. Rev.} {\bfseries
  D50} (1994) 774} [\href{https://arxiv.org/abs/hep-ph/9305275}{{\ttfamily
  hep-ph/9305275}}].

\bibitem{Gavela:1993ts}
M.~B. Gavela, P.~Hernandez, J.~Orloff and O.~Pene, \emph{{Standard model CP
  violation and baryon asymmetry}},
  \href{https://doi.org/10.1142/S0217732394000629}{\emph{Mod. Phys. Lett.}
  {\bfseries A9} (1994) 795}
  [\href{https://arxiv.org/abs/hep-ph/9312215}{{\ttfamily hep-ph/9312215}}].

\bibitem{Huet:1994jb}
P.~Huet and E.~Sather, \emph{{Electroweak baryogenesis and standard model CP
  violation}}, \href{https://doi.org/10.1103/PhysRevD.51.379}{\emph{Phys. Rev.}
  {\bfseries D51} (1995) 379}
  [\href{https://arxiv.org/abs/hep-ph/9404302}{{\ttfamily hep-ph/9404302}}].

\bibitem{Gavela:1994ds}
M.~B. Gavela, M.~Lozano, J.~Orloff and O.~Pene, \emph{{Standard model CP
  violation and baryon asymmetry. Part 1: Zero temperature}},
  \href{https://doi.org/10.1016/0550-3213(94)00409-9}{\emph{Nucl. Phys.}
  {\bfseries B430} (1994) 345}
  [\href{https://arxiv.org/abs/hep-ph/9406288}{{\ttfamily hep-ph/9406288}}].

\bibitem{Gavela:1994dt}
M.~B. Gavela, P.~Hernandez, J.~Orloff, O.~Pene and C.~Quimbay, \emph{{Standard
  model CP violation and baryon asymmetry. Part 2: Finite temperature}},
  \href{https://doi.org/10.1016/0550-3213(94)00410-2}{\emph{Nucl. Phys.}
  {\bfseries B430} (1994) 382}
  [\href{https://arxiv.org/abs/hep-ph/9406289}{{\ttfamily hep-ph/9406289}}].

\bibitem{Kajantie:1995kf}
K.~Kajantie, M.~Laine, K.~Rummukainen and M.~E. Shaposhnikov, \emph{{The
  Electroweak phase transition: A Nonperturbative analysis}},
  \href{https://doi.org/10.1016/0550-3213(96)00052-1}{\emph{Nucl. Phys.}
  {\bfseries B466} (1996) 189}
  [\href{https://arxiv.org/abs/hep-lat/9510020}{{\ttfamily hep-lat/9510020}}].

\bibitem{Kajantie:1996qd}
K.~Kajantie, M.~Laine, K.~Rummukainen and M.~E. Shaposhnikov, \emph{{A
  Nonperturbative analysis of the finite T phase transition in SU(2) x U(1)
  electroweak theory}},
  \href{https://doi.org/10.1016/S0550-3213(97)00164-8}{\emph{Nucl. Phys. B}
  {\bfseries 493} (1997) 413}
  [\href{https://arxiv.org/abs/hep-lat/9612006}{{\ttfamily hep-lat/9612006}}].

\bibitem{Rummukainen:1998as}
K.~Rummukainen, M.~Tsypin, K.~Kajantie, M.~Laine and M.~E. Shaposhnikov,
  \emph{{The Universality class of the electroweak theory}},
  \href{https://doi.org/10.1016/S0550-3213(98)00494-5}{\emph{Nucl. Phys. B}
  {\bfseries 532} (1998) 283}
  [\href{https://arxiv.org/abs/hep-lat/9805013}{{\ttfamily hep-lat/9805013}}].

\bibitem{Trodden:1998ym}
M.~Trodden, \emph{{Electroweak baryogenesis}},
  \href{https://doi.org/10.1103/RevModPhys.71.1463}{\emph{Rev. Mod. Phys.}
  {\bfseries 71} (1999) 1463}
  [\href{https://arxiv.org/abs/hep-ph/9803479}{{\ttfamily hep-ph/9803479}}].

\bibitem{Morrissey:2012db}
D.~E. Morrissey and M.~J. Ramsey-Musolf, \emph{{Electroweak baryogenesis}},
  \href{https://doi.org/10.1088/1367-2630/14/12/125003}{\emph{New J. Phys.}
  {\bfseries 14} (2012) 125003}
  [\href{https://arxiv.org/abs/1206.2942}{{\ttfamily 1206.2942}}].

\bibitem{Carena:1996wj}
M.~Carena, M.~Quiros and C.~Wagner, \emph{{Opening the window for electroweak
  baryogenesis}},
  \href{https://doi.org/10.1016/0370-2693(96)00475-3}{\emph{Phys. Lett. B}
  {\bfseries 380} (1996) 81}
  [\href{https://arxiv.org/abs/hep-ph/9603420}{{\ttfamily hep-ph/9603420}}].

\bibitem{Delepine:1996vn}
D.~Delepine, J.~Gerard, R.~Gonzalez~Felipe and J.~Weyers, \emph{{A Light stop
  and electroweak baryogenesis}},
  \href{https://doi.org/10.1016/0370-2693(96)00921-5}{\emph{Phys. Lett. B}
  {\bfseries 386} (1996) 183}
  [\href{https://arxiv.org/abs/hep-ph/9604440}{{\ttfamily hep-ph/9604440}}].

\bibitem{Carena:1997ki}
M.~Carena, M.~Quiros and C.~Wagner, \emph{{Electroweak baryogenesis and Higgs
  and stop searches at LEP and the Tevatron}},
  \href{https://doi.org/10.1016/S0550-3213(98)00187-4}{\emph{Nucl. Phys. B}
  {\bfseries 524} (1998) 3}
  [\href{https://arxiv.org/abs/hep-ph/9710401}{{\ttfamily hep-ph/9710401}}].

\bibitem{Quiros:1999tx}
M.~Quiros and M.~Seco, \emph{{Electroweak baryogenesis in the MSSM}},
  \href{https://doi.org/10.1016/S0920-5632(99)00860-9}{\emph{Nucl. Phys. B
  Proc. Suppl.} {\bfseries 81} (2000) 63}
  [\href{https://arxiv.org/abs/hep-ph/9903274}{{\ttfamily hep-ph/9903274}}].

\bibitem{Cline:2000kb}
J.~M. Cline and K.~Kainulainen, \emph{{A New source for electroweak
  baryogenesis in the MSSM}},
  \href{https://doi.org/10.1103/PhysRevLett.85.5519}{\emph{Phys. Rev. Lett.}
  {\bfseries 85} (2000) 5519}
  [\href{https://arxiv.org/abs/hep-ph/0002272}{{\ttfamily hep-ph/0002272}}].

\bibitem{Huber:2001xf}
S.~Huber, P.~John and M.~Schmidt, \emph{{Bubble walls, CP violation and
  electroweak baryogenesis in the MSSM}},
  \href{https://doi.org/10.1007/PL00022989}{\emph{Eur. Phys. J. C} {\bfseries
  20} (2001) 695} [\href{https://arxiv.org/abs/hep-ph/0101249}{{\ttfamily
  hep-ph/0101249}}].

\bibitem{Carena:2002ss}
M.~Carena, M.~Quiros, M.~Seco and C.~Wagner, \emph{{Improved Results in
  Supersymmetric Electroweak Baryogenesis}},
  \href{https://doi.org/10.1016/S0550-3213(02)01065-9}{\emph{Nucl. Phys. B}
  {\bfseries 650} (2003) 24}
  [\href{https://arxiv.org/abs/hep-ph/0208043}{{\ttfamily hep-ph/0208043}}].

\bibitem{Lee:2004we}
C.~Lee, V.~Cirigliano and M.~J. Ramsey-Musolf, \emph{{Resonant relaxation in
  electroweak baryogenesis}},
  \href{https://doi.org/10.1103/PhysRevD.71.075010}{\emph{Phys. Rev. D}
  {\bfseries 71} (2005) 075010}
  [\href{https://arxiv.org/abs/hep-ph/0412354}{{\ttfamily hep-ph/0412354}}].

\bibitem{Carena:2008rt}
M.~Carena, G.~Nardini, M.~Quiros and C.~E. Wagner, \emph{{The Effective Theory
  of the Light Stop Scenario}},
  \href{https://doi.org/10.1088/1126-6708/2008/10/062}{\emph{JHEP} {\bfseries
  10} (2008) 062} [\href{https://arxiv.org/abs/0806.4297}{{\ttfamily
  0806.4297}}].

\bibitem{Carena:2008vj}
M.~Carena, G.~Nardini, M.~Quiros and C.~Wagner, \emph{{The Baryogenesis Window
  in the MSSM}},
  \href{https://doi.org/10.1016/j.nuclphysb.2008.12.014}{\emph{Nucl. Phys. B}
  {\bfseries 812} (2009) 243}
  [\href{https://arxiv.org/abs/0809.3760}{{\ttfamily 0809.3760}}].

\bibitem{Cirigliano:2009yd}
V.~Cirigliano, Y.~Li, S.~Profumo and M.~J. Ramsey-Musolf, \emph{{MSSM
  Baryogenesis and Electric Dipole Moments: An Update on the Phenomenology}},
  \href{https://doi.org/10.1007/JHEP01(2010)002}{\emph{JHEP} {\bfseries 01}
  (2010) 002} [\href{https://arxiv.org/abs/0910.4589}{{\ttfamily 0910.4589}}].

\bibitem{Carena:2012np}
M.~Carena, G.~Nardini, M.~Quiros and C.~E. Wagner, \emph{{MSSM Electroweak
  Baryogenesis and LHC Data}},
  \href{https://doi.org/10.1007/JHEP02(2013)001}{\emph{JHEP} {\bfseries 02}
  (2013) 001} [\href{https://arxiv.org/abs/1207.6330}{{\ttfamily 1207.6330}}].

\bibitem{Turok:1990in}
N.~Turok and J.~Zadrozny, \emph{{Dynamical generation of baryons at the
  electroweak transition}},
  \href{https://doi.org/10.1103/PhysRevLett.65.2331}{\emph{Phys. Rev. Lett.}
  {\bfseries 65} (1990) 2331}.

\bibitem{Turok:1990zg}
N.~Turok and J.~Zadrozny, \emph{{Electroweak baryogenesis in the two doublet
  model}}, \href{https://doi.org/10.1016/0550-3213(91)90356-3}{\emph{Nucl.
  Phys. B} {\bfseries 358} (1991) 471}.

\bibitem{Cohen:1991iu}
A.~G. Cohen, D.~Kaplan and A.~Nelson, \emph{{Spontaneous baryogenesis at the
  weak phase transition}},
  \href{https://doi.org/10.1016/0370-2693(91)91711-4}{\emph{Phys. Lett. B}
  {\bfseries 263} (1991) 86}.

\bibitem{Nelson:1991ab}
A.~Nelson, D.~Kaplan and A.~G. Cohen, \emph{{Why there is something rather than
  nothing: Matter from weak interactions}},
  \href{https://doi.org/10.1016/0550-3213(92)90440-M}{\emph{Nucl. Phys. B}
  {\bfseries 373} (1992) 453}.

\bibitem{Cohen:1992yh}
A.~G. Cohen, D.~Kaplan and A.~Nelson, \emph{{Debye screening and baryogenesis
  during the electroweak phase transition}},
  \href{https://doi.org/10.1016/0370-2693(92)91640-U}{\emph{Phys. Lett. B}
  {\bfseries 294} (1992) 57}
  [\href{https://arxiv.org/abs/hep-ph/9206214}{{\ttfamily hep-ph/9206214}}].

\bibitem{Cline:1995dg}
J.~M. Cline, K.~Kainulainen and A.~P. Vischer, \emph{{Dynamics of two Higgs
  doublet CP violation and baryogenesis at the electroweak phase transition}},
  \href{https://doi.org/10.1103/PhysRevD.54.2451}{\emph{Phys. Rev.} {\bfseries
  D54} (1996) 2451} [\href{https://arxiv.org/abs/hep-ph/9506284}{{\ttfamily
  hep-ph/9506284}}].

\bibitem{Cline:1996mga}
J.~M. Cline and P.-A. Lemieux, \emph{{Electroweak phase transition in two Higgs
  doublet models}}, \href{https://doi.org/10.1103/PhysRevD.55.3873}{\emph{Phys.
  Rev. D} {\bfseries 55} (1997) 3873}
  [\href{https://arxiv.org/abs/hep-ph/9609240}{{\ttfamily hep-ph/9609240}}].

\bibitem{Fromme:2006cm}
L.~Fromme, S.~J. Huber and M.~Seniuch, \emph{{Baryogenesis in the two-Higgs
  doublet model}},
  \href{https://doi.org/10.1088/1126-6708/2006/11/038}{\emph{JHEP} {\bfseries
  11} (2006) 038} [\href{https://arxiv.org/abs/hep-ph/0605242}{{\ttfamily
  hep-ph/0605242}}].

\bibitem{Tranberg:2012jp}
A.~Tranberg and B.~Wu, \emph{{Cold Electroweak Baryogenesis in the Two
  Higgs-Doublet Model}},
  \href{https://doi.org/10.1007/JHEP07(2012)087}{\emph{JHEP} {\bfseries 07}
  (2012) 087} [\href{https://arxiv.org/abs/1203.5012}{{\ttfamily 1203.5012}}].

\bibitem{Andersen:2017ika}
J.~O. Andersen, T.~Gorda, A.~Helset, L.~Niemi, T.~V.~I. Tenkanen, A.~Tranberg
  et~al., \emph{{Nonperturbative Analysis of the Electroweak Phase Transition
  in the Two Higgs Doublet Model}},
  \href{https://doi.org/10.1103/PhysRevLett.121.191802}{\emph{Phys. Rev. Lett.}
  {\bfseries 121} (2018) 191802}
  [\href{https://arxiv.org/abs/1711.09849}{{\ttfamily 1711.09849}}].

\bibitem{Curtin:2012aa}
D.~Curtin, P.~Jaiswal and P.~Meade, \emph{{Excluding Electroweak Baryogenesis
  in the MSSM}}, \href{https://doi.org/10.1007/JHEP08(2012)005}{\emph{JHEP}
  {\bfseries 08} (2012) 005} [\href{https://arxiv.org/abs/1203.2932}{{\ttfamily
  1203.2932}}].

\bibitem{Andreev:2018ayy}
{\scshape ACME} collaboration, V.~Andreev et~al., \emph{{Improved limit on the
  electric dipole moment of the electron}},
  \href{https://doi.org/10.1038/s41586-018-0599-8}{\emph{Nature} {\bfseries
  562} (2018) 355}.

\bibitem{Nakai:2016atk}
Y.~Nakai and M.~Reece, \emph{{Electric Dipole Moments in Natural
  Supersymmetry}}, \href{https://doi.org/10.1007/JHEP08(2017)031}{\emph{JHEP}
  {\bfseries 08} (2017) 031}
  [\href{https://arxiv.org/abs/1612.08090}{{\ttfamily 1612.08090}}].

\bibitem{Cesarotti:2018huy}
C.~Cesarotti, Q.~Lu, Y.~Nakai, A.~Parikh and M.~Reece, \emph{{Interpreting the
  Electron EDM Constraint}},
  \href{https://doi.org/10.1007/JHEP05(2019)059}{\emph{JHEP} {\bfseries 05}
  (2019) 059} [\href{https://arxiv.org/abs/1810.07736}{{\ttfamily
  1810.07736}}].

\bibitem{Ishikawa:2014tfa}
K.~Ishikawa, T.~Kitahara and M.~Takimoto, \emph{{Towards a Scale Free
  Electroweak Baryogenesis}},
  \href{https://doi.org/10.1103/PhysRevD.91.055004}{\emph{Phys. Rev. D}
  {\bfseries 91} (2015) 055004}
  [\href{https://arxiv.org/abs/1410.5432}{{\ttfamily 1410.5432}}].

\bibitem{Baldes:2018nel}
I.~Baldes and G.~Servant, \emph{{High scale electroweak phase transition:
  baryogenesis $\&$ symmetry non-restoration}},
  \href{https://doi.org/10.1007/JHEP10(2018)053}{\emph{JHEP} {\bfseries 10}
  (2018) 053} [\href{https://arxiv.org/abs/1807.08770}{{\ttfamily
  1807.08770}}].

\bibitem{Glioti:2018roy}
A.~Glioti, R.~Rattazzi and L.~Vecchi, \emph{{Electroweak Baryogenesis above the
  Electroweak Scale}},
  \href{https://doi.org/10.1007/JHEP04(2019)027}{\emph{JHEP} {\bfseries 04}
  (2019) 027} [\href{https://arxiv.org/abs/1811.11740}{{\ttfamily
  1811.11740}}].

\bibitem{Shelton:2010ta}
J.~Shelton and K.~M. Zurek, \emph{{Darkogenesis: A baryon asymmetry from the
  dark matter sector}},
  \href{https://doi.org/10.1103/PhysRevD.82.123512}{\emph{Phys. Rev.}
  {\bfseries D82} (2010) 123512}
  [\href{https://arxiv.org/abs/1008.1997}{{\ttfamily 1008.1997}}].

\bibitem{Carena:2019xrr}
M.~Carena, M.~Quirós and Y.~Zhang, \emph{{Dark CP violation and gauged lepton
  or baryon number for electroweak baryogenesis}},
  \href{https://doi.org/10.1103/PhysRevD.101.055014}{\emph{Phys. Rev.}
  {\bfseries D101} (2020) 055014}
  [\href{https://arxiv.org/abs/1908.04818}{{\ttfamily 1908.04818}}].

\bibitem{Hall:2019ank}
E.~Hall, T.~Konstandin, R.~McGehee, H.~Murayama and G.~Servant,
  \emph{{Baryogenesis From a Dark First-Order Phase Transition}},
  \href{https://doi.org/10.1007/JHEP04(2020)042}{\emph{JHEP} {\bfseries 04}
  (2020) 042} [\href{https://arxiv.org/abs/1910.08068}{{\ttfamily
  1910.08068}}].

\bibitem{Shu:2006mm}
J.~Shu, T.~M.~P. Tait and C.~E.~M. Wagner, \emph{{Baryogenesis from an Earlier
  Phase Transition}},
  \href{https://doi.org/10.1103/PhysRevD.75.063510}{\emph{Phys. Rev. D}
  {\bfseries 75} (2007) 063510}
  [\href{https://arxiv.org/abs/hep-ph/0610375}{{\ttfamily hep-ph/0610375}}].

\bibitem{Fornal:2017owa}
B.~Fornal, Y.~Shirman, T.~M.~P. Tait and J.~R. West, \emph{{Asymmetric dark
  matter and baryogenesis from $SU(2)_{\ell}$}},
  \href{https://doi.org/10.1103/PhysRevD.96.035001}{\emph{Phys. Rev. D}
  {\bfseries 96} (2017) 035001}
  [\href{https://arxiv.org/abs/1703.00199}{{\ttfamily 1703.00199}}].

\bibitem{Davoudiasl:2016yfa}
H.~Davoudiasl, P.~P. Giardino and C.~Zhang, \emph{{Higgs-like boson at 750 GeV
  and genesis of baryons}},
  \href{https://doi.org/10.1103/PhysRevD.94.015006}{\emph{Phys. Rev. D}
  {\bfseries 94} (2016) 015006}
  [\href{https://arxiv.org/abs/1605.00037}{{\ttfamily 1605.00037}}].

\bibitem{Pati:1974yy}
J.~C. Pati and A.~Salam, \emph{{Lepton Number as the Fourth Color}},
  \href{https://doi.org/10.1103/PhysRevD.10.275}{\emph{Phys. Rev. D} {\bfseries
  10} (1974) 275}.

\bibitem{Mohapatra:1974gc}
R.~Mohapatra and J.~C. Pati, \emph{{A Natural Left-Right Symmetry}},
  \href{https://doi.org/10.1103/PhysRevD.11.2558}{\emph{Phys. Rev. D}
  {\bfseries 11} (1975) 2558}.

\bibitem{Senjanovic:1975rk}
G.~Senjanovic and R.~N. Mohapatra, \emph{{Exact Left-Right Symmetry and
  Spontaneous Violation of Parity}},
  \href{https://doi.org/10.1103/PhysRevD.12.1502}{\emph{Phys. Rev. D}
  {\bfseries 12} (1975) 1502}.

\bibitem{Kim:2017qxo}
J.~E. Kim, \emph{{Naturally realized two dark Z's near the electroweak scale}},
  \href{https://doi.org/10.1103/PhysRevD.96.055033}{\emph{Phys. Rev.}
  {\bfseries D96} (2017) 055033}
  [\href{https://arxiv.org/abs/1703.10925}{{\ttfamily 1703.10925}}].

\bibitem{Beg:1978mt}
M.~Beg and H.-S. Tsao, \emph{{Strong P, T Noninvariances in a Superweak
  Theory}}, \href{https://doi.org/10.1103/PhysRevLett.41.278}{\emph{Phys. Rev.
  Lett.} {\bfseries 41} (1978) 278}.

\bibitem{Mohapatra:1978fy}
R.~N. Mohapatra and G.~Senjanovic, \emph{{Natural Suppression of Strong p and t
  Noninvariance}},
  \href{https://doi.org/10.1016/0370-2693(78)90243-5}{\emph{Phys. Lett. B}
  {\bfseries 79} (1978) 283}.

\bibitem{Hall:2019qwx}
L.~J. Hall and K.~Harigaya, \emph{{Higgs Parity Grand Unification}},
  \href{https://doi.org/10.1007/JHEP11(2019)033}{\emph{JHEP} {\bfseries 11}
  (2019) 033} [\href{https://arxiv.org/abs/1905.12722}{{\ttfamily
  1905.12722}}].

\bibitem{Bertolini:2019out}
S.~Bertolini, A.~Maiezza and F.~Nesti, \emph{{Kaon CP violation and neutron EDM
  in the minimal left-right symmetric model}},
  \href{https://doi.org/10.1103/PhysRevD.101.035036}{\emph{Phys. Rev. D}
  {\bfseries 101} (2020) 035036}
  [\href{https://arxiv.org/abs/1911.09472}{{\ttfamily 1911.09472}}].

\bibitem{Guadagnoli:2010sd}
D.~Guadagnoli and R.~N. Mohapatra, \emph{{TeV Scale Left Right Symmetry and
  Flavor Changing Neutral Higgs Effects}},
  \href{https://doi.org/10.1016/j.physletb.2010.10.027}{\emph{Phys. Lett. B}
  {\bfseries 694} (2011) 386}
  [\href{https://arxiv.org/abs/1008.1074}{{\ttfamily 1008.1074}}].

\bibitem{Mohapatra:2013cia}
R.~N. Mohapatra and Y.~Zhang, \emph{{LHC accessible second Higgs boson in the
  left-right model}},
  \href{https://doi.org/10.1103/PhysRevD.89.055001}{\emph{Phys. Rev. D}
  {\bfseries 89} (2014) 055001}
  [\href{https://arxiv.org/abs/1401.0018}{{\ttfamily 1401.0018}}].

\bibitem{Mohapatra:2019qid}
R.~N. Mohapatra, G.~Yan and Y.~Zhang, \emph{{Ameliorating Higgs induced flavor
  constraints on TeV scale $W_R$}},
  \href{https://doi.org/10.1016/j.nuclphysb.2019.114764}{\emph{Nucl. Phys. B}
  {\bfseries 948} (2019) 114764}
  [\href{https://arxiv.org/abs/1902.08601}{{\ttfamily 1902.08601}}].

\bibitem{Witten:1982fp}
E.~Witten, \emph{{An SU(2) Anomaly}},
  \href{https://doi.org/10.1016/0370-2693(82)90728-6}{\emph{Phys. Lett.}
  {\bfseries B117} (1982) 324}.

\bibitem{Griest:1990kh}
K.~Griest and D.~Seckel, \emph{{Three exceptions in the calculation of relic
  abundances}}, \href{https://doi.org/10.1103/PhysRevD.43.3191}{\emph{Phys.
  Rev. D} {\bfseries 43} (1991) 3191}.

\bibitem{Babu:1988mw}
K.~Babu and R.~N. Mohapatra, \emph{{{CP} Violation in Seesaw Models of Quark
  Masses}}, \href{https://doi.org/10.1103/PhysRevLett.62.1079}{\emph{Phys. Rev.
  Lett.} {\bfseries 62} (1989) 1079}.

\bibitem{Babu:1989rb}
K.~Babu and R.~N. Mohapatra, \emph{{A Solution to the Strong {CP} Problem
  Without an Axion}},
  \href{https://doi.org/10.1103/PhysRevD.41.1286}{\emph{Phys. Rev. D}
  {\bfseries 41} (1990) 1286}.

\bibitem{Hall:2018let}
L.~J. Hall and K.~Harigaya, \emph{{Implications of Higgs Discovery for the
  Strong CP Problem and Unification}},
  \href{https://doi.org/10.1007/JHEP10(2018)130}{\emph{JHEP} {\bfseries 10}
  (2018) 130} [\href{https://arxiv.org/abs/1803.08119}{{\ttfamily
  1803.08119}}].

\bibitem{Kuchimanchi:1995rp}
R.~Kuchimanchi, \emph{{Solution to the strong CP problem: Supersymmetry with
  parity}}, \href{https://doi.org/10.1103/PhysRevLett.76.3486}{\emph{Phys. Rev.
  Lett.} {\bfseries 76} (1996) 3486}
  [\href{https://arxiv.org/abs/hep-ph/9511376}{{\ttfamily hep-ph/9511376}}].

\bibitem{Mohapatra:1995xd}
R.~N. Mohapatra and A.~Rasin, \emph{{Simple supersymmetric solution to the
  strong CP problem}},
  \href{https://doi.org/10.1103/PhysRevLett.76.3490}{\emph{Phys. Rev. Lett.}
  {\bfseries 76} (1996) 3490}
  [\href{https://arxiv.org/abs/hep-ph/9511391}{{\ttfamily hep-ph/9511391}}].

\bibitem{Arkani-Hamed:2020yna}
N.~Arkani-Hamed, R.~Tito~D'agnolo and H.~D. Kim, \emph{{The Weak Scale as a
  Trigger}},  \href{https://arxiv.org/abs/2012.04652}{{\ttfamily 2012.04652}}.

\bibitem{Dolan:1973qd}
L.~Dolan and R.~Jackiw, \emph{{Symmetry Behavior at Finite Temperature}},
  \href{https://doi.org/10.1103/PhysRevD.9.3320}{\emph{Phys. Rev. D} {\bfseries
  9} (1974) 3320}.

\bibitem{Quiros:1999jp}
M.~Quiros, \emph{{Finite temperature field theory and phase transitions}},  in
  \emph{{ICTP Summer School in High-Energy Physics and Cosmology}},
  pp.~187--259, 1, 1999, \href{https://arxiv.org/abs/hep-ph/9901312}{{\ttfamily
  hep-ph/9901312}}.

\bibitem{Dine:1992wr}
M.~Dine, R.~G. Leigh, P.~Y. Huet, A.~D. Linde and D.~A. Linde, \emph{{Towards
  the theory of the electroweak phase transition}},
  \href{https://doi.org/10.1103/PhysRevD.46.550}{\emph{Phys. Rev.} {\bfseries
  D46} (1992) 550} [\href{https://arxiv.org/abs/hep-ph/9203203}{{\ttfamily
  hep-ph/9203203}}].

\bibitem{Arnold:1992rz}
P.~B. Arnold and O.~Espinosa, \emph{{The Effective potential and first order
  phase transitions: Beyond leading-order}},
  \href{https://doi.org/10.1103/physrevd.50.6662.2,
  10.1103/PhysRevD.47.3546}{\emph{Phys. Rev.} {\bfseries D47} (1993) 3546}
  [\href{https://arxiv.org/abs/hep-ph/9212235}{{\ttfamily hep-ph/9212235}}].

\bibitem{Anderson:1991zb}
G.~W. Anderson and L.~J. Hall, \emph{{The Electroweak phase transition and
  baryogenesis}}, \href{https://doi.org/10.1103/PhysRevD.45.2685}{\emph{Phys.
  Rev.} {\bfseries D45} (1992) 2685}.

\bibitem{Linde:1980ts}
A.~D. Linde, \emph{{Infrared Problem in Thermodynamics of the Yang-Mills Gas}},
  \href{https://doi.org/10.1016/0370-2693(80)90769-8}{\emph{Phys. Lett.}
  {\bfseries 96B} (1980) 289}.

\bibitem{Gross:1980br}
D.~J. Gross, R.~D. Pisarski and L.~G. Yaffe, \emph{{QCD and Instantons at
  Finite Temperature}},
  \href{https://doi.org/10.1103/RevModPhys.53.43}{\emph{Rev. Mod. Phys.}
  {\bfseries 53} (1981) 43}.

\bibitem{Weinberg:1974hy}
S.~Weinberg, \emph{{Gauge and Global Symmetries at High Temperature}},
  \href{https://doi.org/10.1103/PhysRevD.9.3357}{\emph{Phys. Rev. D} {\bfseries
  9} (1974) 3357}.

\bibitem{Arnold:1994bp}
P.~B. Arnold, \emph{{The Electroweak phase transition: Part 1. Review of
  perturbative methods}},  in \emph{{8th International Seminar on High-energy
  Physics (Quarks 94) Vladimir, Russia, May 11-18, 1994}}, pp.~71--86, 1994,
  \href{https://arxiv.org/abs/hep-ph/9410294}{{\ttfamily hep-ph/9410294}}.

\bibitem{Linde:1980tt}
A.~D. Linde, \emph{{Fate of the False Vacuum at Finite Temperature: Theory and
  Applications}},
  \href{https://doi.org/10.1016/0370-2693(81)90281-1}{\emph{Phys. Lett.}
  {\bfseries 100B} (1981) 37}.

\bibitem{Funakubo:2009eg}
K.~Funakubo and E.~Senaha, \emph{{Electroweak phase transition, critical
  bubbles and sphaleron decoupling condition in the MSSM}},
  \href{https://doi.org/10.1103/PhysRevD.79.115024}{\emph{Phys. Rev.}
  {\bfseries D79} (2009) 115024}
  [\href{https://arxiv.org/abs/0905.2022}{{\ttfamily 0905.2022}}].

\bibitem{Fuyuto:2014yia}
K.~Fuyuto and E.~Senaha, \emph{{Improved sphaleron decoupling condition and the
  Higgs coupling constants in the real singlet-extended standard model}},
  \href{https://doi.org/10.1103/PhysRevD.90.015015}{\emph{Phys. Rev. D}
  {\bfseries 90} (2014) 015015}
  [\href{https://arxiv.org/abs/1406.0433}{{\ttfamily 1406.0433}}].

\bibitem{Carrington:1993ng}
M.~E. Carrington and J.~I. Kapusta, \emph{{Dynamics of the electroweak phase
  transition}}, \href{https://doi.org/10.1103/PhysRevD.47.5304}{\emph{Phys.
  Rev. D} {\bfseries 47} (1993) 5304}.

\bibitem{Joyce:1994zn}
M.~Joyce, T.~Prokopec and N.~Turok, \emph{{Nonlocal electroweak baryogenesis.
  Part 1: Thin wall regime}},
  \href{https://doi.org/10.1103/PhysRevD.53.2930}{\emph{Phys. Rev. D}
  {\bfseries 53} (1996) 2930}
  [\href{https://arxiv.org/abs/hep-ph/9410281}{{\ttfamily hep-ph/9410281}}].

\bibitem{Weldon:1982aq}
H.~Weldon, \emph{{Covariant Calculations at Finite Temperature: The
  Relativistic Plasma}},
  \href{https://doi.org/10.1103/PhysRevD.26.1394}{\emph{Phys.\ Rev.\ D}
  {\bfseries 26} (1982) 1394}.

\bibitem{Weldon:1982bn}
H.~A. Weldon, \emph{{Effective Fermion Masses of Order gT in High Temperature
  Gauge Theories with Exact Chiral Invariance}},
  \href{https://doi.org/10.1103/PhysRevD.26.2789}{\emph{Phys. Rev.} {\bfseries
  D26} (1982) 2789}.

\bibitem{Moore:1997sn}
G.~D. Moore, C.-r. Hu and B.~Muller, \emph{{Chern-Simons number diffusion with
  hard thermal loops}},
  \href{https://doi.org/10.1103/PhysRevD.58.045001}{\emph{Phys. Rev. D}
  {\bfseries 58} (1998) 045001}
  [\href{https://arxiv.org/abs/hep-ph/9710436}{{\ttfamily hep-ph/9710436}}].

\bibitem{Bodeker:1999gx}
D.~Bodeker, G.~D. Moore and K.~Rummukainen, \emph{{Chern-Simons number
  diffusion and hard thermal loops on the lattice}},
  \href{https://doi.org/10.1103/PhysRevD.61.056003}{\emph{Phys. Rev. D}
  {\bfseries 61} (2000) 056003}
  [\href{https://arxiv.org/abs/hep-ph/9907545}{{\ttfamily hep-ph/9907545}}].

\bibitem{Moore:2000ara}
G.~D. Moore, \emph{{Do we understand the sphaleron rate?}},  in \emph{{Strong
  and electroweak matter. Proceedings, Meeting, SEWM 2000, Marseille, France,
  June 13-17, 2000}}, pp.~82--94, 6, 2000,
  \href{https://arxiv.org/abs/hep-ph/0009161}{{\ttfamily hep-ph/0009161}},
  \href{https://doi.org/10.1142/9789812799913\_0007}{DOI}.

\bibitem{DOnofrio:2014rug}
M.~D'Onofrio, K.~Rummukainen and A.~Tranberg, \emph{{Sphaleron Rate in the
  Minimal Standard Model}},
  \href{https://doi.org/10.1103/PhysRevLett.113.141602}{\emph{Phys. Rev. Lett.}
  {\bfseries 113} (2014) 141602}
  [\href{https://arxiv.org/abs/1404.3565}{{\ttfamily 1404.3565}}].

\bibitem{Harvey:1990qw}
J.~A. Harvey and M.~S. Turner, \emph{{Cosmological baryon and lepton number in
  the presence of electroweak fermion number violation}},
  \href{https://doi.org/10.1103/PhysRevD.42.3344}{\emph{Phys. Rev. D}
  {\bfseries 42} (1990) 3344}.

\bibitem{Weldon:1989ys}
H.~A. Weldon, \emph{{Dynamical Holes in the Quark - Gluon Plasma}},
  \href{https://doi.org/10.1103/PhysRevD.40.2410}{\emph{Phys. Rev. D}
  {\bfseries 40} (1989) 2410}.

\bibitem{Braaten:1992gd}
E.~Braaten and R.~D. Pisarski, \emph{{Calculation of the quark damping rate in
  hot QCD}}, \href{https://doi.org/10.1103/PhysRevD.46.1829}{\emph{Phys. Rev.}
  {\bfseries D46} (1992) 1829}.

\bibitem{Petitgirard:1991mf}
E.~Petitgirard, \emph{{Massive fermion dispersion relation at finite
  temperature}}, \href{https://doi.org/10.1007/BF01559497}{\emph{Z. Phys. C}
  {\bfseries 54} (1992) 673}.

\bibitem{Moore:1995si}
G.~D. Moore and T.~Prokopec, \emph{{How fast can the wall move? A Study of the
  electroweak phase transition dynamics}},
  \href{https://doi.org/10.1103/PhysRevD.52.7182}{\emph{Phys. Rev. D}
  {\bfseries 52} (1995) 7182}
  [\href{https://arxiv.org/abs/hep-ph/9506475}{{\ttfamily hep-ph/9506475}}].

\bibitem{Moore:1995ua}
G.~D. Moore and T.~Prokopec, \emph{{Bubble wall velocity in a first order
  electroweak phase transition}},
  \href{https://doi.org/10.1103/PhysRevLett.75.777}{\emph{Phys. Rev. Lett.}
  {\bfseries 75} (1995) 777}
  [\href{https://arxiv.org/abs/hep-ph/9503296}{{\ttfamily hep-ph/9503296}}].

\bibitem{Konstandin:2014zta}
T.~Konstandin, G.~Nardini and I.~Rues, \emph{{From Boltzmann equations to
  steady wall velocities}},
  \href{https://doi.org/10.1088/1475-7516/2014/09/028}{\emph{JCAP} {\bfseries
  09} (2014) 028} [\href{https://arxiv.org/abs/1407.3132}{{\ttfamily
  1407.3132}}].

\bibitem{Kozaczuk:2015owa}
J.~Kozaczuk, \emph{{Bubble Expansion and the Viability of Singlet-Driven
  Electroweak Baryogenesis}},
  \href{https://doi.org/10.1007/JHEP10(2015)135}{\emph{JHEP} {\bfseries 10}
  (2015) 135} [\href{https://arxiv.org/abs/1506.04741}{{\ttfamily
  1506.04741}}].

\bibitem{Tanabashi:2018oca}
{\scshape Particle Data Group} collaboration, M.~Tanabashi et~al.,
  \emph{{Review of Particle Physics}},
  \href{https://doi.org/10.1103/PhysRevD.98.030001}{\emph{Phys. Rev. D}
  {\bfseries 98} (2018) 030001}.

\bibitem{Salvioni:2009mt}
E.~Salvioni, G.~Villadoro and F.~Zwirner, \emph{{Minimal Z-prime models:
  Present bounds and early LHC reach}},
  \href{https://doi.org/10.1088/1126-6708/2009/11/068}{\emph{JHEP} {\bfseries
  11} (2009) 068} [\href{https://arxiv.org/abs/0909.1320}{{\ttfamily
  0909.1320}}].

\bibitem{Aaboud:2017buh}
{\scshape ATLAS} collaboration, M.~Aaboud et~al., \emph{{Search for new
  high-mass phenomena in the dilepton final state using 36 fb$^{−1}$ of
  proton-proton collision data at $ \sqrt{s}=13 $ TeV with the ATLAS
  detector}}, \href{https://doi.org/10.1007/JHEP10(2017)182}{\emph{JHEP}
  {\bfseries 10} (2017) 182}
  [\href{https://arxiv.org/abs/1707.02424}{{\ttfamily 1707.02424}}].

\bibitem{Aad:2019fac}
{\scshape ATLAS} collaboration, G.~Aad et~al., \emph{{Search for high-mass
  dilepton resonances using 139 fb$^{-1}$ of $pp$ collision data collected at
  $\sqrt{s}=$13 TeV with the ATLAS detector}},
  \href{https://doi.org/10.1016/j.physletb.2019.07.016}{\emph{Phys. Lett. B}
  {\bfseries 796} (2019) 68}
  [\href{https://arxiv.org/abs/1903.06248}{{\ttfamily 1903.06248}}].

\bibitem{Aad:2019wvl}
{\scshape ATLAS} collaboration, G.~Aad et~al., \emph{{Search for a heavy
  charged boson in events with a charged lepton and missing transverse momentum
  from $pp$ collisions at $\sqrt{s} = 13$ TeV with the ATLAS detector}},
  \href{https://doi.org/10.1103/PhysRevD.100.052013}{\emph{Phys. Rev. D}
  {\bfseries 100} (2019) 052013}
  [\href{https://arxiv.org/abs/1906.05609}{{\ttfamily 1906.05609}}].

\bibitem{ATL-PHYS-PUB-2018-044}
{\scshape ATLAS Collaboration} collaboration, \emph{{Prospects for searches for
  heavy $Z^\prime$ and $W^\prime$ bosons in fermionic final states with the
  ATLAS experiment at the HL-LHC}},  Tech. Rep. ATL-PHYS-PUB-2018-044, CERN,
  Geneva, Dec, 2018.

\bibitem{Achard:2001qw}
{\scshape L3} collaboration, P.~Achard et~al., \emph{{Search for heavy neutral
  and charged leptons in $e^{+} e^{-}$ annihilation at LEP}},
  \href{https://doi.org/10.1016/S0370-2693(01)01005-X}{\emph{Phys. Lett. B}
  {\bfseries 517} (2001) 75}
  [\href{https://arxiv.org/abs/hep-ex/0107015}{{\ttfamily hep-ex/0107015}}].

\bibitem{Sirunyan:2019ofn}
{\scshape CMS} collaboration, A.~M. Sirunyan et~al., \emph{{Search for
  vector-like leptons in multilepton final states in proton-proton collisions
  at $\sqrt{s}$ = 13 TeV}},
  \href{https://doi.org/10.1103/PhysRevD.100.052003}{\emph{Phys. Rev. D}
  {\bfseries 100} (2019) 052003}
  [\href{https://arxiv.org/abs/1905.10853}{{\ttfamily 1905.10853}}].

\bibitem{Bhattiprolu:2019vdu}
P.~N. Bhattiprolu and S.~P. Martin, \emph{{Prospects for vectorlike leptons at
  future proton-proton colliders}},
  \href{https://doi.org/10.1103/PhysRevD.100.015033}{\emph{Phys. Rev. D}
  {\bfseries 100} (2019) 015033}
  [\href{https://arxiv.org/abs/1905.00498}{{\ttfamily 1905.00498}}].

\bibitem{Abdullah:2016avr}
M.~Abdullah, J.~L. Feng, S.~Iwamoto and B.~Lillard, \emph{{Heavy bino dark
  matter and collider signals in the MSSM with vectorlike fourth-generation
  particles}}, \href{https://doi.org/10.1103/PhysRevD.94.095018}{\emph{Phys.
  Rev. D} {\bfseries 94} (2016) 095018}
  [\href{https://arxiv.org/abs/1608.00283}{{\ttfamily 1608.00283}}].

\bibitem{Aaboud:2019trc}
{\scshape ATLAS} collaboration, M.~Aaboud et~al., \emph{{Search for heavy
  charged long-lived particles in the ATLAS detector in 36.1 fb$^{-1}$ of
  proton-proton collision data at $\sqrt{s} = 13$ TeV}},
  \href{https://doi.org/10.1103/PhysRevD.99.092007}{\emph{Phys. Rev. D}
  {\bfseries 99} (2019) 092007}
  [\href{https://arxiv.org/abs/1902.01636}{{\ttfamily 1902.01636}}].

\bibitem{Peskin:1991sw}
M.~E. Peskin and T.~Takeuchi, \emph{{Estimation of oblique electroweak
  corrections}}, \href{https://doi.org/10.1103/PhysRevD.46.381}{\emph{Phys.
  Rev. D} {\bfseries 46} (1992) 381}.

\bibitem{Skiba:2010xn}
W.~Skiba, \emph{{Effective Field Theory and Precision Electroweak
  Measurements}},  in \emph{{Theoretical Advanced Study Institute in Elementary
  Particle Physics}: {Physics of the Large and the Small}}, pp.~5--70, 2011,
  \href{https://arxiv.org/abs/1006.2142}{{\ttfamily 1006.2142}},
  \href{https://doi.org/10.1142/9789814327183\_0001}{DOI}.

\bibitem{an2018improved}
V.~Andreev, D.~G. Ang, D.~DeMille, J.~M. Doyle, G.~Gabrielse, J.~Haefner
  et~al., \emph{Improved limit on the electric dipole moment of the electron},
  \href{https://doi.org/10.1038/s41586-018-0599-8}{\emph{Nature} {\bfseries
  562} (2018) 355}.

\bibitem{Brdar:2019fur}
V.~Brdar, L.~Graf, A.~J. Helmboldt and X.-J. Xu, \emph{{Gravitational Waves as
  a Probe of Left-Right Symmetry Breaking}},
  \href{https://doi.org/10.1088/1475-7516/2019/12/027}{\emph{JCAP} {\bfseries
  12} (2019) 027} [\href{https://arxiv.org/abs/1909.02018}{{\ttfamily
  1909.02018}}].

\bibitem{Caprini:2015zlo}
C.~Caprini et~al., \emph{{Science with the space-based interferometer eLISA.
  II: Gravitational waves from cosmological phase transitions}},
  \href{https://doi.org/10.1088/1475-7516/2016/04/001}{\emph{JCAP} {\bfseries
  04} (2016) 001} [\href{https://arxiv.org/abs/1512.06239}{{\ttfamily
  1512.06239}}].

\bibitem{Fujikura:2018duw}
K.~Fujikura, K.~Kamada, Y.~Nakai and M.~Yamaguchi, \emph{{Phase Transitions in
  Twin Higgs Models}},
  \href{https://doi.org/10.1007/JHEP12(2018)018}{\emph{JHEP} {\bfseries 12}
  (2018) 018} [\href{https://arxiv.org/abs/1810.00574}{{\ttfamily
  1810.00574}}].

\bibitem{Caprini:2019egz}
C.~Caprini et~al., \emph{{Detecting gravitational waves from cosmological phase
  transitions with LISA: an update}},
  \href{https://doi.org/10.1088/1475-7516/2020/03/024}{\emph{JCAP} {\bfseries
  03} (2020) 024} [\href{https://arxiv.org/abs/1910.13125}{{\ttfamily
  1910.13125}}].

\bibitem{Turner:1990rc}
M.~S. Turner and F.~Wilczek, \emph{{Relic gravitational waves and extended
  inflation}}, \href{https://doi.org/10.1103/PhysRevLett.65.3080}{\emph{Phys.
  Rev. Lett.} {\bfseries 65} (1990) 3080}.

\bibitem{Kosowsky:1991ua}
A.~Kosowsky, M.~S. Turner and R.~Watkins, \emph{{Gravitational radiation from
  colliding vacuum bubbles}},
  \href{https://doi.org/10.1103/PhysRevD.45.4514}{\emph{Phys. Rev.} {\bfseries
  D45} (1992) 4514}.

\bibitem{Turner:1992tz}
M.~S. Turner, E.~J. Weinberg and L.~M. Widrow, \emph{{Bubble nucleation in
  first order inflation and other cosmological phase transitions}},
  \href{https://doi.org/10.1103/PhysRevD.46.2384}{\emph{Phys. Rev.} {\bfseries
  D46} (1992) 2384}.

\bibitem{Kosowsky:1992vn}
A.~Kosowsky and M.~S. Turner, \emph{{Gravitational radiation from colliding
  vacuum bubbles: envelope approximation to many bubble collisions}},
  \href{https://doi.org/10.1103/PhysRevD.47.4372}{\emph{Phys. Rev.} {\bfseries
  D47} (1993) 4372} [\href{https://arxiv.org/abs/astro-ph/9211004}{{\ttfamily
  astro-ph/9211004}}].

\bibitem{Jinno:2016vai}
R.~Jinno and M.~Takimoto, \emph{{Gravitational waves from bubble collisions: An
  analytic derivation}},
  \href{https://doi.org/10.1103/PhysRevD.95.024009}{\emph{Phys. Rev.}
  {\bfseries D95} (2017) 024009}
  [\href{https://arxiv.org/abs/1605.01403}{{\ttfamily 1605.01403}}].

\bibitem{Jinno:2017fby}
R.~Jinno and M.~Takimoto, \emph{{Gravitational waves from bubble dynamics:
  Beyond the Envelope}},
  \href{https://doi.org/10.1088/1475-7516/2019/01/060}{\emph{JCAP} {\bfseries
  1901} (2019) 060} [\href{https://arxiv.org/abs/1707.03111}{{\ttfamily
  1707.03111}}].

\bibitem{Hindmarsh:2015qta}
M.~Hindmarsh, S.~J. Huber, K.~Rummukainen and D.~J. Weir, \emph{{Numerical
  simulations of acoustically generated gravitational waves at a first order
  phase transition}},
  \href{https://doi.org/10.1103/PhysRevD.92.123009}{\emph{Phys. Rev.}
  {\bfseries D92} (2015) 123009}
  [\href{https://arxiv.org/abs/1504.03291}{{\ttfamily 1504.03291}}].

\bibitem{Bodeker:2009qy}
D.~Bodeker and G.~D. Moore, \emph{{Can electroweak bubble walls run away?}},
  \href{https://doi.org/10.1088/1475-7516/2009/05/009}{\emph{JCAP} {\bfseries
  0905} (2009) 009} [\href{https://arxiv.org/abs/0903.4099}{{\ttfamily
  0903.4099}}].

\bibitem{Bodeker:2017cim}
D.~Bodeker and G.~D. Moore, \emph{{Electroweak Bubble Wall Speed Limit}},
  \href{https://doi.org/10.1088/1475-7516/2017/05/025}{\emph{JCAP} {\bfseries
  1705} (2017) 025} [\href{https://arxiv.org/abs/1703.08215}{{\ttfamily
  1703.08215}}].

\bibitem{Ellis:2019oqb}
J.~Ellis, M.~Lewicki, J.~M. No and V.~Vaskonen, \emph{{Gravitational wave
  energy budget in strongly supercooled phase transitions}},
  \href{https://doi.org/10.1088/1475-7516/2019/06/024}{\emph{JCAP} {\bfseries
  1906} (2019) 024} [\href{https://arxiv.org/abs/1903.09642}{{\ttfamily
  1903.09642}}].

\bibitem{Mancha:2020fzw}
M.~Barroso~Mancha, T.~Prokopec and B.~Swiezewska, \emph{{Field-theoretic
  derivation of bubble-wall force}},
  \href{https://doi.org/10.1007/JHEP01(2021)070}{\emph{JHEP} {\bfseries 01}
  (2021) 070} [\href{https://arxiv.org/abs/2005.10875}{{\ttfamily
  2005.10875}}].

\bibitem{Hoeche:2020rsg}
S.~H\"oche, J.~Kozaczuk, A.~J. Long, J.~Turner and Y.~Wang, \emph{{Towards an
  all-orders calculation of the electroweak bubble wall velocity}},
  \href{https://doi.org/10.1088/1475-7516/2021/03/009}{\emph{JCAP} {\bfseries
  03} (2021) 009} [\href{https://arxiv.org/abs/2007.10343}{{\ttfamily
  2007.10343}}].

\bibitem{Vanvlasselaer:2020niz}
A.~Azatov and M.~Vanvlasselaer, \emph{{Bubble wall velocity: heavy physics
  effects}}, \href{https://doi.org/10.1088/1475-7516/2021/01/058}{\emph{JCAP}
  {\bfseries 01} (2021) 058}
  [\href{https://arxiv.org/abs/2010.02590}{{\ttfamily 2010.02590}}].

\bibitem{Espinosa:2010hh}
J.~R. Espinosa, T.~Konstandin, J.~M. No and G.~Servant, \emph{{Energy Budget of
  Cosmological First-order Phase Transitions}},
  \href{https://doi.org/10.1088/1475-7516/2010/06/028}{\emph{JCAP} {\bfseries
  1006} (2010) 028} [\href{https://arxiv.org/abs/1004.4187}{{\ttfamily
  1004.4187}}].

\bibitem{Ellis:2018mja}
J.~Ellis, M.~Lewicki and J.~M. No, \emph{{On the Maximal Strength of a
  First-Order Electroweak Phase Transition and its Gravitational Wave Signal}},
  \href{https://doi.org/10.1088/1475-7516/2019/04/003}{\emph{JCAP} {\bfseries
  04} (2019) 003} [\href{https://arxiv.org/abs/1809.08242}{{\ttfamily
  1809.08242}}].

\bibitem{Cutting:2019zws}
D.~Cutting, M.~Hindmarsh and D.~J. Weir, \emph{{Vorticity, kinetic energy, and
  suppressed gravitational wave production in strong first order phase
  transitions}},
  \href{https://doi.org/10.1103/PhysRevLett.125.021302}{\emph{Phys. Rev. Lett.}
  {\bfseries 125} (2020) 021302}
  [\href{https://arxiv.org/abs/1906.00480}{{\ttfamily 1906.00480}}].

\bibitem{Ellis:2020awk}
J.~Ellis, M.~Lewicki and J.~M. No, \emph{{Gravitational waves from first-order
  cosmological phase transitions: lifetime of the sound wave source}},
  \href{https://doi.org/10.1088/1475-7516/2020/07/050}{\emph{JCAP} {\bfseries
  07} (2020) 050} [\href{https://arxiv.org/abs/2003.07360}{{\ttfamily
  2003.07360}}].

\bibitem{Caprini:2009yp}
C.~Caprini, R.~Durrer and G.~Servant, \emph{{The stochastic gravitational wave
  background from turbulence and magnetic fields generated by a first-order
  phase transition}},
  \href{https://doi.org/10.1088/1475-7516/2009/12/024}{\emph{JCAP} {\bfseries
  0912} (2009) 024} [\href{https://arxiv.org/abs/0909.0622}{{\ttfamily
  0909.0622}}].

\bibitem{Binetruy:2012ze}
P.~Binetruy, A.~Bohe, C.~Caprini and J.-F. Dufaux, \emph{{Cosmological
  Backgrounds of Gravitational Waves and eLISA/NGO: Phase Transitions, Cosmic
  Strings and Other Sources}},
  \href{https://doi.org/10.1088/1475-7516/2012/06/027}{\emph{JCAP} {\bfseries
  1206} (2012) 027} [\href{https://arxiv.org/abs/1201.0983}{{\ttfamily
  1201.0983}}].

\bibitem{Klein:2015hvg}
A.~Klein et~al., \emph{{Science with the space-based interferometer eLISA:
  Supermassive black hole binaries}},
  \href{https://doi.org/10.1103/PhysRevD.93.024003}{\emph{Phys. Rev. D}
  {\bfseries 93} (2016) 024003}
  [\href{https://arxiv.org/abs/1511.05581}{{\ttfamily 1511.05581}}].

\bibitem{Yagi:2011wg}
K.~Yagi and N.~Seto, \emph{{Detector configuration of DECIGO/BBO and
  identification of cosmological neutron-star binaries}},
  \href{https://doi.org/10.1103/PhysRevD.83.044011}{\emph{Phys. Rev. D}
  {\bfseries 83} (2011) 044011}
  [\href{https://arxiv.org/abs/1101.3940}{{\ttfamily 1101.3940}}].

\bibitem{Kudoh:2005as}
H.~Kudoh, A.~Taruya, T.~Hiramatsu and Y.~Himemoto, \emph{{Detecting a
  gravitational-wave background with next-generation space interferometers}},
  \href{https://doi.org/10.1103/PhysRevD.73.064006}{\emph{Phys. Rev. D}
  {\bfseries 73} (2006) 064006}
  [\href{https://arxiv.org/abs/gr-qc/0511145}{{\ttfamily gr-qc/0511145}}].

\bibitem{Kuroyanagi:2014qaa}
S.~Kuroyanagi, S.~Tsujikawa, T.~Chiba and N.~Sugiyama, \emph{{Implications of
  the B-mode Polarization Measurement for Direct Detection of Inflationary
  Gravitational Waves}},
  \href{https://doi.org/10.1103/PhysRevD.90.063513}{\emph{Phys. Rev. D}
  {\bfseries 90} (2014) 063513}
  [\href{https://arxiv.org/abs/1406.1369}{{\ttfamily 1406.1369}}].

\end{thebibliography}\endgroup
%%%%%%%%%%%%%%%%%%%%%%%%%%%%%%%%%%%%%%%%%%%%%%%%

\end{document}